\documentclass[aps,rmp,reprint,amsmath,amssymb,graphicx,floatfix,longbibliography,superscriptaddress]{revtex4-1}

\usepackage{graphicx}
\usepackage{color}
\usepackage{bm}
\usepackage{amsmath}
\usepackage{subfigure}
\usepackage{epstopdf}
\usepackage{verbatim}

\newcommand{\bir}{\affiliation{School of Physics and Astronomy, University of Birmingham, Birmingham
B15 2TT, UK}}

\newcommand{\rcnp}{\affiliation{Research Center for Nuclear Physics (RCNP), Osaka University, Osaka 567-0047, Japan}}

\newcommand{\iias}{\affiliation{International Institute for Advanced Studies, Kizugawa 619-0225, Japan}}

\newcommand{\kyoto}{\affiliation{Department of Physics, Kyoto University, Kyoto 606-8502, Japan}}

\newcommand{\msu}{\affiliation{Facility for Rare Isotope Beams and Department~of~Physics and Astronomy, Michigan State University, East Lansing, MI~48824, USA}}

\newcommand{\ncsu}{\affiliation{Department~of~Physics, North~Carolina~State~University,
Raleigh, NC~27695, USA}}

\newcommand{\bonn}{\affiliation{Helmholtz-Institut f\"ur Strahlen- und Kernphysik (Theorie) and Bethe Center for Theoretical Physics,
 Universit\"at Bonn, D-53115 Bonn, Germany}}

\newcommand{\fzj}{\affiliation{Forschungszentrum J\"ulich, Institute for Advanced Simulation (IAS-4),
Institut f\"ur Kernphysik (IKP-3), J\"ulich Center for Hadron Physics and JARA-HPC, D-52425 J\"ulich, Germany}}

\begin{document}

\title{Microscopic Clustering in Light Nuclei}

\author{Martin Freer}\email{m.freer@bham.ac.uk }\bir
%%\affiliation{School of Physics and Astronomy, University of Birmingham, Birmingham
%B15 2TT UK}

\author{Hisashi Horiuchi}\email{horiuchi@rcnp.osaka-u.ac.jp}\rcnp\iias
%\affiliation{Research Center for Nuclear Physics (RCNP), Osaka University, Osaka 567-0047, Japan}
%\affiliation{International Institute for Advanced Studies, Kizugawa 619-0225, Japan}

\author{Yoshiko Kanada-En'yo}\email{yenyo@ruby.scphys.kyoto-u.ac.jp}\kyoto
%\affiliation{Department of Physics, Kyoto University, Kyoto 606-8502, Japan}

\author{Dean Lee}\email{leed@frib.msu.edu}\msu\ncsu
%\affiliation{Department~of~Physics, North~Carolina~State~University, Raleigh, NC~27695, USA}

\author{Ulf-G. Mei{\ss}ner}\email{meissner@hiskp.uni-bonn.de}\bonn\fzj
%\affiliation{Helmholtz-Institut f\"ur Strahlen- und Kernphysik (Theorie)
%and Bethe Center for Theoretical Physics,
% Universit\"at Bonn, D-53115 Bonn, Germany}
%\affiliation{Forschungszentrum J\"ulich, Institute for Advanced Simulation (IAS-4),
%Institut f\"ur Kernphysik (IKP-3), J\"ulich Center for Hadron Physics, JARA-HPC and
%JARA-FAME,
%             D-52425 J\"ulich, Germany}

%\thispagestyle{empty}

\begin{abstract}
We review recent experimental and theoretical progress in understanding the microscopic details of clustering in light nuclei.  We discuss recent experimental results on $\alpha$-conjugate systems, molecular structures in neutron-rich nuclei, and constraints for {\it ab initio} theory.  We then examine nuclear clustering in a wide range of theoretical methods, including the resonating group and generator coordinate methods, antisymmetrized molecular dynamics, Tohsaki-Horiuchi-Schuck-R{\"o}pke
wave function and container model, no-core shell model methods, continuum quantum Monte Carlo, and lattice effective field theory.
\end{abstract}

\maketitle

%\pagebreak

\tableofcontents

%%%%%%%%%%%%%%%%%%%%%%%%%%%%%%%%%%%%%%%%%%%%%%%%%%%%%%%%%%%%%%%%%%%%%%%
%
% \input{introduction_section.tex}

\section{Introduction}
\def\theequation{\arabic{section}.\arabic{equation}}

Nuclear clustering describes the emergence of molecular-like structures in nuclear
physics.  In molecules there is a rich phenomenology of different chemical
bonds, complex rotational and vibrational excitations, and intricate structural
geometries.  Could there be a similar level of complexity in nuclear
systems? The possibilities are certainly there with strong binding among the four nucleons in an $\alpha$-particle and the consequences of a nearly bound di-neutron channel.  However, the underlying physics is made more challenging by the
democracy of particles involved in nuclear binding.  Instead of heavy ions
surrounded by light electrons, the protons and neutrons have nearly equal
masses, and the clustering structures emerge from a delicate balance among
repulsive short-range forces and Pauli blocking effects, attractive medium-range
nuclear forces, and long-range Coulomb repulsion among protons.

The study of nuclear clustering really began with Rutherford's discovery by
alpha radiation \cite{Rutherford:1899} and  the development of quantum mechanics.
Gamow \cite{Gamow:1928} and, independently, Gurney and Condon \cite{Gurney:1928}
described the $\alpha$-particle as undergoing quantum-mechanical tunneling
from inside the decaying nucleus. About a decade later, Wheeler \cite{Wheeler:1937zz}
developed the resonating group method to describe $\alpha$-clusters and other
cluster groupings within nuclei, while allowing protons and neutrons to maintain
their fermionic quantum statistics. Afterwards came the work of Hafstad and
Teller, which described even-even $N=Z$ nuclei in terms of an $\alpha$-particle
model with bonds connecting clusters \cite{Hafstad:1938}.   Following along
the same lines, Dennsion proposed a model of the low-lying states $^{16}$O
in terms of four  $\alpha$-clusters at the vertices of a regular tetrahedron
\cite{Dennison:1940,Dennison:1954zz}. At a more microscopic level, Margenau
used a Slater determinant wave function for $\alpha$-clusters to compute
an effective $\alpha$-$\alpha$ interaction \cite{Margenau:1941}.

Some years later, Morinaga suggested that non-spherical and even linear chains
of $\alpha$-clusters could describe some states of $\alpha$-like nuclei \cite{Morinaga:1956zza}.
One of the candidates for such a description was the second $0^+$ state of
$^{12}$C postulated by Hoyle \cite{Hoyle:1954zz} as responsible for enhancing
the triple-$\alpha$ reaction in stars and experimentally observed soon after \cite{Cook:1957}.
 Concurrent with these theoretical developments,  new experiments provided
high-quality data on elastic $\alpha$-$\alpha$ scattering \cite{Heydenburg:1956zza,Nilson:1958zz,Afzal:1969zz}.
This is in turn led to  the development of an effective $\alpha$-$\alpha$
interaction \cite{Ali:1966}.

At around the same time, Brink used Margneau's Slater determinant wave function
for the $\alpha$-cluster and the generator coordinate method to simplify calculations
that were difficult in the more general formalism of the resonating group
method \cite{Brink:1966}.  The equivalence of the generator coordinate method
and resonating group method was later clarified by Horiuchi \cite{Horiuchi:1970}.
On the topic of $\alpha$-decays, Clark and Wang computed the probability of
$\alpha$-clusters to form near the surface of heavy nuclei \cite{Clark:1966}.
 Meanwhile Ikeda, Takigawa, and Horiuchi noticed that $\alpha$-clustering appeared close to $\alpha$-decay thresholds, and these were denoted
schematically with the so-called Ikeda diagrams \cite{Ikeda:1968}.  Following these
same concepts, the study of clustering has been extended to proton-rich and neutron-rich systems with nearby open thresholds.
The  corresponding states are weakly-bound systems of clusters and excess neutrons or protons.

There have been a number of reviews on clustering in nuclei \cite{Akaishi:1986gm,vonOertzen:2006a,Freer:2007,Beck:2010a,Beck:2012a,Horiuchi:2012a,Beck:2014a,Funaki:2015uya}.
The purpose of this review is to give a broad overview of the exciting developments
in the past few years. Due to space limitations, it is not possible to cover
all areas of research in depth. Nevertheless, we try to give a balanced view of
the field as seen by a team of practitioners covering a range of methods
and expertise.  In the review of theoretical methods, we
focus on microscopic clustering where clusters emerge from nucleonic
degrees of freedom.  As the field is dynamic and evolving, several  key issues
are not resolved at present, and there are disagreements among different methods.  Furthermore,
some of the most interesting results will likely come in the
near future.  This is to be expected in a growing field with important
open questions and active research  being pursued by many.

It is useful to briefly summarize the strengths and challenges of the various
theoretical approaches.  Most of the  methods we discuss are variational
calculations using some prescribed ansatz for the nuclear wave function.  
These include antisymmetrized molecular dynamics,
fermionic molecular dynamics, the Tohsaki-Horiuchi-Schuck-R{\"o}pke
wave function and container model,
and microscopic cluster models using the resonating group or generator coordinate methods.  These variational approaches often yield good agreement with experimental data as well as an intuitive picture of the underlying nuclear wave functions.  The main challenges
are to incorporate first principles nuclear forces and remove systematic errors associated with the choice of variational
basis states.

Some variational methods have also been combined with Monte Carlo techniques.  Variational Monte Carlo uses stochastic sampling to compute overlap integrals.  It is also often used as a starting point for diffusion or Green's function\ Monte Carlo simulations.
These calculations have used first principles nuclear forces, and the systematic errors can be estimated by allowing unrestricted evolution of the quantum wave function.  The major challenge for these calculations is that the computational effort increases exponentially with the number of particles. Another method called Monte Carlo shell model
uses auxiliary-field Monte Carlo to select optimized variational basis states.   As with other variational methods, the challenges are systematic errors due to the choice of basis states.

No-core shell model with continuum calculations start from first principles
nuclear forces described by chiral effective field theory and have shown impressive agreement for the continuum properties of light nuclei. Similar to Green's function\ Monte Carlo, the challenge for this method is the exponential scaling of effort when treating larger systems. The symmetry-adapted no-core shell model provides some very promising ideas for mobilizing computational resources in an efficient manner based on symmetries.  Nevertheless difficulties remain in reaching larger systems accurately with first principles nuclear forces.

Nuclear lattice effective field theory uses chiral effective field theory and lattice Monte Carlo techniques to determining nuclear structure, scattering, and reactions. It has the advantage of relatively  mild scaling with system size and a common platform in which to treat few-body and many-body systems at zero and nonzero temperature. However there is the added difficulty of working on a lattice with broken rotational symmetry, and the lattice spacing must be decreased to reduce systematic errors.

We also mention several other recent studies.  In one recent work the states of $^{12}$C are
considered in a Skyrme model \cite{Lau:2014baa}.  While the calculations
produce good agreement with the measured experimental spectrum, the detailed
connection to the underlying nuclear forces is not yet fully realized.
While the inadequacies of the shell model in describing cluster structures have been known since the early years, the explanation of nuclear clustering as an emergent collective phenomenon near open thresholds is provided in Ref.~\cite{Okolowicz:2012kv} by treating the nucleus as an open quantum system coupling through nearby continuum states.  

The review begins with an account of recent experimental results and future directions. We then discuss several theoretical approaches, including the resonating group and generator coordinate methods, antisymmetrized molecular dynamics, Tohsaki-Horiuchi-Schuck-R{\"o}pke wave function and container model, no-core shell model methods, continuum quantum Monte Carlo, and lattice effective field theory.  We then conclude with a summary and outlook for the future.

%%%%%%%%%%%%%%%%%%%%%%%%%%%%%%%%%%%%%%%%%%%%%%%%%%%%%%%%%%%%%%%%%%%%%%%
% \input{RMP_mf_section.tex}
%
%
%

\section{Recent experimental results}
\subsection{Experimental observables}
The experimental study of the role of clustering in nuclei dates back to
the earliest observations of $\alpha$-decay of heavy nuclei. In the early
models of nuclei, it was assumed by many that the $\alpha$-particle may play
an important role, e.g. the paper by Hafstad and Teller in 1938~\cite{Hafstad:1938}
nicely describes the possible structures of nuclei such as $^8$Be, $^{12}$C
and $^{16}$O as constructed from $\alpha$-particles.  This early work also
speculated on the existence of molecular structures in light nuclei, where
neutrons, or even neutron holes, might be exchanged among $\alpha$-particle
cores. These basic ideas remain the drivers for much of the present experimental
program. The ``modern'' era of nuclear clustering was catalyzed by the
ideas of Morinaga in 1956, who had suggested that the 7.65 MeV Hoyle state
in $^{12}$C, which had recently been experimentally measured, might be a
linear arrangement of 3$\alpha$-particles~\cite{Morinaga:1956zza}. The concept
that linear chain structures might exist in nuclei has stuck with the subject
until the present and remains to be resolved. Experiment has been substantially
motivated by the desire to provide evidence for the types of structures envisaged
by Morinaga and those calculated by Brink using the Bloch-Brink Alpha Cluster
Model~\cite{Brink:1967zz,Bri08}. For example, in the case of $^{12}$C, the
$\alpha$-cluster model finds two structures. The first is an equilateral triangular
arrangement which historically has been associated with the ground-state,
and the second is a linear arrangement (or chain).

The ability of experiments to elucidate the cluster structures of light and
heavy nuclei is determined by the range of experimental observables that
may be extracted. { From a simplistic starting point}, the moment of inertia of a rotating
nucleus gives an insight into the deformation which can be at least shown
to be consistent with a cluster structure, even if not direct evidence. If
$^8$Be is used as an example, then the ground-state rotational band has 0$^+$,
2$^+$ and 4$^+$ states at 0, 3.06 and 11.35 MeV. The ratio of the 4$^+$ to
2$^+$ energy is 3.7, very close to that one would expect for a rotational
nucleus, 3.33. The moment of inertia that one extracts from $E_{\rm rot}=J(J+1)\hbar^2/2\mathcal{I}$
is commensurate with that found in {\it ab initio} Green's function Monte
Carlo (GFMC) calculations, which strongly reveal the cluster structure \cite{Wiringa:2000gb}.
We discuss calculations using Green's function Monte Carlo  in subsection
\ref{GFMC}. As a simple guide, the value of $\hbar^2/2\mathcal{I}$ associated with the
2$^+$ state is 0.51~MeV, which even in a simple calculation yields a separation
of two $\alpha$-particles by twice the $\alpha$-particle radius. The observation
of a series of states which lie on a rotational sequence is not watertight
evidence of either clustering or deformation. Here  measurements of electromagnetic
transition strengths provide tests of the overlaps of initial and final-state structures
and the degree of collectivity. For the case of $^8$Be, a measurement
of the $B(E2)$ transition strength from the 4$^+$ to the 2$^+$ state provides
a consistent description with both the rotational picture and the GFMC calculations~\cite{Datar:2013pbd}.

{ However, and as noted above, this simplistic interpretation needs to be treated with care. Firstly, all of the
states in $^8$Be are unbound and hence are embedded in the continuum and hence will have continuum contributions. Second,
the widths of the states are significant (see section~\ref{sec:8Be}), and correspondingly the lifetimes short, and thus an understanding
of what collectivity means on such short timescales is unclear. Finally, many calculations use bound-state approximations and hence cannot be
completely accurate. There is an interesting discussion of the meaning of rotational bands where the resonances are embedded in the continuum, with a focus
on $^8$Be by Garrido \emph{et al.}~\cite{Garrido:2013rta}. The conclusion is that rotational bands embedded in the continuum may still be a meaningful concept, but that the continuum
affects properties such as transition probabilities and hence here the continuum needs to be treated carefully. This is particularly important for the comparison with \emph{ab initio} methods.}

The width of a state reveals a significant amount of detail regarding the
structure and the decay. The greater the overlap of initial structure with
the decay partition then the shorter the lifetime and the greater the width.
In the case of the 2$^+$ excitation of $^8$Be, the width is tabulated as
1.5~MeV. The decay width is also affected by the barrier through which the
decay must proceed, but if the Coulomb and centrifugal barriers are removed,
then the reduced width may be compared with the Wigner limit. This is the
value the reduced width should adopt if the $\alpha$-particles are fully preformed.
For this particular state, it is found that the experimental width is very
close to the Wigner limit, again indicating the existence of the cluster structure~\cite{Overway:1981zdf,Cerny:1974bs}.
A further signature, not available to the decay of the example states in
$^8$Be, is the measurement of the dominant decay channel. States with strong
cluster-like properties should preferentially decay by cluster emission as
opposed to proton or neutron decay, for example. In reactions, this structural
similarity would be described in terms of a spectroscopic factor or an asymptotic
normalization coefficient (ANC).

In the following sections we explore many of the recent developments in the
experimental study of nuclear clustering. In many cases the recent work builds
on significant historical work. There are many review articles which describe
the development of the subject and we refer the reader to the following references:~\cite{Fre97,vonOertzen:2006a,Freer:2007,Beck:2010a,Beck:2012a,Beck:2014a,Freer:2014qoa}.

\subsection{Status of studies of light nuclei}
\subsubsection{Alpha-conjugate systems; $N$-alpha structures and chains}
\label{sec:8Be}
By far the most experimental attention has been devoted to the study of the
cluster structure of $\alpha$-conjugate nuclei. Here the challenges have
been to first provide a deeper insight into the nature of the cluster structures
and ultimately to determine if the chain-states really exist in light nuclei
or not. The eventual aim is to determine experimental characteristics such
that they may be tested against {\it ab initio} or other microscopic calculations.  \\

{\it $^8$Be}\\
{As already described,}
one of the best examples of the comparison between {\it ab initio} theory and experiment, is the measurement of the gamma decay of the 4$^+$ state in $^8$Be to the
2$^+$ state~\cite{Datar:2013pbd}. This was a tour-de-force where a gamma decay
branch of $\sim 10^{-7}$ was observed. The experiment involved the use of
a helium gas-jet target, and the 4$^+$ state was resonantly populated with
a $^4$He beam. The emitted gamma-ray and the subsequent emission of the two
$\alpha$-particles from the decay of the 2$^+$ state were detected in a triple
coincidence. A cross section of 165(54)~nb was observed which translated
to a $B(E2)$ of $25\pm8$ $e^2$fm$^4$. This is remarkably close to the value
most recently calculated in the GFMC approach of $26.0\pm0.6$ $e^2$fm$^4$~\cite{Datar:2013pbd}
(see reference [16] in this paper and \cite{Wiringa:2000gb}). These latter calculations had famously found the ground state of $^8$Be to be highly clustered and predicted with
significant precision the excitation energy spectrum \cite{Wiringa:2000gb}. Given that the $B(E2)$ is
sensitive to both the overlap of the charge distribution and the collective
behavior, such a result could be taken as evidence of both the cluster
and collective behaviors. However, that being the case, this raises a rather
interesting conundrum.\footnote{W. Nazarewicz, private communication at the 2015 Gordon
Research Conference, New Hampshire, USA} The widths of both the 2$^+$ and
4$^+$ states are large (1.5 and 3.5~MeV, respectively). From the uncertainty
principle, these would correspond to lifetimes of the order of 10$^{-22}$~seconds. 
This is the transit time of a nucleon with the Fermi energy to cross
the nucleus. How is it possible for collective processes
to develop and for rotational behavior to occur given the apparent mismatch
in timescales, and what do rotations mean in such systems \cite{Fossez:2015cma}? It is therefore possible
that what is observed experimentally are simply patterns more generally linked to
the underlying symmetry of a dumbbell-like structure. {When it comes to precisely describing the properties
of such states embedded in the continuum, the influence of the continuum on transition properties need to be fully accounted for~\cite{Garrido:2013rta}, and it is
vital that {\it ab initio} methods be developed for such unbound systems.}\\

{\it $^{12}$C ground-state and rotational band}\\

\begin{figure}
\vspace{0cm}
\includegraphics[angle=-0.0,width=0.4\textwidth]{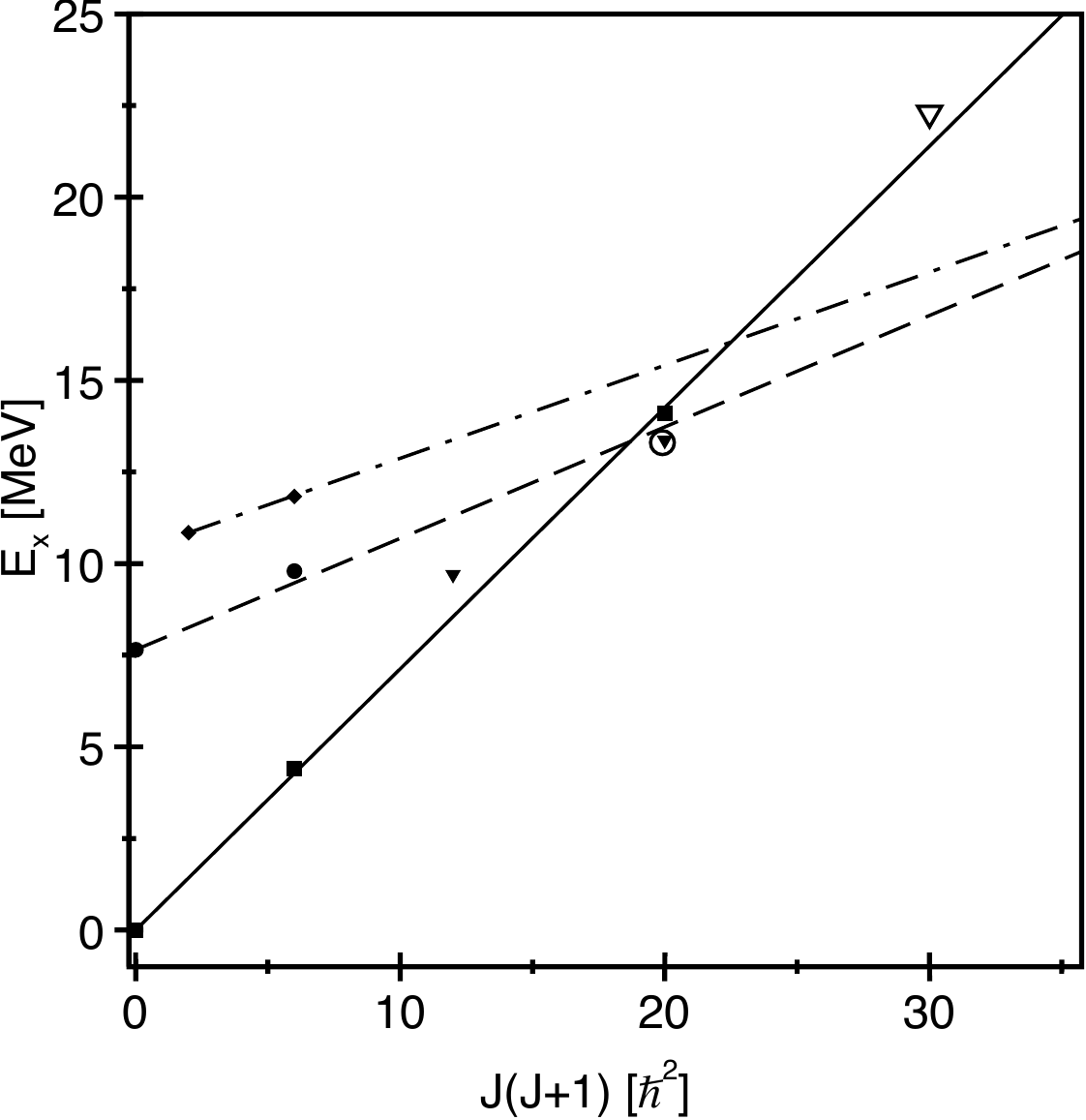}\\
\includegraphics[width=0.4\textwidth]{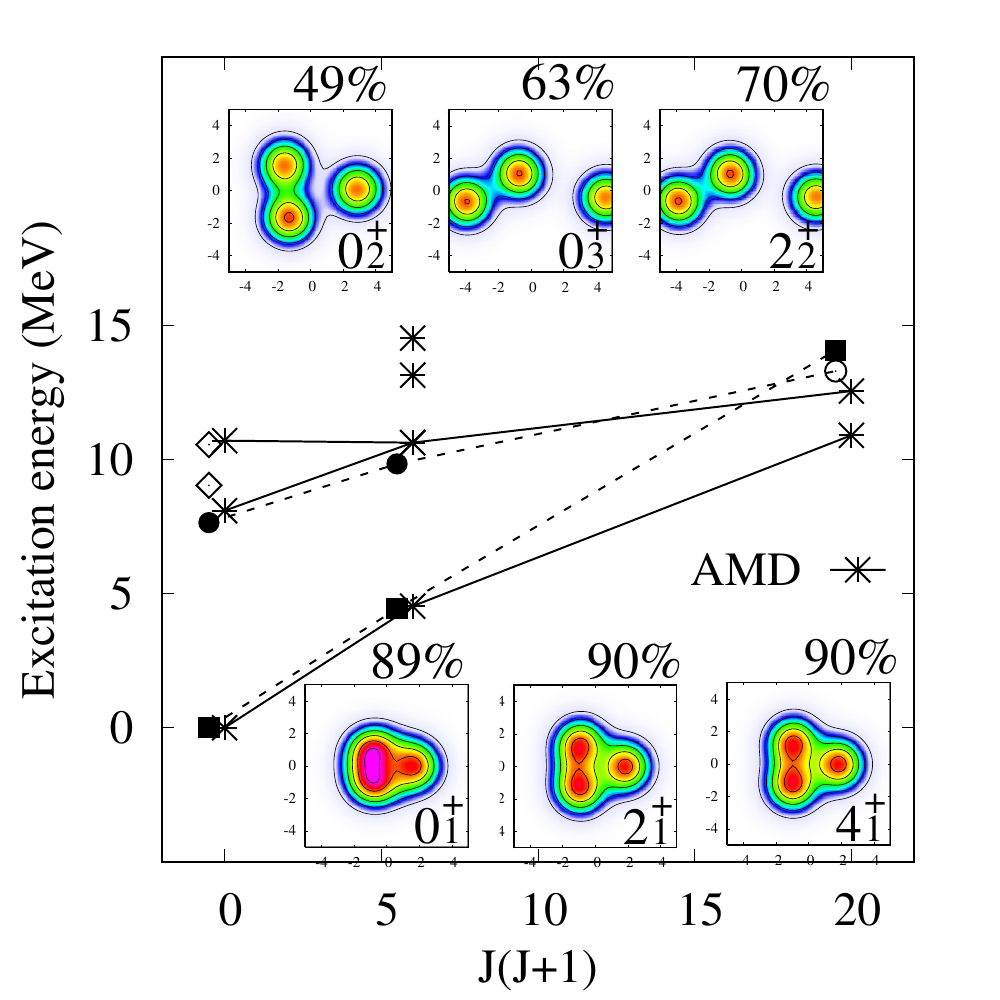}
\caption{\label{fig:exj}(Color online)
(upper) Experimental energy spin systematics of states in $^{12}$C.
Filled
symbols are strong assignments, open symbols are tentative assignments which
are yet to be confirmed. The squares correspond to the ground-state rotational
band, 0$^+$, 2$^+$ and 4$^+$. Triangles are the 3$^-$, 4$^-$ and 5$^-$ states.
Circles are states associated with the Hoyle-band (0$^+$, 2$^+$ and 4$^+$),
and diamonds are the 1$^-$ and 2$^-$ states. The various lines correspond to
best fits to the rotational systematics.
(lower)
Energy levels of the $0^+$, $2^+$, and $4^+$ states  in $^{12}$C and matter density distributions obtained
by antisymmetrized molecular dynamics (AMD) with variation after projection using the MV1 force \cite{KanadaEnyo:2006ze} (asterisk symbols) are compared with
the experimental energy spectra from Ref.~\cite{AjzenbergSelove:1990zh,Freer:2007zz,Itoh:2011zz,Freer:2011zza}.
The intrinsic density distributions are shown together with percentages of the dominant component in the final wave functions.
The $4^+_2$ state has dominantly the same intrinsic component as that of the $2^+_2$ state.
The states with strong $E2$ transitions are connected by solid lines. Dashed lines correspond to the tentative
assignments of experimental levels in the upper panel.}
\end{figure}

Similar questions are pertinent for the next $\alpha$-conjugate system, $^{12}$C. { The effect of the continuum on the rotational bands in $^{12}$C is
discussed in Ref.~\cite{Garrido:2016lqq}. Here the transitions between states are found to be consistent with the rotational picture.}
For $^8$Be all the states lie above the $\alpha$-decay threshold and hence,
by the definition for the emergence of clustering developed by Ikeda, have the ingredients for the formation of clusters~\cite{Ikeda:1968}. However,
the ground state of $^{12}$C lies $\sim 7.3$~MeV below the decay threshold,
and hence the cluster structure would be suppressed. However, as shown in Fig.~\ref{fig:exj}, antisymmetrized
molecular dynamics (AMD) calculations indicate \definecolor{purple2}{rgb}{0.7, 0.0, 1.0}
that
states above the decay threshold (Hoyle-band) clearly have a cluster
structure, but even within the ground state this component may not be insignificant~\cite{KanadaEnyo:2006ze}.
This is supported by recent calculations using nuclear lattice simulations~\cite{Epelbaum:2012qn}.

%\begin{figure}
%\vspace{0cm}
%\includegraphics[angle=-90.0,width=0.50\textwidth]{Martin/amd_12c.pdf}
%\vspace{0cm}
%\caption{\label{fig:AMD1}
%AMD densities of the ground state band, 0$^+_1$:
%a1, b1 and c1, Hoyle-band, 0$^+_2$ a2, b2, c2 and 0$^+_3$ band a3, b3, c3,
%from Ref.~\cite{KanadaEnyo:2006ze}.}
%\end{figure}

We discuss AMD methods in some detail in section~\ref{AMDsection} and lattice methods
in section~\ref{sec:nleft}. The experimental $B(E2)$ for the transition from the first 2$^+_1$ state
at 4.4~MeV to the ground state has been determined to be $7.6 \pm 0.4$~$e^2$fm$^4$,
which compares favourably with that calculated within the AMD framework of
8.5~$e^2$fm$^4$~\cite{KanadaEnyo:2006ze}. The calculated value for the transition
from the 4$^+_1$ state is similar with a value of 16~$e^2$fm$^4$. To date,
there is no experimental measurement, but this would in principle confirm
if these states are rotationally linked.

This raises the question whether it might be experimentally possible to
observe the intrinsic cluster structure shown in the AMD calculations for
the $^{12}$C ground state. {One possibility might be via ultra-relativistic $^{12}$C+$^{208}$Pb collisions
where differences between the $\alpha$-clustered and uniform $^{12}$C nucleus may be visible in quantities such as the triangular flow,
event-by-event fluctuations, or the correlations of the elliptic and triangular flows~\cite{Broniowski:2013dia}. A similar approach, e.g. examination of the
properties of the fragmentation of $^{12}$C at high energy have been explored in~\cite{Artemenkov:2017kzs}.}
Another possibility is via $\alpha$-particle knockout
from the ground state. The measurement of the $^{12}$C($p$,$p\alpha$) reaction
using polarized beams found analyzing powers which were strongly indicative
of $\alpha$-particles being preformed in the ground state~\cite{Mabiala:2009zz}.
This provides no information on any geometric arrangement or otherwise. Alternatively,
it may be possible to exploit the dynamical symmetries associated with the
triangular arrangement of the three $\alpha$-particles. The early work of
Hafstad and Teller~\cite{Hafstad:1938} paved the way for the more recent
work of Bijker and Iachello~\cite{Bijker:2014tka}. The dynamical symmetries
of the 3$\alpha$-system correspond to a spinning top with a triangular point
symmetry ($D_{3h}$). The rotational properties of these states are given
by
\begin{equation}
E_{J,K}=\frac{\hbar^2 J(J+1)}{2\mathcal{I}_{\rm Be}}-\frac{\hbar^2 K^2}{4\mathcal{I}_{\rm Be}},
\end{equation}
where $\mathcal{I}_{\rm Be}$ is the moment of inertia corresponding to two touching
$\alpha$-particles, which can be determined from the $^8$Be ground-state
rotational band~\cite{Hafstad:1938}. $K$ is the projection of the angular
momentum onto the symmetry axis of the 3$\alpha$ system. One would expect
that there should be a number of rotational bands with different values of
$K$. For $K^\pi=0^+$, the rotations will be around an axis which lies in
the plane of the three $\alpha$-particles, generating a series of states
0$^+$, 2$^+$, 4$^+, \ldots\,$. These correspond to the rotation of a $^8$Be nucleus, the rotation axis passing through the center of the third $\alpha$-particle.
The next set of rotations corresponds to the rotation around an axis perpendicular
to the plane of the triangle, with each $\alpha$-particle having one unit
on angular momentum, thereby giving $L=3\times1\hbar$; $K^\pi=3^-$. Rotations
around this axis and those parallel to the plane combine to give a series
of states 3$^-$, 4$^-$, 5$^-, \ldots$~.

The ground-state band described above, with 2$^+$ and 4$^+$ states at 4.4
and 14.1~MeV, would correspond to the $K^\pi=0^+$ rotational band. A candidate
for the $K^\pi=3^-$ band head is the 9.6~MeV 3$^-$ state. The 13.3~MeV state
presently tentatively labelled with $J^\pi=2^-$ in tabulations has recently
been shown to almost certainly have $J^\pi=4^-$~\cite{Freer:2007zz,Kirsebom:2010zz}.
Moreover, the 3$^-$ state has been shown to have a reduced $\alpha$-width
which indicates a cluster structure~\cite{Kokalova:2013gna}. The observation
of a candidate for a 5$^-$ state at 22.5 MeV~\cite{Marin-Lambarri:2014zxa},
would appear to complete the systematics and are also consistent with the
AMD calculations~\cite{KanadaEnyo:2006ze}.
{Widths of the negative-parity states have not been
calculated with AMD. However, Uegaki's $3\alpha$GCM calculations describe well the widths of the $3^-$ at 9.64~MeV
and $4^-$ at 13.35~MeV \cite{uegaki3}. The former and the latter are dominated by the
$^8{\rm Be}(0^+) +\alpha$ and $^8{\rm Be}(2^+) + \alpha$ partial decay widths.
The width of the $5^-$ may be dominated by the $^8{\rm Be}(2^+) +\alpha$ partial decay width, but as yet there are no calculations to confirm this. }

{
As with $^8$Be, the widths of the unbound states in $^{12}$C influence the possible collective interpretation.
The 14.1~MeV, 4$^+$, state has a width of 270~keV and the 9.6~MeV $3^-$ state has a width of 46~keV, both of which may not affect the collective
timescale. However, the states associated with the Hoyle state (see below) have large widths of the order of MeV or greater and
a simple rotational picture may be an over simplification.
}\\

{\it The Hoyle state and collective excitations} \\

The Hoyle state in $^{12}$C is one of the best known states in nuclei given
its rather crucial role in the synthesis of carbon through the triple-$\alpha$
process. The recent review of this state~\cite{Freer:2014qoa} provides a
comprehensive description of its role in synthesis and its experimental
properties. Suffice to say, from an experimental perspective those properties
have been well characterized. On the other hand, its structure is less well
understood.

The fact that no-core shell model calculations fail to reproduce the
energy of {the Hoyle state~\cite{Navratil:2000ww,Navratil:2007we}}, without resorting to a
significantly expanded harmonic oscillator basis, indicates already that
the structure lies beyond that described readily by the shell model. The
first {\em ab initio} calculation of the Hoyle state was performed only a few
years ago in Ref.~\cite{Epelbaum:2011md}. {These latter calculations were able to
explicitly capture the $\alpha$-clusterization that appears in this state.
{ The AMD calculations, Fig.~\ref{fig:exj},} indicate
that the Hoyle state is an extended three $\alpha$-system and that the associated 2$^+$ and 4$^+$ excited states are
not rigid, rotational, excitations and that a loose assembly of $\alpha$-particles, an $\alpha$-gas, may be a better
description. A similar conclusion was reached in the fermionic molecular dynamics (FMD) calculations for the same states~\cite{Neff:2014iha}. Here
it was suggested that the 2$^+$ and 4$^+$ resonances
might be considered as members of a rotational band built on the $^8$Be ground state with the third $\alpha$-particle
orbiting around the $^8$Be nucleus with relative orbital angular momentum 2 or 4, respectively. The origin of nuclear clustering with
relevance to the formation of the Hoyle state is also discussed by Okolowicz \emph{et al.}~\cite{Okolowicz:2012kv}.}

  It was
observed by Barker and Treacy~\cite{Barker1962} that in order to reproduce
the width of the Hoyle state, one has to use an unusually large radius: with
a radius of $1.6~{\rm fm}\,A^{1/3}$, a width of 9.3~eV corresponds to a dimensionless
reduced width, $\theta^2={\gamma_{\lambda}}^2 M_{red} R^2/ \hbar^2$, as large
as 1.5. Hence, the width of the Hoyle state is very large; this can only
be understood if there is a large degree of $\alpha$-clustering. The presence
of this cluster structure enhances the $\alpha$-capture cross section. But
its existence within the Gamow window results in the overall
capture cross section being boosted by a factor 10$^8$. Without the precise
location of this state the abundance of carbon-12 would be greatly reduced,
and thus it is intimately related to the existence of organic life. The rather
deep question is if this is a happy accident, or if there is some reason
why states with strongly-developed cluster structure should exist close to
the corresponding decay thresholds~\cite{Okolowicz:2012kv,Freer:2014qoa,
Epelbaum:2012iu,Epelbaum:2013wla}.

Beyond the fact that the Hoyle state has a 3$\alpha$-cluster structure, the
nature of that structure remains to be resolved. { The AMD calculations in
 Fig.~\ref{fig:exj}} indicate a dominance of $^8{\rm Be}+\alpha$ configurations
in a loose assembly such that the 2$^+$ and 4$^+$ excitations do not possess
a clear rotational behavior. The fermionic molecular dynamics (FMD) calculations
of the Hoyle state yield similar conclusions~\cite{Chernykh:2007zz}. An extension
of these ideas is that the state may be described by a gas/condensate of
$\alpha$-particles~\cite{Funaki:2009fc}. In principle, it may be possible
to gain an insight into the structure through the decay properties of the
state. In this instance there are two decay modes open; sequential and direct.
In the latter the system does not decay through the $^8$Be ground-state.
An upper limit for non-sequential $\alpha$-decay of 4\% was first determined
in 1994~\cite{Freer:1994zz}. { Subsequently, a measurement of the $^{40}{\rm Ca}+{^{12}{\rm C}}$ reaction at 25 MeV/nucleon
suggested that the branching ratio was in fact higher at $7.5\pm4\%$. This was challenged by further measurements where}
 upper limits as low as $5\times 10^{-3}$
(95\% C.L.)  \cite{Kirsebom:2012zza,Manfredi:2012zz} and 9(2)$\times$10$^{-3}$
has been put forward~\cite{Rana:2013hka}. This was improved
to be 0.2$\%$~\cite{Itoh:2014mwa}. These measurements have now reached a
sensitivity at which the phase space effects cease to be the dominant factor
and it may be possible to probe the structure {with limits of 0.047$\%$~\cite{Smith:2017} and 0.043$\%$~\cite{DellAquila:2017}, compared with the predicted
phase space limit of 0.06$\%$~\cite{Smith:2017}.}

\begin{figure}
  \vspace{-0.20cm}
   \begin{center}
    \hspace{-.6cm}
    \includegraphics[angle=-00.0,trim={0cm 3cm 0 0},width=0.45\textwidth]{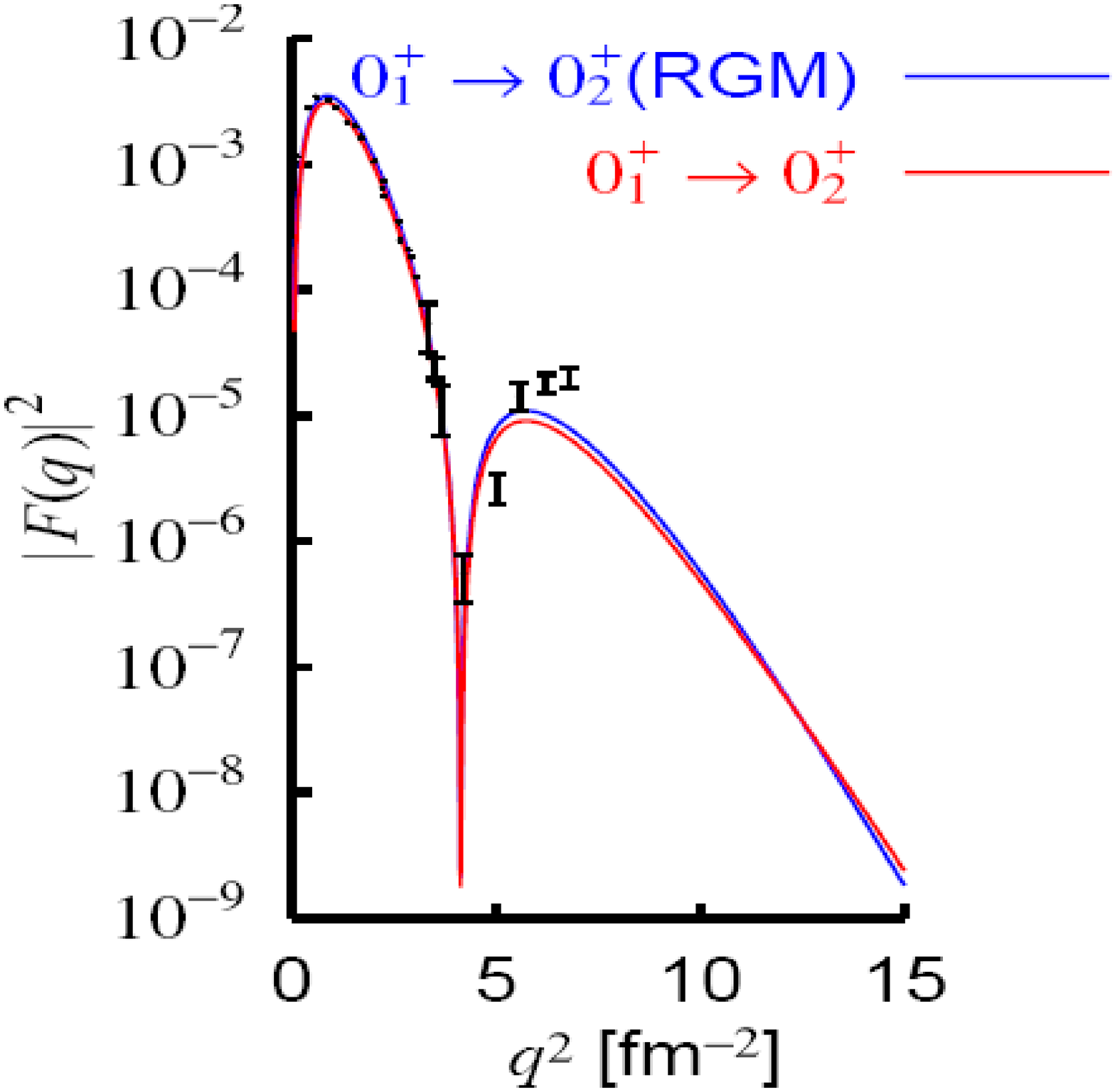}
     \vspace{-0.3cm}
     \end{center}
      \caption{(Color online) The calculated inelastic form factor for electron
      inelastic scattering from the $0^+_1$ ground state to the $0^+_2$
      excited state~\cite{Funaki:2006gt} for the BEC approach (red), compared
with the experimental data from Ref.~\cite{Sick:1970ma,Horikawa:1971oau,Horikawa:1971oau2,Strehl:1968pik}.
}
    \label{fig:BEC}
 \end{figure}

A second approach is to probe the charge distribution through electron inelastic
scattering~\cite{Sick:1970ma,Horikawa:1971oau,Horikawa:1971oau2,Strehl:1968pik}.
In such measurements the transition form factor is determined, which probes
the overlap of the ground state with the Hoyle state. To interpret such measurements
a model is required which can describe both the ground and excited states.
Both the condensate~\cite{Funaki:2006gt} and FMD descriptions~\cite{Chernykh:2007zz}
indicate that the Hoyle state is associated with a radius larger than that
of the ground state by a factor of 1.35 to 1.60 (depending on the model used
to analyze the data), which would correspond to an increase in volume by a
factor of 2.5 to 4. Fig.~\ref{fig:BEC} shows the calculated electron inelastic
scattering distribution for the condensate model~\cite{Funaki:2006gt}.

\begin{figure}
   \begin{center}
    \hspace{-.6cm}
    \includegraphics[angle=-00.0,width=0.4\textwidth]{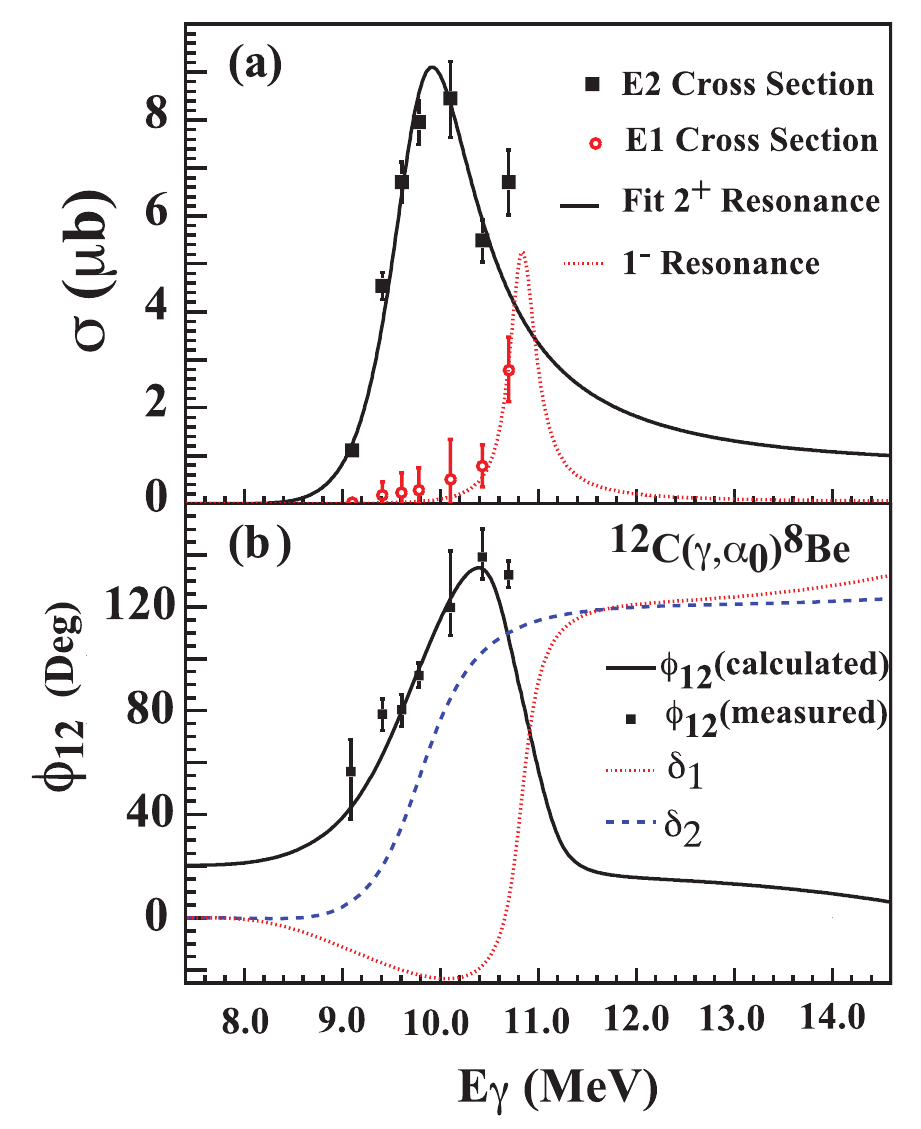}
     \end{center}
      \caption{(Color online) (a) The measured E1 and E2 cross
sections of the $^{12}$C($\gamma$,$\alpha_0$)$^8$Be reaction. (b) The measured
E1-E2 relative phase angle ($\phi_{12}$) together with the phase angle
calculated from a two-resonance model~\cite{Zimmerman:2013cxa}. }
    \label{fig:Zim2}
 \end{figure}

A third approach to deduce the structure of the Hoyle state is to search
for collective excitations, in particular the 2$^+$ excitation.
Inelastic scattering measurements~\cite{Freer:2009zz,Zimmerman:2011zz,Itoh:2011zz}
were the first to provide evidence for such an excitation. A common analysis
of the evidence for a 2$^+$ resonance from the proton- and $\alpha$-particle
scattering data is given in Ref.~\cite{Freer:2012se}, and a discussion of the
impact of these measurements is given in Ref.~\cite{Fynbo2011}. The 2$^+$ lineshape,
which is found in the inelastic scattering measurements $^{12}$C($\alpha,\alpha'$)
and $^{12}$C($p,p'$)~\cite{Freer:2012se}, determined the properties to be
$E_x$ = 9.75(0.15)~MeV with a width of 750(150)~keV. The existence of the
2$^+$ resonance  was confirmed  by a measurement of the $^{12}$C($\gamma,3\alpha$)
reaction at the HI$\gamma$S facility~\cite{Zimmerman:2013cxa}.  The excitation
function for these measurements are shown in Fig.~\ref{fig:Zim2} and gives
resonant parameters of $E_x$ = 10.13(6)~MeV and $\Gamma$ = 2.1(3)~MeV~\cite{Zimmerman2014}.

These measurements have now been extended to higher energies and continue
the expected trend for the 2$^+$ excitation.\footnote{M. Gai, private communication}
If the state has a rotational behavior then there should also be a 4$^+$
state close to 14~MeV. There exists tentative evidence for such a state at
13.3~MeV, with a width of 1.7~MeV ~\cite{Freer:2011zza,Jyfl2013,Ogloblin:2014}.
The existence of this latter state has yet to be definitively confirmed.
It appears to decay strongly to the $^8$Be ground state as opposed to the
2$^+$ excited state, which might provide an insight into the way the angular
momentum is constructed, i.e., through the orbiting of the $\alpha$-particle
around a $^{8}$Be($0^+$) core. Although much progress has been made in terms
of understanding the structure of $^{12}$C, the measurements are typically
challenging and often far from unambiguous. As such, the need for detailed
spectroscopy continues. Here the approach of the Aarhus group~\cite{Kirsebom:2014zra}
in measuring electromagnetic properties points the way for those future studies.
The $p+{}^{11}$B capture reaction is used to resonantly populate states
in $^{12}$C, and their decay after emitting an unobserved gamma decay is recorded
through the subsequent charged particle channel.

The Hoyle state, though extended, is not consistent with a linear
chain structure arrangement that would require the 2$^+$ state to lie $\sim 1$~MeV
lower than observed experimentally. Guidance from theory~\cite{KanadaEnyo:2006ze}
suggests that the 10.3~MeV, 0$^+_3$ state is the best possibility. This state
has a width of 3~MeV and a 2$^+$ state corresponding to a linear chain structure
would be expected close to 11.5~MeV and would have a very large width. As
yet, such a state remains to be observed.
{Recently, the possibility of two $0^+$ states around 10~MeV was experimentally reported by Itoh
{\it et al.} \cite{Itoh:2011zz} and supported by the extended Tohsaki-Horiuchi-Schuck-R{\"o}pke (THSR) calculation \cite{funaki15,Funaki:2015uya}.}

\begin{figure*}
\begin{center}
\includegraphics[width=0.8\textwidth]{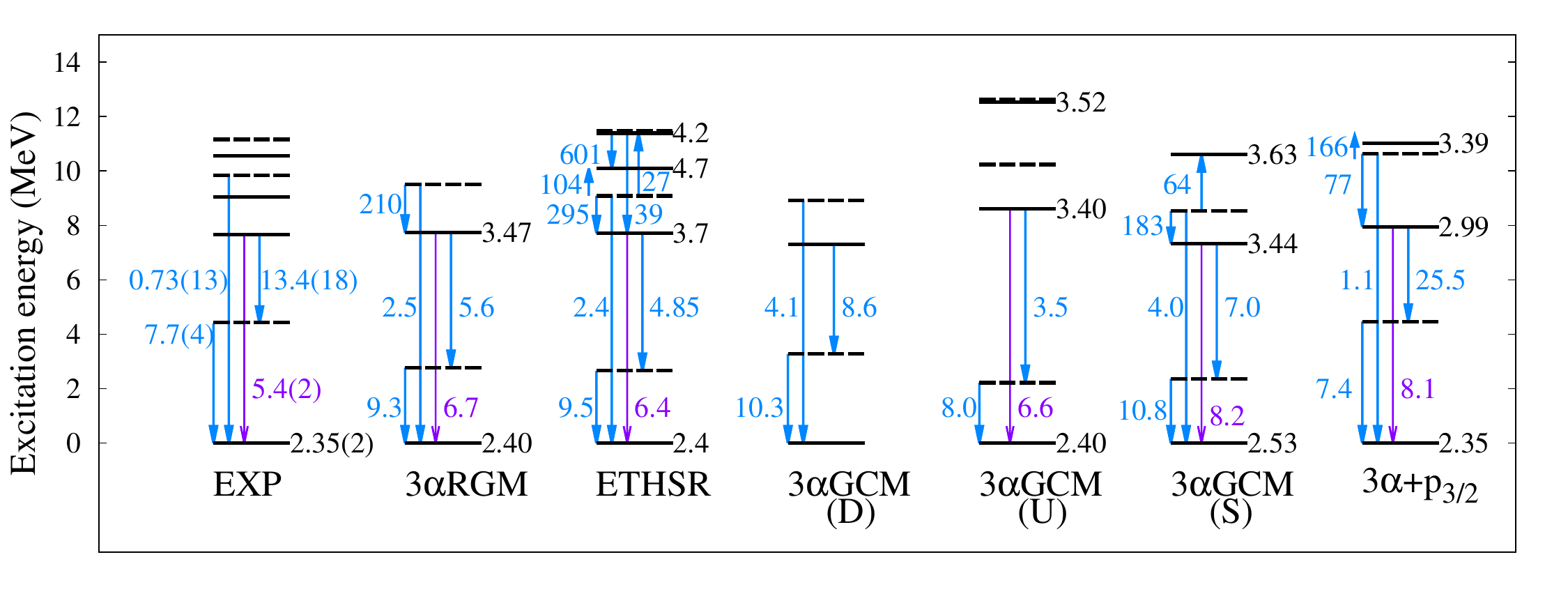}\\
\includegraphics[width=0.8\textwidth]{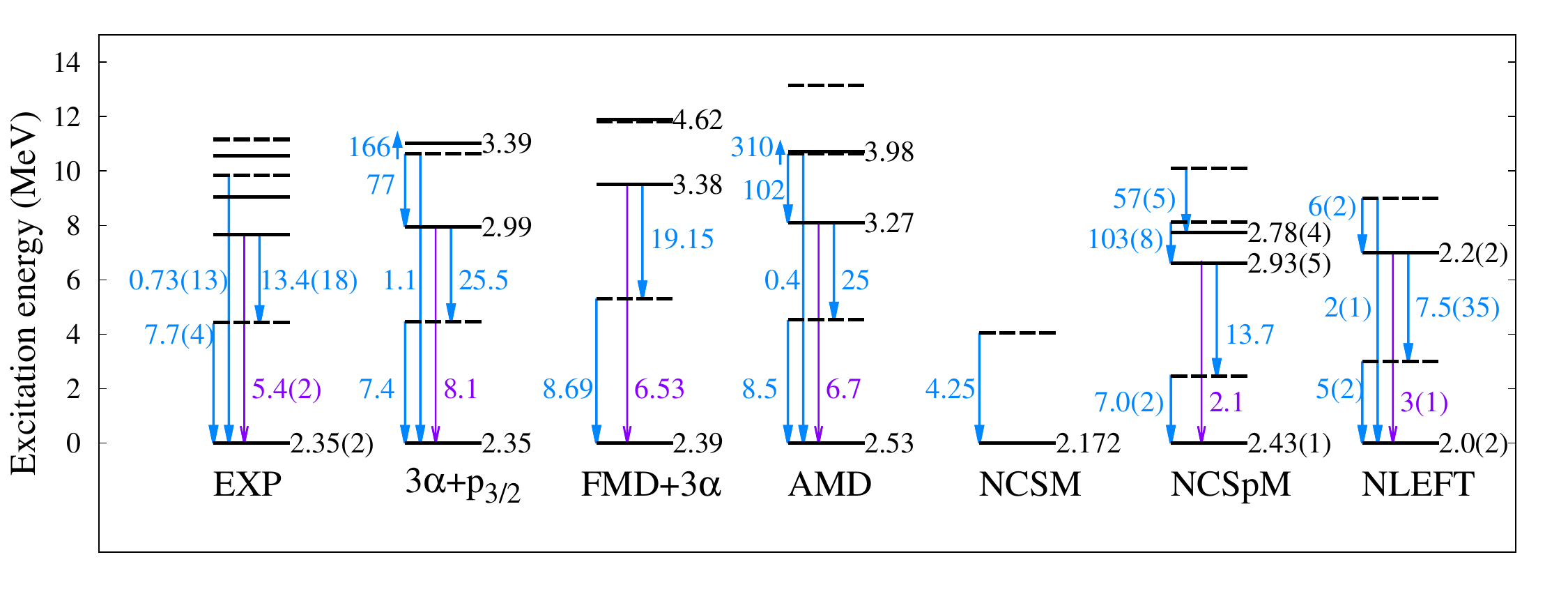}
\caption{\label{fig:c12-comparison}
{
(Color online)
Theoretical and experimental energy levels, $E0$ and $E2$ transitions, radii of  $^{12}$C.
The energy levels of  $0^+$ and $2^+$ states are shown by solid and dashed lines, respectively.
$B(E2)$  values ($e^2$fm$^4$) are shown by the corresponding arrows (cyan text and arrows).
$M(E0)$ values for $0^+_2\to 0^+_1$ are shown by the arrows from $0^+_2$ to $0^+_1$ (purple text and arrows).
Root-mean-square radii of point-proton distribution in $0^+$ states are shown at the right side
of the energy levels (black text). Experimental data are from Refs.~\cite{AjzenbergSelove:1990zh,ANGELI201369,Chernykh:2007zz,Itoh:2011zz,Zimmerman:2013cxa}.
Theoretical results of the microscopic $3\alpha$ models,  $3\alpha$RGM\cite{Kamimura:1981oxj}, the extended THSR\cite{funaki15,Funaki:2015uya},
$3\alpha$GCM(D) in Ref.~\cite{Descouvemont:1987zzb}, $3\alpha$GCM(U) in Ref.~\cite{uegaki3},  and
$3\alpha$GCM(S) in Ref.~\cite{Suhara:2014wua} are shown in the upper panel,
and those of  the $3\alpha$GCM+$p_{3/2}$ \cite{Suhara:2014wua},
the FMD+$3\alpha$ \cite{Chernykh:2007zz}, AMD \cite{KanadaEnyo:1998rf,KanadaEnyo:2006ze},
NCSM \cite{Navratil:2007we}, NCSpM \cite{Dreyfuss:2012us}, and NLEFT \cite{Epelbaum:2012qn} calculations
are shown in the lower panel. For comparison, the $3\alpha$GCM+$p_{3/2}$ results are shown also in the upper panel.
The  NLEFT energies have about 2~MeV errors and can be improved in future work using new methods as described in Ref.~\cite{Lahde:2014sla}.
}
}
\end{center}
\end{figure*}

{
Figure~\ref{fig:c12-comparison} shows the compilation of theoretical spectra and transitions for
$0^+$ and $2^+$ states compared with the experimental data.
Although there are many non- and semi-microscopic $3\alpha$ calculations,
we only show microscopic calculations with fully antisymmetrized wave functions and
nucleon-nucleon interactions.  It is difficult to directly compare the reproduction quality of
microscopic calculations with non-microscopic calculations where interactions (or the Hamiltonian) are
usually phenomenologically adjusted to fit the energy spectra of $^{12}$C.
It should be also noted that we should not discuss {\it ab initio} calculations obtained from the
realistic nuclear forces on the same footing with the
calculations using phenomenological effective nuclear interactions.
Details of the theoretical frameworks and  interactions are explained in later sections.
In the $3\alpha$RGM \cite{Kamimura:1981oxj}, extended THSR \cite{funaki15,Funaki:2015uya},
$3\alpha$GCM \cite{Descouvemont:1987zzb,uegaki3,Suhara:2014wua}, and
$3\alpha$+$p_{3/2}$ \cite{Suhara:2014wua} calculations,
phenomenological effective nuclear interactions of the Volkov forces \cite{Volkov:1965zz} are used.
The interaction parameters of the Volkov forces are tuned to reproduce $\alpha$-$\alpha$ scattering, though
there are minor differences in the parameters among these calculations.
The AMD results \cite{KanadaEnyo:1998rf,KanadaEnyo:2006ze} are obtained by using the MV1 force\cite{Ando:1980hp}, which is a
phenomenological effective nuclear interactions modified from the Volkov force to describe the saturation properties,
whereas the FMD+$3\alpha$ results \cite{Chernykh:2007zz} are obtained
based on the realistic Argonne V18 potential with phenomenological tuning.
For the NCSM \cite{Navratil:2007we} and nuclear lattice effective field theory (NLEFT) \cite{Epelbaum:2012qn} calculations, the results
obtained with the realistic $NN$ and $NNN$ forces derived
from the chiral effective theory are shown.
In the no-core symplectic model (NCSpM) calculation \cite{Dreyfuss:2012us},
a simplified effective Hamiltonian is used.

In general, the $3\alpha$ calculations describe well the energy spectra of cluster states
above the $3\alpha$ threshold and the electron scattering form factors for the $0^+_1$ state
and $0^+_2\to 0^+_1$ transition,
but they are not sufficient in describing some properties of low-lying states
such as the $0^+_1$-$2^+_1$ level spacing, $E2$ transition strength for $2^+_1\to 0^+_1$ and
$0^+_2\to 2^+_1$.
Hybrid calculations of the $3\alpha$+$p_{3/2}$ and FMD+$3\alpha$ models as well as the
AMD can reasonably describe the ground band properties and excited spectra for cluster states.
The NCSM calculation fails to describe
the excited cluster states above threshold since those states are
beyond the model space, whereas the NCSpM, which contains higher shell configurations
for cluster excitations, and the NLEFT calculations describe cluster structures in excited
states above the threshold. The {\it ab initio} calculations (NCSpM and NLEFT) tend to
much underestimate the size of the ground state, and also give
small values of the size and $E0$ matrix element for the Hoyle state.
The $\alpha$-decay widths are calculated in the $3\alpha$RGM\cite{Kamimura:1981oxj}
and $3\alpha$GCM(D) in Ref.~\cite{Descouvemont:1987zzb} by solving $^8{\rm Be}+\alpha$
scattering, and evaluated in the extended THSR \cite{funaki15,Funaki:2015uya},
$3\alpha$GCM(U) in Ref.~\cite{uegaki3}, and AMD \cite{KanadaEnyo:2006ze}
within bound state approximations using reduced width amplitudes.
The available data for the $\alpha$-decay widths are reproduced quantitatively or qualitatively by theoretical
calculations.
}

Though much progress has been made in understanding the structure of $^{12}$C,
it is apparent there is both need and scope for measurements to more precisely
constrain the properties of the states presented in Fig.~\ref{fig:exj}. In
particular, this requires the measurement of electromagnetic
transition rates where possible.\\

{\it States in $^{16}$O, dynamical symmetries and chains }\\

One way of further testing our understanding of cluster correlations and
the structure of $^{12}$C is through the extension of that understanding
to $^{16}$O, which now is within the reach of {\it ab initio} approaches,
see e.g., Ref.~\cite{Epelbaum:2013paa}. Though
much work has been done in both experiment and theory for this
nucleus, here we provide some historical perspective first and then reflect on the most recent developments.

\begin{figure}
   \begin{center}
    \hspace{-.6cm}
    \includegraphics[angle=-00.0,width=0.45\textwidth]{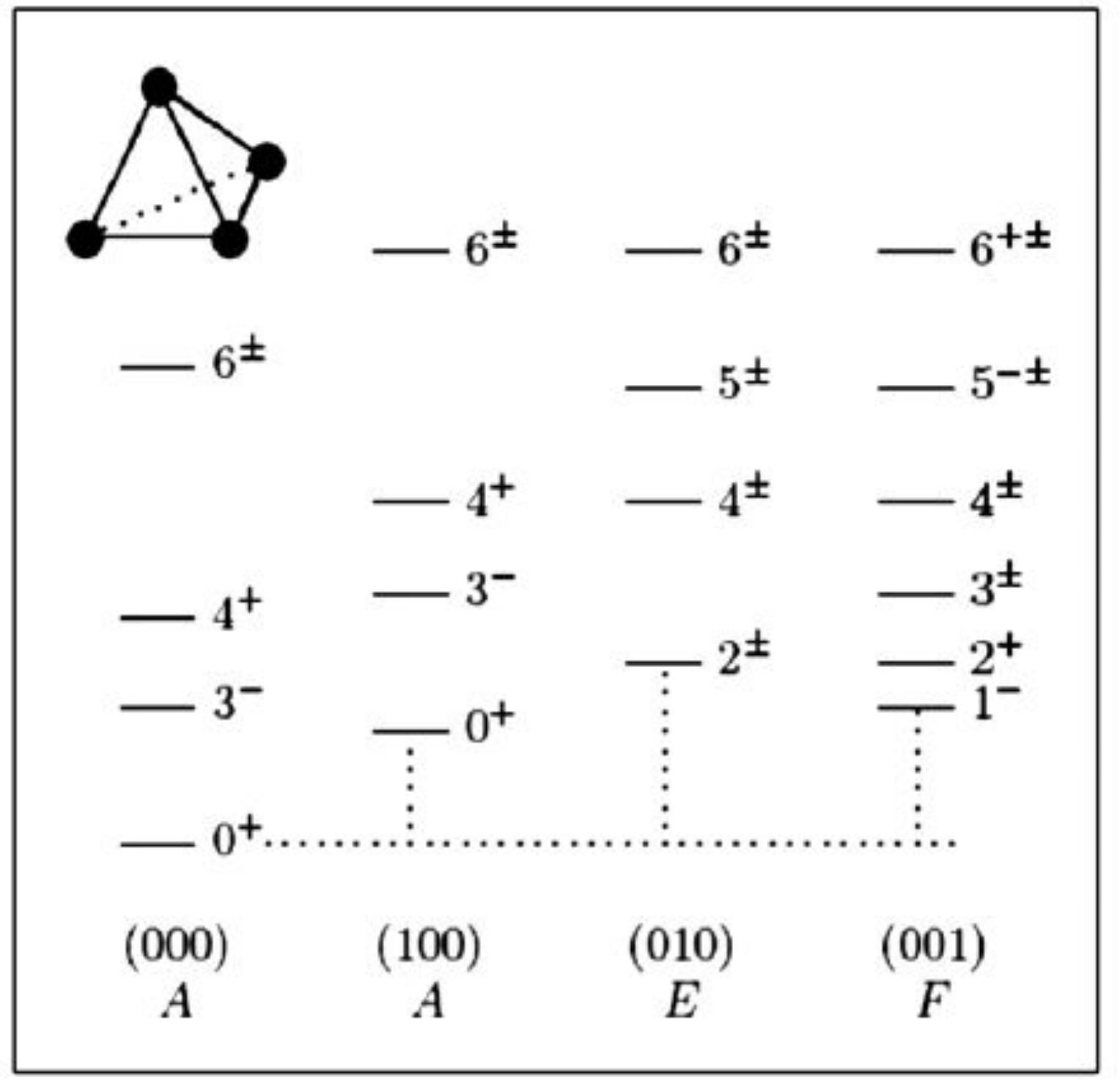}
     \end{center}
      \caption{The calculated spectrum of $^{16}$O states assuming a $T_d$
dynamical symmetry, obtained using the algebraic cluster model~\cite{Bijker:2014tka}.
}
    \label{fig:Dyn}
 \end{figure}

 The work by Hafstad and Teller~\cite{Hafstad:1938} indicates the collective
properties of the 4$\alpha$ system should be described by the tetrahedral
symmetry group, $T_d$. Here the characteristics are those of a spherical
top, with equal moments of inertia and independent of rotation axis. If
one assumes the separation between the $\alpha$-particles is that which is associated
with the $^8$Be ground state, $\mathcal{I}_{\rm Be}$, then the rotational energies
are given by
 \begin{equation}
 E_J=\hbar^2\frac{J(J+1)}{4\mathcal{I}_{\rm Be}}.
 \end{equation}
The rotation of the tetrahedral structure corresponds to the equivalent
rotation of two $^8$Be nuclei around their symmetry axis, and hence the ${4\mathcal{I}_{\rm Be}}$
in the denominator.  The symmetry then dictates that all values of $J$ are
permitted except $J=1$, 2 and 5; states with $J= 0$, 4 and 8 have even parity
and $J=3$, 7 and 11 have negative parity. A key feature of this structure
would be degenerate 6$^+$ and 6$^-$ states. A similar conclusion can be found
in the recent work of Bijker and Iachello~\cite{Bijker:2014tka} as shown in Fig.~\ref{fig:Dyn},
which was triggered by related work performed using lattice simulations in Ref.~\cite{Epelbaum:2013paa}. 
The algebraic cluster model of Bijker and Iachello~\cite{Bijker:2014tka}  generates energy eigenvalues
and eigenvectors obtained by diagonalizing a finite-dimensional matrix, rather than by solving a set of coupled differential equations in coordinate
space~\cite{Bijker:2015}. The model then describes the relative motion of the clusters.

The potential similarity between the structural properties of $^{12}$C and
$^{16}$O and the underlying dynamical symmetries is compelling. However there
are other models, e.g., the $\alpha$-cluster model~\cite{Bauhoff:1984zza} that
provide a good description of the energy spectrum of $^{16}$O states. Hence,
it is important to examine the decay properties.
The experimentally observed states at 6.130~MeV, 3$^-$; 10.356~MeV, 4$^+$ {($\Gamma$=26 keV)}; and
21.052~MeV, 6$^+$ {($\Gamma$=205 keV)} have been linked in the work of Ref.~\cite{Bijker:2014tka}
to the collective excitations of the tetrahedral structure. These same calculations
predicted states at 6.132, 10.220 and 21.462~MeV and electromagnetic transition
strengths $B(E3)$ and $B(E4)$ of 181 and 338~$e^2$fm$^{2L}$ compared with
experimental values of 205(10) and 378(133)~$e^2$fm$^{2L}$. The comparison
between experiment and theory is clearly compelling.  { The widths of the unbound 4$^+$ and 6$^+$ states are similar to those
for the ground-state band in $^{12}$C. Nevertheless, caution is required when interpreting transition rates for such states.}

The alternative theoretical approach provided by the $\alpha$-cluster model
(ACM) calculations of Bauhoff, Shultheis and Shultheis~\cite{Bauhoff:1984zza}
offer a different perspective. These calculations identify a number of cluster
structures, including a tetrahedral arrangement of the four $\alpha$-particles
in the ground-state. In addition, a planar arrangement of $\alpha$-particles
is found for the first excited 0$^+$ state, which can be associated with
a $^{12}{\rm C}+\alpha$ structure.  We note the similarity to the lattice results
from Ref.~\cite{Epelbaum:2013paa}.
The main difference between the ACM and algebraic
cluster model (ACM') of Ref.~\cite{Bijker:2014tka} is evident in the assignment
of the 10.356~MeV 4$^+$ state to rotational bands. The ACM assigns it to
the planar rotational structure, whereas the ACM' links it
to the tetrahedral ground-state.

The algebraic cluster model reproduces the $B(E4)$ for the 10.356~MeV to
ground state transition, while the $\alpha$-cluster model would place this state
in a different band. This is clearly contradictory.  What is clear from measurements
of the $\alpha$-decay branching ratios for decay to the $^{12}$C ground state
and first excited states is that the states in the ACM planar band, above
the $\alpha$-decay threshold, all have very similar decay properties. They
predominantly decay to the ground state~\cite{Tilley:1993zz,Wheldon:2011zz}. Moreover there
is also a negative parity band built on the 7.12~MeV, 1$^-$ state with very
similar decay properties. This similar structure of this group of states
conflicts with the tetrahedral interpretation and indicates a collective
excitation built around a $^{12}{\rm C}+\alpha$ cluster structure where the total
angular momentum of the state is generated by the orbital motion of the $\alpha$-particle
around the $^{12}$C core. These two different perspectives on the nature
of the low-lying states in $^{16}$O need to be resolved. Precision measurements
of the \emph{complete} electromagnetic decay patterns are likely to be the
way forward. Measurements of the ANCs of the states close to the decay threshold
in $^{16}$O have recently been reported~\cite{Avila:2014xsa}. ANCs provide
a model-independent assessment of the cluster structure and as such are also
a key ingredient in refining the understanding of $^{16}$O as a test case for nuclear theory.

One of the longest-standing questions related to $^{16}$O is the existence
of the 4$\alpha$ chain-state. The 4$\alpha$ decay threshold is at 14.4~MeV
and thus a chain state would exist close to or above this energy. The 15.1~MeV
0$^+$ state is a potential candidate, though the state has been identified
as the analogue of the Hoyle state~\cite{Funaki:2009fc}.  The proximity of
this state to the decay threshold means that the 15.1~MeV state cannot decay
strongly to the 4$\alpha$ final state. There are, however, a number of resonances
that decay to the ${}^8{\rm Be}+{}^8{\rm Be}$ or $^{12}{\rm C(Hoyle)}+\alpha$ final states.
The pioneering measurements of Chevallier {\it et al.}~\cite{Chevallier:1967zz}
revealed both the excitation energy and dominant angular momenta of the $^{16}$O
resonances that decayed to ${}^8{\rm Be}+{}^8{\rm Be}$ as populated in the $^{12}$C($^4$He,$^8$Be)$^8$Be
reaction. Remarkably, the energy-spin systematics of selected narrow resonances
fell onto a $J(J+1)$ trajectory with moment of inertia commensurate with
a structure where the $\alpha$-particles are arranged in a linear fashion;
an $\alpha$-particle chain. This work was published in 1967 and until the present
has been held up as an example of extreme $\alpha$-clustering.
\begin{figure}
   \begin{center}
    \hspace{-.6cm}
    \includegraphics[angle=-0.0,width=0.5\textwidth]{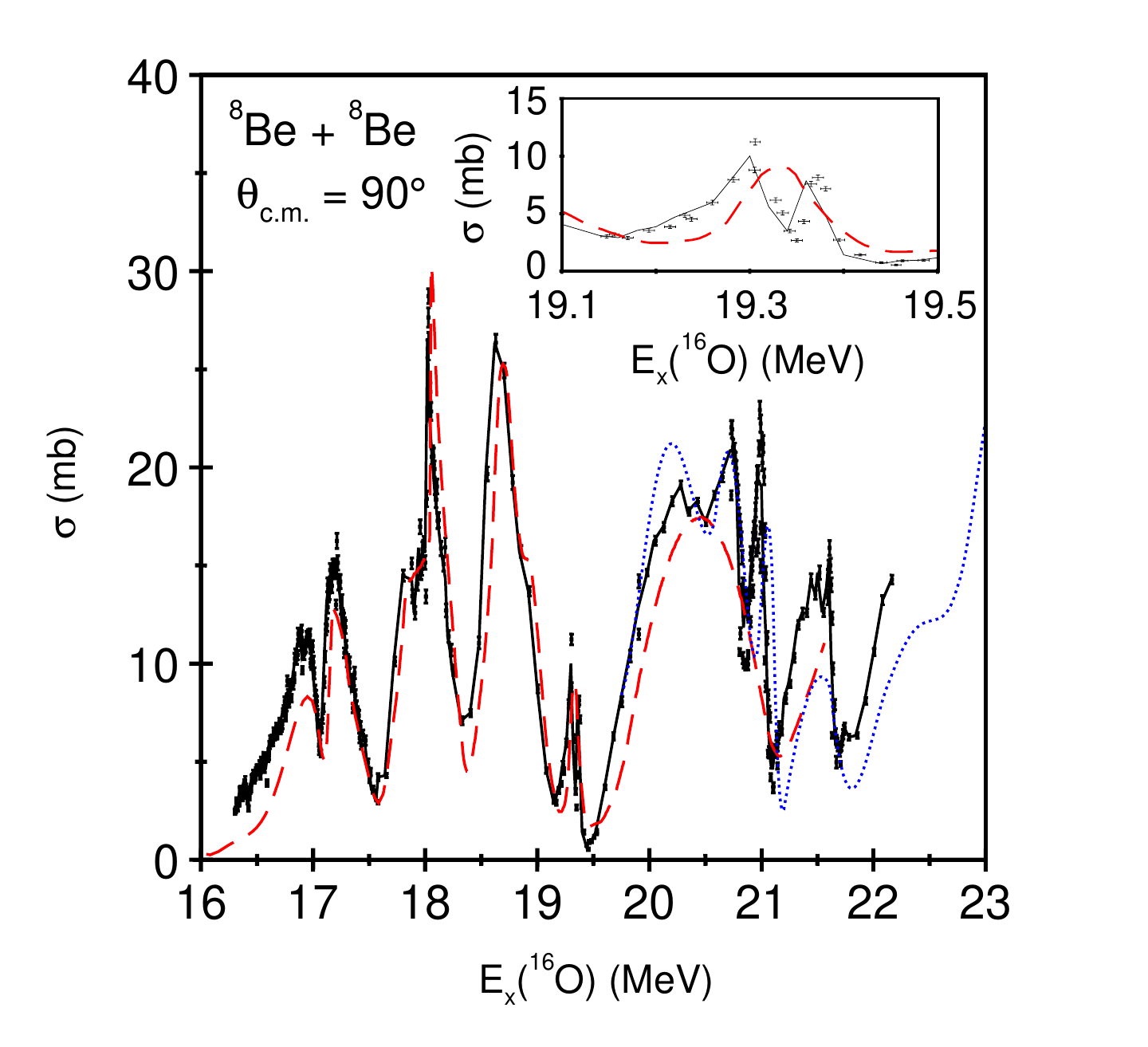}
     \end{center}
      \caption{(Color online) Measurements of the $^{12}$C($^4$He,$^8$Be)$^8$Be reaction
for $\theta_{c.m.}=90^\circ$~\cite{Curtis:2013sxa} data points. The dashed
and dotted lines correspond to the measurements of Chevallier \emph{ et al.}~\cite{Chevallier:1967zz}
and Brochard \emph{ et al.}~\cite{Brochard:1976zz}. See Ref.~\cite{Curtis:2013sxa}
for more details.  }
    \label{fig:16O}
 \end{figure}
Confirmation of such an exotic structure is clearly vital. There are a number
of possible approaches. One is to confirm the details of the excitation
function, and the second is to search for higher spin members of the 4 $\alpha$-particle
chain band. The band was only observed up to spin~6. Subsequent measurements
by Brochard {\it et al.}~\cite{Brochard:1976zz} found no evidence for the 8$^+$
member. The measurements of Chevalier {\it et al.} have been revisited~\cite{Curtis:2013sxa},
as displayed in Fig.~\ref{fig:16O}. The highly detailed excitation functions for the $^{12}$C($\alpha$,$^8$Be)$^8$Be
and $^{12}$C($\alpha$,$^{12}$C[7.65 MeV])$\alpha$ reactions presented in
Ref.~\cite{Curtis:2013sxa} show that the original structure that was
interpreted as resonances in the earlier work \cite{Chevallier:1967zz} was
more complex and that no evidence for an 8$^+$ state could be identified. This
most recent study contained over 400 measurements at different energies,
with significant coverage of the angular distributions which should permit
the components from resonances and transfer-like processes to be disentangled.
A measurement of  $^{13}$C($\alpha$,$^8$Be+$^8$Be)$n$ has recently been
published which provides some insight as to what are resonant features in
the $^{12}{\rm C}+\alpha$ excitation function \cite{Curtis:2016suk}.
With the existence of excitation functions for $^{12}{\rm C}+\alpha$ leading
to 4$\alpha$ unbound final-states as well  as bound states~\cite{Ames:1982zz},
there is in principle sufficient data to perform a complete R-matrix analysis
of the data to constrain states with enhanced 4$\alpha$ reduced decay widths.
This is likely to be a key component in constraining the structure of
$^{16}$O above the 4$\alpha$-decay threshold. Recent measurements of $\alpha$ inelastic scattering populating 0$^+$ states
in this region also indicates the spectrum of states may be more complicated than has been previously been concluded~\cite{Li2017}.\\

\begin{comment}  %Removed from paper
\subsubsection{Alpha-particle states in heavy systems}

The existence of nuclei which exhibit $\alpha$-gas-like structures continues
to be of interest to the field. At the rather extreme end of the experimental
spectrum, the measurements of inelastic $\alpha$-scattering experiment on $^{56}$Ni,
performed using an active target at GANIL, show a very high multiplicity
for
$\alpha$-particle emission  which  cannot  be  explained  by
 means  of  the  statistical  decay  model~\cite{Aki13}. This has been associated
with an $\alpha$-gas-like structure at high excitation. The appearance of
clustering in heavy nuclei in the ground state or low-lying states is less well
documented.
Clearly, many heavy nuclei undergo $\alpha$-decay, but preformation factors
are only typically 10$\%$. This does not require a cluster model prescription
to obtain such factors.
The best case may $^{212}$Po which may be described as an $\alpha$-particle
orbiting a $^{208}$Pb core~\cite{Xu:2015pvv}. Experimentally the measurement
of enhanced dipole transitions in this system appears to be consistent with
this picture~\cite{Astier:2009bs}.
\end{comment}

\subsubsection{Molecular structures in neutron-rich nuclei}
\begin{figure}
  \vspace{-0.20cm}
   \begin{center}
    \hspace{-.6cm}
    \includegraphics[angle=-00.0,width=0.4\textwidth]{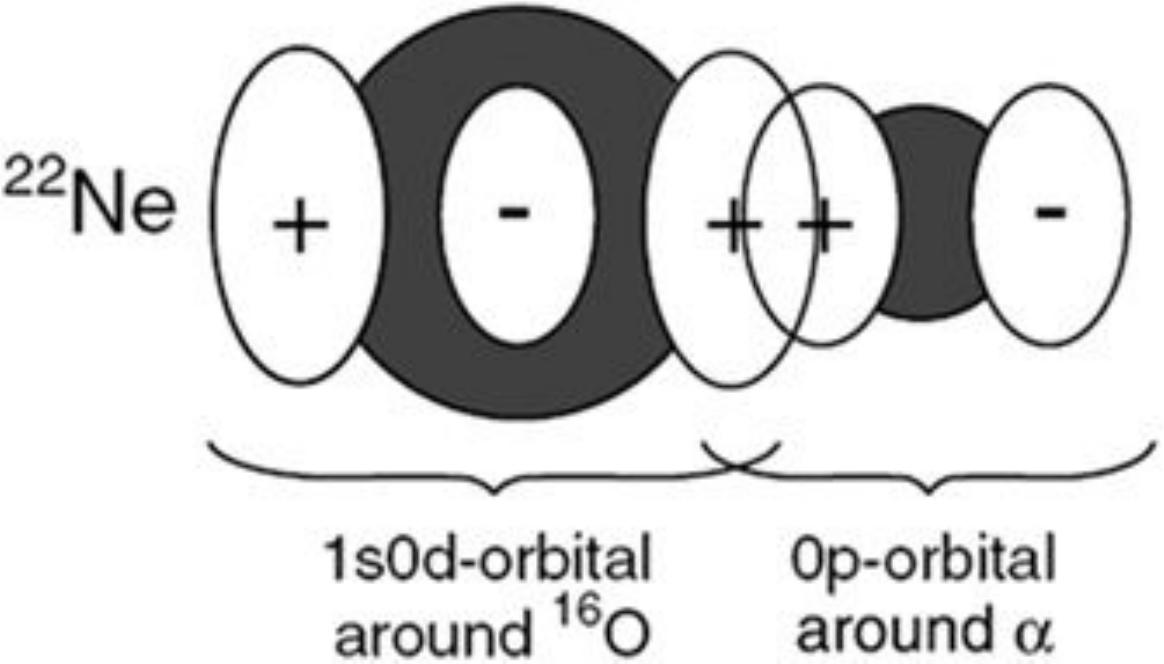}
     \vspace{-0.3cm}
     \end{center}
      \caption{Illustration of the formation of molecular orbitals for neon
isotopes from the valence orbitals of neutrons around the cores of $^{16}$O
and $^4$He ~\cite{Kimura:2007kz}. }
    \label{fig:MO}
 \end{figure}

The idea that light nuclei might have a molecular structure where typically
the valence neutron is exchanged between $\alpha$-particle cores has been
explored extensively~\cite{Sey81,vonOertzen:1996,voe97,voe97_2,Itagaki:1999vm}.
In essence, it is possible to form linear combinations of the neutron wave function
around the $\alpha$-particle cores and obtain, for example, two-centered molecules
with delocalized neutrons in $\pi$ and $\sigma$-orbitals~\cite{vonOertzen:2006a,Freer:2007}.
Here the single-center orbitals both have $p$-type character. It is also
possible to build more complex molecular structures with non-identical cores,
for example, in nuclei such as $^{21,22}$Ne~\cite{vonOertzen:2006a}.  This is illustrated in Fig.~\ref{fig:MO} \cite{Kimura:2003ue}.

\begin{figure}
  \vspace{-0.20cm}
   \begin{center}
    \hspace{-.6cm}
    \includegraphics[angle=-00.0,width=0.45\textwidth]{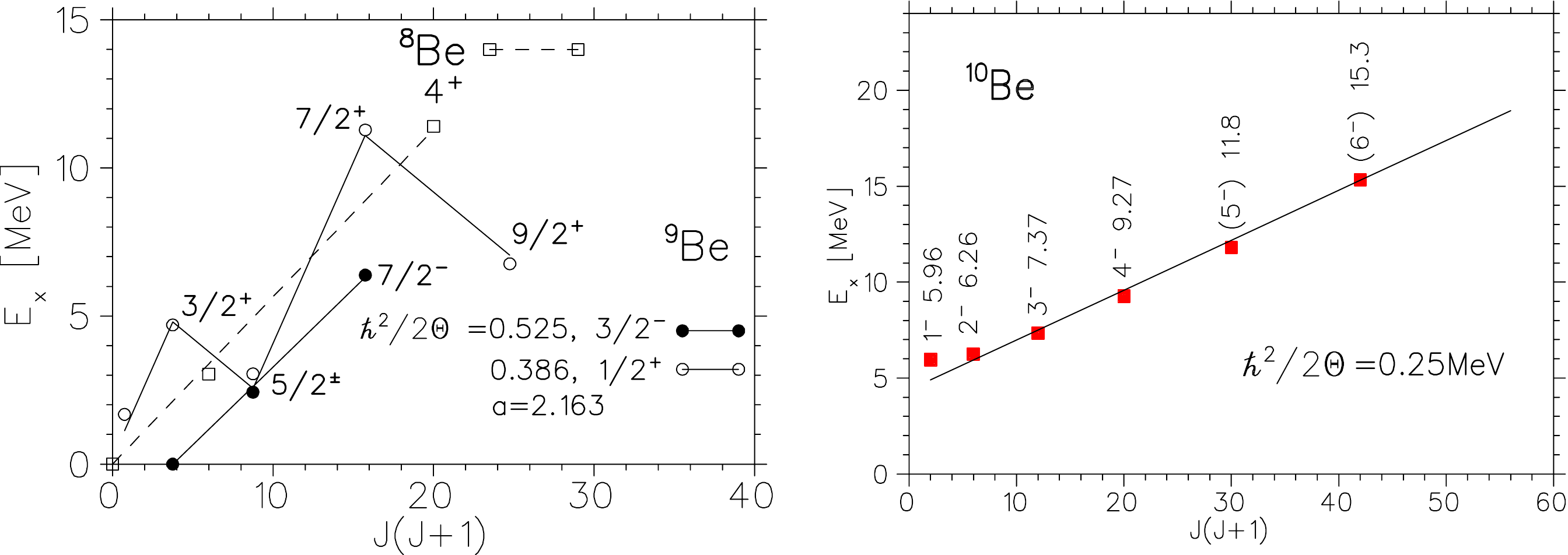}
     \vspace{-0.3cm}
     \end{center}
      \caption{Rotational bands of $^8$Be, $^9$Be (left panel) and $^{10}$Be
(right panel). The
excitation energies are plotted as a function of angular momentum $J(J+1)$.
The
Coriolis decoupling parameter, $a$, for the $K = 1/2$ band is indicated.
From Ref.
~\cite{Freer:2007}. }
    \label{fig:Be_bands}
 \end{figure}

 The simplest example of this molecular behavior is found in the rotational
bands of $^9$Be. The ground state band ($K^\pi=3/2^-$) is well-understood
in terms of its $\pi$-type characteristics. The  1/2$^+$ excited state at
1.68~MeV has a sequence of positive parity states ($3/2^+$, $5/2^+$, $7/2^+$...)
which may be connected to $\sigma$-type molecular structures, as shown in Fig.~\ref{fig:Be_bands}.
These two bands have spin and parity values consistent with molecular structures. Furthermore, as indicated in the figure, the moments of inertia extracted from the gradients of the bands are similar to the moment of inertia found for $^8$Be, i.e., indicating the $\alpha$-$\alpha$ core structure
is largely preserved. There are few ways to observe directly the ground state
structure of the nucleus $^9$Be, however  measurements of the decay correlations
of $^7$Be and $^6$Li nuclei following the interaction with a $^9$Be target
showed strong and unexpected alignment. This was concluded to be evidence
for the $\pi$-type molecular structure of the $^9$Be ground-state~\cite{Charity:2015vca}.
The right hand panel of Fig.~\ref{fig:Be_bands} shows the systematics of
a negative parity band in $^{10}$Be. Here this would correspond to a mixed
$\pi$-$\sigma$ configuration for the valence neutrons. Again the deformation
is found to be consistent with the molecular picture, though it is apparent
the moment of inertia has increased. Part of the origin of this effect is
the proximity of this band to the $\alpha$-decay threshold, such that, as
in the case of $^{12}$C, the cluster structure is enhanced.

The most pronounced example of molecular behavior studied to date is that
associated with a series of states close to the $\alpha$- and neutron-decay
thresholds in $^{10}$Be. The 6.179~MeV, 0$^+_2$ state has a suppressed gamma decay,
with a lifetime of the order of 1~ps. This isomeric behavior does
not arise due to the lack of possible decay paths, but may be
understood in terms of the small overlap of its structure and that of the
more compact lower energy, 3.36~MeV, 2$^+_1$ state. This already signals
an unusual structure, in analogy to the Hoyle state in $^{12}$C. The excited
state at 7.542~MeV, 2$^+_2$, is believed to be a collective excitation of
this state. This state lies above, but very close to, the $\alpha$-decay
threshold (7.409~MeV), and thus its decay to this channel is strongly suppressed
by the Coulomb and centrifugal barriers. Nevertheless, the $\alpha$-decay
has been found to correspond to a very large reduced width~\cite{Liendo:2002gx},
representative of the large degree of clusterization, although there is disagreement in the absolute value~\cite{Milin:2005txl}. The 4$^+$ member of
the band has been identified to lie at 10.15~MeV~\cite{Milin:2005txl}.
An unambiguous measurement of the spin and parity of the state was found
in the resonant scattering of $^6{\rm He}+{}^4{\rm He}$~\cite{Freer:2006zz}. This result
has also been confirmed through a second resonant scattering measurement
performed at Notre Dame~\cite{Suzuki:2013mga}. If a collective model is applied, the moment of inertia associated with the rotational band would indicate
that the state has a rather extreme deformation associated with the two valence
neutrons occupying $\sigma$-like orbitals, with a density maximum between the two
$\alpha$-particles which, via the Pauli exclusion principle, forces an increased separation
of the two $\alpha$-particles. There are also indications from as yet unpublished
measurements of a possible 6$^+$ state at higher energy.\footnote{G. Rogachev,
private communication} Such a structure would be the analogue of the 3$\alpha$-chain
state, but with two proton holes.

{
Valence neutrons in $\pi$- and $\sigma$-orbitals
play an important role also in structure change of the ground states
along Be isotopes.
Because of the lowering mechanism of
the $\sigma_{1/2}$-orbital in a well-clusterized $2\alpha$ system,
the $N=8$ shell gap vanishes in neutron-rich Be.
As a result,  the ground states of $^{11}$Be and $^{12}$Be
have $\sigma$-type molecular structures characterized by
intruder configurations with large deformation (enhanced clustering),
as supported by experimental observations such as
Gamov-Teller and $E2$ transitions
as well as the low-lying energy spectra
\cite{Suzuki:1997zza,Iwasaki:2000gh,Iwasaki:2000gp,Navin:2000zz,Shimoura200331,Pain:2005xw,Imai:2009zza,Meharchand:2012zz}.
Contrary to the enhanced clustering in $^{11}$Be and $^{12}$Be, the ground state of $^{10}$Be has
weak clustering because of the attractive role of the $\pi$-orbital neutrons.
The systematic change of cluster structures along the Be isotope chain is reflected in the
$N$ dependence of  charge radii, which have been precisely determined by isotope shift measurements
\cite{Nortershauser:2008vp,Krieger:2012jx}.
The charge radius is smallest at $N=6$ for $^{10}$Be indicating a possible new magic number at $N=6$
instead of $N=8$. This trend is described well by
the weakening and enhancement of the cluster structures in AMD and FMD calculations
\cite{Krieger:2012jx,KanadaEnyo:2014toa}.
}

The experimental efforts to extend the systematics from dimers to trimers
has seen a focus on trying to understand the systematics of three-centered
molecules. Milin and von Oertzen had performed some pioneering work which
established a set of candidate bands in $^{13}$C~\cite{Mil02} and $^{14}$C~\cite{voe04}.
In the case of $^{13}$C the experimental situation remains unclear as the
rotational systematics proposed in Ref.~\cite{Mil02} are inconsistent with
measurements of $^9{\rm Be}+\alpha$ resonant scattering~\cite{Freer:2012zz}.\footnote{Also
measurements from the Naples group, yet unpublished}. { There are other studies of the $^{13}$C system, e.g.~\cite{Soic:2003ht,Rodrigues:2010}, but these
are inconclusive in terms of the molecular structure of this nucleus.} There have been a number
of studies of $^{10}{\rm Be}+\alpha$ resonant scattering which populate resonances
above the $\alpha$-decay threshold
(12~MeV)~\cite{Freer:2014gza,Fritsch:2016vcq,Yamaguchi201711}.
%(12~MeV)~\cite{Freer:2014gza,Fritsch:2016vcq}\footnote{H.
%Yamaguchi, \emph{et al.}, to be published, private communication}
This is higher energy than the 0$^+$ band head identified by von Oertzen and co-workers~\cite{voe04},
9.75~MeV, and as such resonances may be associated with higher nodal cluster
structures. Nevertheless, these latest measurements provide some tentative
evidence for linear chain structures in $^{14}$C,
{ as the level spacing and relative energies to the $^{10}{\rm Be}+\alpha$ threshold
of the observed states agree well with the AMD prediction \cite{Suhara:2010ww}.}
However, it is clear that
a definitive conclusion has yet to be reached here.

\subsubsection{Key measurements that constrain {\it ab initio} theory}
Clustering reveals much about the nature of the force through which the constituent
components of the nucleus interact and the symmetries that result. This provides
a crucial connection with {\it ab initio} theory. The nuclear strong interaction
is clearly complex and this is revealed in the details of the unbound and
bound light nuclei. The $\alpha$-particle is one of the most highly bound
light nuclei with a very high-lying, $\sim 20$~MeV, first excited state. And
here the array of correlations include not only $n-n$ and $p-p$ but
 also $n-p$ to maximize the binding energy. The tendency of other nuclei to optimize
their own binding by generating spatial and momentum correlations induces
the formation of clusters. This is responsible for clustering in $\alpha$-conjugate
nuclei, Borromean and molecular systems, alike.

Nuclei that display extreme or exotic behavior where the effects of the
correlations are maximal are an excellent test of theory. Good examples of
this are the ground and excited states of $^8$Be and the Hoyle state in $^{12}$C,
which have both been described above. To be useful in constraining
theory and providing discrimination between approaches, high precision measurements
are often required. One of the best examples of this is the study of the $T=1$
analogue states in $^{10}$Be, $^{10}$C and $^{10}$B$^*$.

Precision measurements of the lifetime of first 2$^+$ state in $^{10}$C using
the Doppler Shift Attenuation Method deduced a lifetime of
$\tau$ = 219~$\pm$(7)stat $\pm$(10)sys~fs,
corresponding to a $B(E2)$ of 8.8(3)~$e^2$fm$^4$~\cite{McCutchan:2012tw}.
Similar measurements of the same transition in
$^{10}$Be found a $B(E2)$ of 9.2(3)~$e^2$fm$^4$~\cite{McCutchan:2009th}.
The ground and 2$^+$ states of these nuclei are believed to possess a molecular
structure where two valence particles (2 neutrons or 2 protons) orbit the
2$\alpha$-particle cores. These measurements were compared with both the
Green's function Monte Carlo and no-core shell model (NCSM) calculations. The reproduction
of the experimental results, especially the GFMC calculations, was not satisfactory and showed
significant sensitivity to the details of the 2-- and 3--body forces employed. A subsequent measurement of the $B(E2)$ for
the transition from the $J = 2,$ $T = 1$ state at 5.164~MeV to the $J = 0,$
$T = 1$
state at 1.740~MeV in $^{10}$B found a value of 6.1(22)~e$^2$fm$^4$ ~\cite{McCutchan:2012wx}.
This is much lower than the simple average of the $^{10}$Be and $^{10}$C
measurement, which may not simply be understood and stands as an important
test of {\it ab initio} theory.
\begin{figure}
  \vspace{-0.20cm}
   \begin{center}
    \hspace{-.6cm}
    \includegraphics[angle=90.0,width=0.4\textwidth]{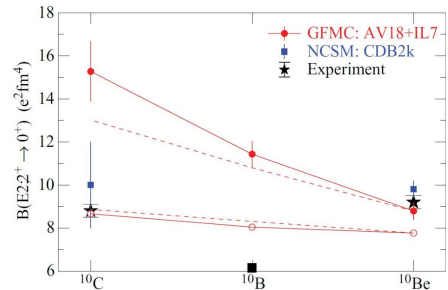}
     \vspace{-0.3cm}
     \end{center}
      \caption{(Color online) Comparison between experimental $B(E2)$ values and GFMC and
NCSM calculations using the AV18 potential with the IL7 three-nucleon interactions. The dashed lines show the corresponding isospin-symmetric results.  See  Ref.~\cite{McCutchan:2012tw} for the full details. {The black square at $^{10}$B illustrates the experimental value of 6.1~e$^2$fm$^4$ ~\cite{McCutchan:2012wx}.}  }
    \label{fig:Lister}
 \end{figure}

This set of measurements is a fine example of the need for precision experimental
data to properly understand the nature of the strong interactions in light
nuclei and the ability of first principles approaches to reproduce
experimental properties. This must be a significant area of effort for experiment
and theory over the next decade.

%%%%%%%%%%%%%%%%%%%%%%%%%%%%%%%%%%%%%%%%%%%%%%%%%%%%%%%%%%%%%%%%%%%%%%%
%
% \input{enyo_section.tex}
%

\section{Microscopic cluster models and antisymmetrized molecular dynamics}
\label{AMDsection}
\newcommand{\bvec}{\boldsymbol}

Various cluster phenomena in stable nuclei have been theoretically
investigated using microscopic cluster models such as resonating group methods (RGM) and generator coordinate methods (GCM).
In the progress of understanding the physics of unstable nuclei, cluster models have been extended
to deal with cluster structures with valence neutrons. Moreover,
more flexible methods such as antisymmetrized molecular dynamics
(AMD) \cite{Ono:1991uz,Ono:1992uy,Ono2004501,KanadaEnyo:1995tb,KanadaEnyo:2001qw,KanadaEnyo:2012bj}
and fermionic molecular dynamces (FMD) \cite{Feldmeier:1989st,Feldmeier:1994he,Feldmeier:2000cn,Neff:2002nu,Neff:2003ib}
have been developed which do not rely on the {\it a priori} assumption of
existence of clusters.
The AMD and FMD wave functions are based not on cluster degrees of freedom
but on nucleon degrees of freedom. In this sense, these models are not
cluster models. Nevertheless, since they
can express various kinds of cluster structures as well as shell-model features,
they are powerful approaches for the study of cluster features in general nuclei.

{
In this section we give a brief overview of cluster models, the AMD method, and their extensions.
Then we discuss some topics focusing on how these models describe the coexistence of
cluster and mean-field aspects. It should be commented that the AMD and FMD methods were originally
developed  in the time-dependent form for nuclear reaction studies. However, we here
 describe the models for structure studies.
 For details of the AMD method for nuclear structure and reaction studies,
see Refs.~\cite{KanadaEnyo:2001qw,KanadaEnyo:2012bj,Kimura:2016fce}
and references therein.}

\subsection{Overview of microscopic cluster models}

Since the 1960's, microscopic cluster models have been applied to investigate
cluster phenomena such as nuclear scattering and cluster structures.
In the early history, scattering between light nuclei
such as $\alpha+\alpha$ scattering has been
intensively studied with the RGM \cite{Wheeler:1937zza,Wheeler:1937zz,WILDERMUTH1958150,Tang:1978zz}.

Despite of  the success of the RGM in microscopic description of relative
motion between composite particles,
practical application of the RGM
is limited to the light mass region because of the computational costs of antisymmetrization in the
treatment of
the norm and Hamiltonian kernels.
We discuss some recent developments in subsection~\ref{ncsmc} in {\it ab initio} no-core shell model calculations.

Since 1970's,  owing to application of the GCM \cite{Hill:1952jb,Griffin:1957zza}
using the Bloch-Brink cluster wave function \cite{Brink},
further progress of microscopic studies of cluster phenomena has been made
for
heavy mass and many-cluster systems as well as unstable nuclei \cite{Fujiwara-supp}.
The RGM and GCM are microscopic cluster models, in which
antisymmetrization of all nucleons composing clusters
are fully taken into account, and the Hamiltonian
is composed of nucleon kinetic energies and
nucleon-nucleon interactions based on nucleon degrees of freedom.
Clusters are usually written in terms of simple
shell-model configurations
with/without excitation, and the inter-cluster motion is solved within the
model wave functions.

The model wave function of the RGM for a single-channel case of two clusters
$C_1$ and $C_2$
is given as
\begin{equation}
\Psi_\textrm{RGM}={\cal A}\left\{ \phi(C_1)\phi(C_2) \chi(\bvec{\xi}) \right\},
\end{equation}
where ${\cal A}$ is the nucleon antisymmetrizer,
$\phi(C_i)$ is the internal wave function of the $C_i$-cluster, and
$\bvec{\xi}$ is the relative coordinate between the centers of mass of the
clusters.
The inter-cluster wave function $\chi(\bvec{\xi})$ is determined by solving
the RGM equation derived from the projection of the Schr\"odinger equation onto
the RGM model space. Distortion of clusters and multi-channel systems can
be
taken into account in the RGM by extending the single-channel to coupled-channel
problems.

To describe the inter-cluster motion with the GCM approach,
Brink adopted the following multi-center cluster wave function (called the
Bloch-Brink
cluster wave function) as a basis wave function \cite{Brink},
\begin{equation}
\Phi_\textrm{BB}(\bvec{S}_1,\ldots,\bvec{S}_k)=n_0{\cal A}\left\{\psi(C_1;\bvec{S}_1)
\cdots \psi(C_k;\bvec{S}_k) \right\},
\end{equation}
where the $i$th cluster ($C_i$) is localized around $\bvec{S}_i$,
and $n_0$ is a normalization constant. The wave function
$\psi(C_i;\bvec{S}_i)$ for the $i$th cluster is written in terms of the
harmonic oscillator shell-model wave function
located at $\bvec{S}_i$. When the
clusters are far from each other and feel weak antisymmetrization effects
between clusters,
the parameter $\bvec{S}_i$ indicates the mean center position of the cluster,
and hence,
the spatial configuration of the parameters $\{\bvec{S}_1,\ldots,\bvec{S}_k
\}$
specifies the geometry of cluster structures. It means that
the single Bloch-Brink cluster wave function
expresses a cluster wave function, in which
centers of clusters are localized around certain positions.
In the small distance ($|\bvec{S}_i|$) case
that clusters largely overlap with each other,
the Bloch-Brink cluster wave function becomes a specific shell-model wave
function
of the SU(3) shell model
because of antisymmetrization of nucleons among clusters.

For the detailed description of
inter-cluster motion, the superposition of the Bloch-Brink wave functions
is considered
by adopting the cluster center parameters $\{\bvec{S}_1,\ldots,\bvec{S}_k
\}$
as generator coordinates in the GCM approach,
\begin{eqnarray}
\Psi_\textrm{GCM}&=&\int d\bvec{S}_1,\ldots,d\bvec{S}_k
f(\bvec{S}_1,\ldots,\bvec{S}_k) \nonumber \\
&&\times P^{J\pi}_{MK}\Phi_\textrm{BB}(\bvec{S}_1,\ldots,\bvec{S}_k),
\end{eqnarray}
where $P^{J\pi}_{MK}$ is the total-angular momentum and parity projection
operator, and
coefficients $f(\bvec{S}_1,\ldots,\bvec{S}_k)$ are determined by solving
the Hill-Wheeler equation \cite{Hill:1952jb}.
In principle, the GCM with full model space of the basis Bloch-Brink wave
functions
is equivalent to the RGM \cite{Horiuchi:1970}.
With the GCM approach it became possible to practically calculate
heavy mass systems and also many-cluster systems microscopically.

%Note that the THSR wave function is given by a superposition of the Bloch-Brink
%wave functions for multi-$\alpha$ systems with a Gaussian weight and it
%is able to
%express the localized and non-localized clustering states.

For scattering problems, the RGM can be applied rather straightforwardly
because
the inter-cluster wave function is explicitly treated.
On the other hand, in the application of the GCM to scattering problems,
it is necessary to connect the basis wave functions in the internal region
with continuum states in the asymptotic region at a chosen channel radius \cite{Kamimura01021977,Descouvemont:2010cx}.

Based on the GCM, Bay and Descouvemont studied various
low-energy reactions of astrophysical interest \cite{Descouvemont:2010cx}.
Following the progress in the physics of unstable nuclei, the microscopic
cluster approaches have been extended and applied to study cluster structures
of unstable nuclei.
One of the main interests in the study of unstable nuclei are properties
of valence neutrons surrounding one core or two clusters in neutron-rich
nuclei.
Microscopic three-body calculations for two valence neutrons around a core
nucleus
have been achieved by many groups to investigate the neutron halo and
two-neutron correlation in drip-line nuclei such as $^6$He, $^{11}$Li, and
$^{14}$Be \cite{Varga:1993wp,Arai:1999zz,Descouvemont:1995zz,Descouvemont:1997zkg}.
%[14,15,70,71,78,269,274,276].
Baye and Descouvemont have studied cluster features of the Be isotopes using
a GCM
approach with the Bloch-Brink wave functions of two $\alpha$-clusters and
valence neutrons \cite{Descouvemont:2002mnw}.
They have also applied the coupled-channel GCM of
$^6{\rm He}+{}^6{\rm He}$ and $^8{\rm He}+\alpha$ channels
to study cluster structures of $^{12}$Be \cite{Descouvemont:2001wek,Dufour:2010dmf}.

 Ito {\it et al.} have applied a more generalized approach of the
coupled-channel GCM to $^{10}$Be and $^{12}$Be \cite{Ito:2003px,Ito:2008zza,Ito14-rev}.
The method is successful in describing
gradual changes of valence neutron configurations
from strong-coupling clustering with a molecular orbital structure
to
weak-coupling clustering in the asymptotic region with the increase of
the $\alpha$-$\alpha$
distance.
Varga and his collaborators have performed
accurate calculations for many cluster systems in unstable $p$-shell nuclei
with the stochastic variational method (SVM) \cite{Varga:1993wp,Varga:1994mg,Varga:1995dm,Arai01032001}.
The SVM is a microscopic cluster model with the RGM-type cluster wave function
written as
a linear combination of stochastically chosen basis wave functions.
Because of the stochastic procedure in choosing the basis wave functions,
it is a powerful approach to treat many cluster systems. For instance, it
has been applied to
accurately solve four-cluster problems in the study of unstable $p$-shell nuclei
such as the $2\alpha+2n$ system of $^{10}$Be \cite{Varga:1994mg,Ogawa:1998et,Arai01032001,Arai:2004yf}.
%[16,220,275,276].

To understand cluster structures of low-lying states of neutron-rich Be
isotopes, molecular orbitals (MO) for surrounding neutrons around the $2\alpha$
core were
proposed \cite{vonOertzen:2006a,SEYA,OERTZENa,Itagaki:1999vm,Ito14-rev}.
Microscopic MO models \cite{OKABEa,OKABEb}
have been developed and applied to $^{10}$Be by Itagaki {\it et al.} \cite{Itagaki:1999vm}.
The model is based on the GCM
for $2\alpha+2n$ using a truncated model space. Neutron configurations are
restricted to the MOs, which are covalent bond orbitals written
as linear combinations of $p$-orbits around each $\alpha$-cluster, whereas
the $\alpha$-$\alpha$ distance is treated as the generator coordinate. The
molecular orbital models
have been also applied to neutron-rich C isotopes with the $3\alpha$ core
and valence neutrons
\cite{Itagaki:2001mb}.

\begin{figure}
\begin{center}
\includegraphics[width=0.3\textwidth]{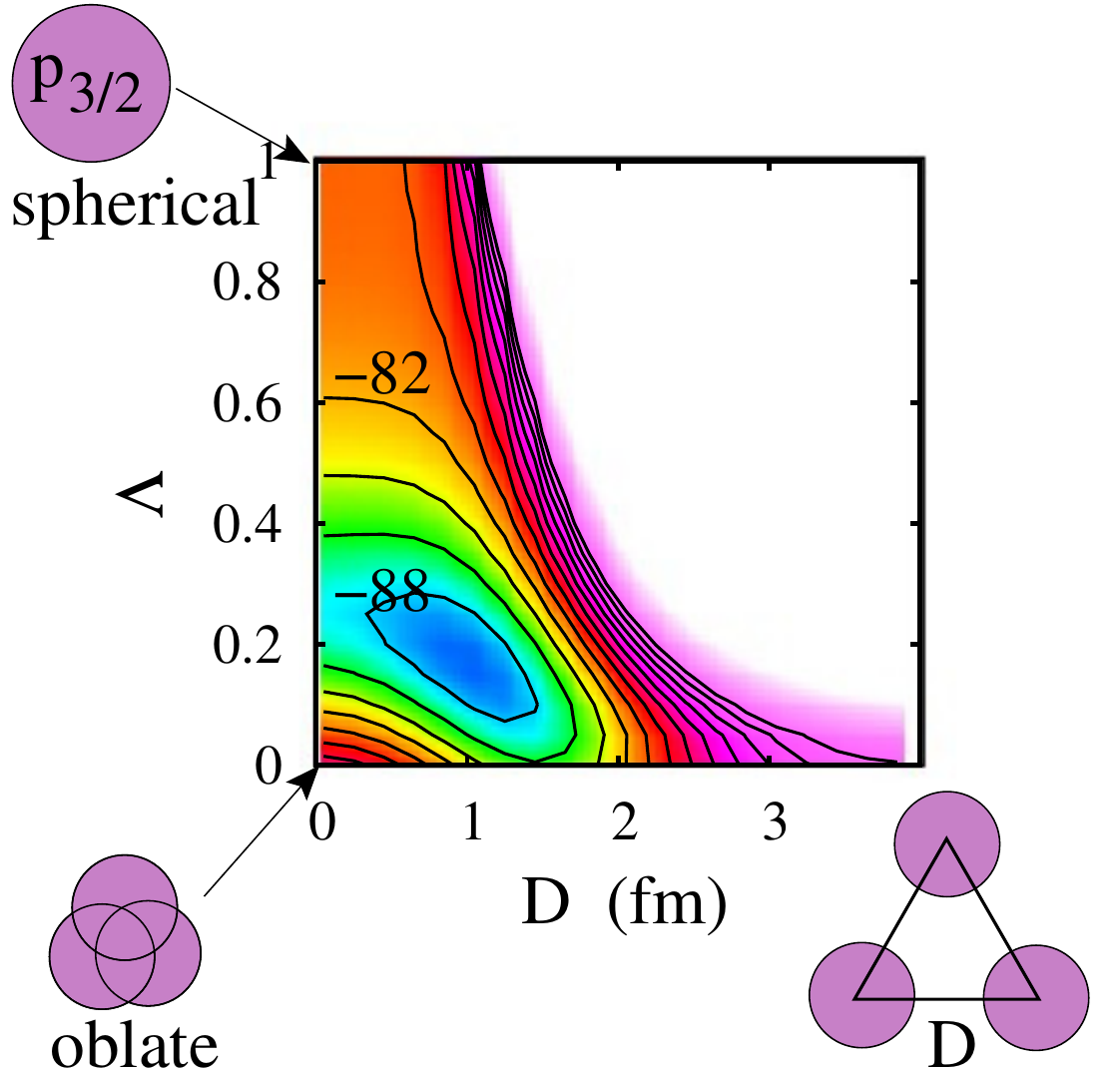}
\caption{\label{fig:ls-jj}(Color online) $0^+$-projected energy surface on the $\Lambda$-$D$
plane for  $^{12}$C calculated by the AQCM. The interaction and width parameters are same as those
in Ref.~\cite{Suhara:2013mja}.
}
\end{center}
\end{figure}

{
The relation of the cluster wave functions with shell model ones was described
based on the harmonic oscillator basis expansion and discussed from the SU(3) group symmetry
\cite{WILDERMUTH1958150,BAYMAN1958596,Elliott:1958zj,Elliott:1958yc}. The concept has
been followed by symmetry-adapted models such as
symplectic (no-core) shell models \cite{Rowe:1980rs,Rowe:2010a,Draayer:1983fni,Dytrych:2007sv}
and algebraic cluster models \cite{Bijker:2002ac,Cseh:1992zhb,Cseh:2014qwa}.

To describe competition between the cluster and $jj$-coupling shell model states,
Itagaki {\it et al.} extended the Bloch-Brink Alpha cluster  model wave function
by adding spin-dependent imaginary parts to the cluster center parameters.  This is essential for spin-orbit interactions in the $jj$-coupling shell-model \cite{Itagaki:2005sy,Suhara:2013mja}.
The model is called antisymmetrized quasi-cluster model (AQCM) and can efficiently describe the
smooth transition from the $\alpha$-cluster wave function
to the $jj$-coupling shell model wave function in $^{12}$C
with the cluster breaking parameter $\Lambda$ from $\Lambda=0$ to $\Lambda=1$
as shown in Fig.~\ref{fig:ls-jj}.}

\subsection{Antisymmetrized molecular dynamics method}

 The AMD method is an approach which treats nucleon degrees of freedom independently
without assuming any clusters. Nevertheless, the AMD can describe various
cluster structures
because the Bloch-Brink cluster wave functions for any cluster channels are
contained
in the AMD model space.

An AMD wave function is given by a Slater determinant of single-nucleon
Gaussian wave functions,
\begin{equation}
 \Phi_{\rm AMD}({\bvec{Z}}) = \frac{1}{\sqrt{A!}} {\cal{A}} \{
  \varphi_1,\varphi_2,...,\varphi_A \},\label{eq:slater}
\end{equation}
where the $i$th single-particle wave function
$\varphi_i$ is written by a product of
spatial, spin, and isospin
wave functions as
\begin{eqnarray}
 \varphi_i&=& \phi_{{\bvec{X}}_i}\chi_i\tau_i,\\
 \phi_{{\bvec{X}}_i}({\bvec{r}}) & = &  \left(\frac{2\nu}{\pi}\right)^{4/3}
\exp\bigl\{-\nu({\bvec{r}}-\bvec{X}_i)^2\bigr\},
\label{eq:spatial}\\
 \chi_i &=& \left(\frac{1}{2}+\xi_i\right)\chi_{\uparrow}
 + \left(\frac{1}{2}-\xi_i\right)\chi_{\downarrow}.
\end{eqnarray}
$\phi_{{\bvec{X}}_i}$ and $\chi_i$ are the spatial and spin functions, respectively,
and
$\tau_i$ is the isospin
function fixed to be up (proton) or down (neutron).
The width parameter $\nu$ is fixed to be an optimized value
for each nucleus.

In the AMD wave function, the $i$th single-particle wave function is expressed
by
the Gaussian wave packet localized around the position $\bvec{X}_i$.
The Gaussian center positions $\bvec{X}_i$ and  the
intrinsic-spin orientations $\xi_i$ for all nucleons are treated independently
as
variational parameters which are determined by energy variation.
{
It should be noted that the AMD wave function can be interpreted as an extended version of the
Bloch-Brink wave function
in a sense that all clusters are resolved completely to single nucleons.
The AMD model covers the Bloch-Brink cluster model space
as well as the AQCM. Indeed, }
by choosing a specific configuration of the Gaussian center
positions
$\{\bvec{X}_i\}$, the AMD wave function can express the Bloch-Brink
and AQCM wave functions.
If a system favors a specific cluster structure, that structure is automatically
obtained in the AMD model space after energy variation.
The AMD wave function can also describe shell-model configurations
because of the antisymmetrization between nucleons.
It means that
formation and dissolution of clusters are taken into account
owing to the flexibility of the model wave function.
This is a great advantage superior to cluster models in description of both
cluster and mean-field features
in the ground and excited states of exotic nuclei.
As an extension of the AMD wave function,
triaxially-deformed Gaussian wave packets were proposed by Kimura {\it
et al.}
instead of the spherical Gaussian
wave packets \cite{Kimura:2003uf}.
The deformed basis AMD is efficient  to describe
the coexistence of deformed mean-field states and cluster states
in $sd$- and $pf$-shell nuclei. See Refs.~\cite{KanadaEnyo:2012bj,Kimura:2016fce}
and references therein.
{It should be commented that deformed Gaussian wave packets
have been proposed in Ref.~\cite{Bauhoff:1985zz} for
the time-dependent cluster model (TDCM) \cite{DROZDZ1982145,CAURIER1982150}.
}

For structure studies in the AMD framework,  the energy variation is performed
after
parity projection. For the angular-momentum projection, the variation
before
the projection (VBP) is performed in the simple AMD \cite{KanadaEnyo:1995tb},
whereas the variation is performed
after the projection (VAP) in the AMD+VAP method \cite{KanadaEnyo:1998rf}.

For the description of excited states, the
AMD wave functions obtained by the energy variation are superposed. For instance,
mixing of different basis AMD wave functions (multiconfiguration mixing)
is usually done
in the AMD+VAP method.
In the AMD+GCM method, many AMD wave functions are superposed
by means of the GCM with
constraint parameters as generator coordinates.
In the $\beta$- and $\beta\gamma$-constraint AMD  \cite{Kimura:2003uf,Suhara:2009jb},
the energy variation is done under the
constraints on the deformation parameters $\beta$ and $(\beta,\gamma)$,
respectively.
In the $d$-constraint AMD \cite{Taniguchi:2004zz},
the constraint for the distance between two (or three) centers of subgroups
is adopted.
After the energy variation with the constraints, the obtained AMD wave functions
are superposed with the GCM treatment. Namely, coefficients of wave functions
are determined by solving
the Hill-Wheeler equation, i.e., by diagonalizing the norm and Hamiltonian
matrices
with respect to the adopted basis AMD wave functions.
In the AMD+GCM, large amplitude dynamics along the generator coordinates
are microscopically taken into account.

{
Although the AMD+GCM is useful for large amplitude collective motion, it is not efficient
to describe single-particle excitations on a mean-field state because the lowest state is chosen
in the energy variation procedure. To overcome this problem,
the shifted basis AMD {(sAMD)} \cite{KanadaEnyo:2015vkg,Kanada-Enyo:2015knx,Chiba:2015khu}
has been constructed to describe small amplitude modes on top of the ground state. 
A small shift of the Gaussian center position of each single-particle
wave function is prepared on the ground state AMD wave function and all the shifted bases wave functions
are superposed to describe linear combinations of one-particle and one-hole (1p-1h) excitations.
The method combined with the cluster GCM was applied to monopole and dipole excitations in light nuclei
and described coexistence of low-energy cluster modes and high-energy giant resonances.
}

Since the basis AMD wave function is written as a Slater determinant
of single-particle wave functions, the simplest case of the
single AMD wave function without projections can be regarded as
a Hartree-Fock approach simplified in the restricted model space.
However, because of the linear superpositions as well as the
parity and angular momentum projections
of AMD wave functions, higher correlations beyond mean-field approaches
are taken into account even in the ground state in the AMD framework. As
mentioned previously, the AMD
model contains mean-field states as well as various cluster states in its
model space
and therefore it is able to describe the coexistence of mean-field and cluster
aspects
in the ground and excited states of nuclear systems.

{
\subsection{Time-dependent antisymmetrized molecular dynamics method}

The AMD wave function was originally used for nuclear reaction studies
in a time-dependent framework \cite{Ono:1991uz,Ono:1992uy,Ono2004501}.
In the time-dependent AMD, the spin functions $\chi_i$ are usually fixed to be
$\chi_\uparrow$ or $\chi_\downarrow$, and time evolution of a system is
described by the time-dependent Gaussian center positions $\bvec{X}_i$
determined by the time-dependent variational principle as
\begin{eqnarray}
&&i\hbar\sum_{j\rho} C_{i\sigma, j\rho} \frac{d X_{j\rho}}{dt}=\frac{\partial}
{\partial X^*_{i\sigma}}
\frac{\langle \Phi_ {\rm AMD}({\bvec{Z}})|H| \Phi_ {\rm AMD}({\bvec{Z}}) \rangle}
{\langle \Phi_ {\rm AMD}({\bvec{Z}})|\Phi_ {\rm AMD}({\bvec{Z}}) \rangle}\nonumber\\
&&C_{i\sigma, j\rho}\equiv\frac{\partial^2}
{\partial X^*_{i\sigma}\partial X_{j\rho}}{\rm In}
{\langle \Phi_ {\rm AMD}({\bvec{Z}})|\Phi_ {\rm AMD}({\bvec{Z}}) \rangle},
\end{eqnarray}
where $\sigma,\rho=x,y,z$.
The time-dependent AMD can be regarded as an extended version of the TDCM
\cite{DROZDZ1982145,CAURIER1982150,Bauhoff:1985zz} in the sense that all clusters are resolved completely into single nucleons.

In applications of the AMD and extended versions to heavy-ion collisions, the
stochastic two-nucleon collision term is added to the equation of motion. The model
successfully described multifragmentations at intermediate energy.
Feldmeier has proposed a wave function quite similar to the AMD wave function
for nuclear reactions and structure studies \cite{Feldmeier:1989st,Feldmeier:1994he,Feldmeier:2000cn}
and named it fermionic molecular dynamics (FMD).
The model wave function of the FMD is also given by a Slater determinant
of single-nucleon
Gaussian wave packets.
The major difference in the wave function
between the FMD and AMD is that the width parameter $\nu$ can be independently chosen
for each nucleon as $\nu_i$ and treated as variational parameters in the FMD, whereas  it
is common for all nucleons in the AMD. Instead, the diffusion and the deformation of wave
packets are stochastically incorporated in an extended version (AMD+Vlasov) for
reaction studies \cite{Ono:1995uw}.

In structure studies the flexible treatment of the width parameters in the
FMD
is efficient, for example, for the neutron-halo structure of neutron-rich
nuclei. The variation of width parameters in the time-dependent FMD
is also effective in description of the giant monopole resonance \cite{Furuta:2010ad}, whereas
the giant resonances are described with superposition of shifted single-particle Gaussian wave packets in the sAMD framework.}

%Another difference in the structure calculations between the AMD
%and FMD is
%the effective nuclear interactions, as explained below.

\subsection{Effective nuclear interactions}

In the cluster model and the AMD calculations,
phenomenological effective nuclear interactions composed of the two-body
central and spin-orbit ($ls$) forces
are usually used.
The Hamiltonian consists of the kinetic energies, the effective nuclear interactions,
and
Coulomb interaction as
\begin{eqnarray}
&H=\displaystyle\sum_i t_i -T_\textrm{c.m.} +\displaystyle\sum_{i<j} v^\textrm{nuclear}_{ij}+\displaystyle\sum_{i<j}
v_{ij}^\textrm{Coulomb}, \nonumber\\
&v^\textrm{nuclear}_{ij}=v^\textrm{central}_{ij}+v^{ls}_{ij},
\end{eqnarray}
where the center-of-mass kinetic energy $T_\textrm{c.m.}$ is subtracted.
Note that the center-of-mass motion can be easily separated from the wave
functions
in the cluster and AMD models when a common width parameter is used.
For the central forces of the effective nuclear interactions, Gaussian
finite-range interactions with and without
the zero-range density-dependent term (or the zero-range three-body term)
are adopted in most cases.
The central forces are supplemented with the finite-range or zero-range $ls$
forces.

For light mass nuclei, density-independent interactions such as
the Minnesota \cite{THOMPSON197753} and Volkov \cite{Volkov:1965zz} interactions are
often used.
The Minnesota force is originally
adjusted to fit the $S$-wave nucleon-nucleon scattering
as well as scattering between light nuclei.
For the Volkov force,
the standard parameter set reproduces $\alpha$-$\alpha$ scattering.
The Volkov force can be adjusted to fit
the $S$-wave nucleon-nucleon scattering lengths
by tuning parameters for the Bartlett and Heisenberg terms.
In general, these density-independent effective interactions cannot
describe the saturation properties of nuclear matter and
have  an overbinding problem in heavy mass nuclei. Therefore,
interaction parameters are sometimes readjusted to reproduce energies
for the mass number region of interest, though the original parameter sets
reproduce the properties of nucleon-nucleon and $\alpha$-$\alpha$ scattering.

To overcome the overbinding problem, the central forces
with the zero-range density dependent or zero-range three-body term are
used for heavy mass nuclei.
Examples are the Gogny forces \cite{Berger:1991zza} and the Modified Volkov
(MV) forces \cite{Ando:1980hp}.
The central force of the Gogny forces consists of finite-range
two-body terms and a zero-range density dependent term, whereas
that of the MV forces contains a zero-range three-body term instead of
a zero-range density-dependent term.
These interactions systematically reproduce the binding energies
over a wide mass number region. However, they cannot quantitatively
reproduce the scattering and structure properties of very light systems
such as the nucleon-nucleon and $\alpha$-$\alpha$ scattering as well as
the size of the $\alpha$-particle.

There are many cluster model calculations for light nuclei
using the Minnesota and Volkov forces.
In the AMD calculations for $p$-shell, $sd$-shell, and $pf$-shell nuclei,
the Volkov, Modified Volkov No.1 (MV1), and Gogny forces are used.
As already mentioned,
these interactions used in the cluster model and AMD calculations are
effective nuclear interactions that are phenomenologically adjusted to
properties of nuclear structures and/or scattering.

In the FMD calculations, effective nuclear interactions
derived from the realistic nuclear interactions are usually
used with the unitary correlation operator method (UCOM), in which
the short-range and tensor correlations are taken into account
in the interaction operator of the Hamiltonian \cite{Feldmeier:1997zh,Neff:2002nu,Roth:2010bm}.
Therefore the FMD+UCOM calculation is a first principles method starting
from realistic nuclear
interactions.

{
\subsection{Description of cluster and mean-field aspects in AMD models}

\subsubsection{Cluster breaking effects on $3\alpha$ structures in $^{12}$C}
Despite the success of $3\alpha$-cluster models for many excited states $^{12}$C,
microscopic $3\alpha$-cluster models are not sufficient to describe
the large level spacing between the $0^+_1$ and $2^+_1$ states
because $\alpha$-cluster breaking is not taken into account in the models.
Moreover, it is difficult to confirm the $3\alpha$ cluster formation
in the 12-nucleon dynamics
because clusters are {\it a priori} assumed in the models.
These problems have been overcome by the AMD and FMD models.
In the AMD and FMD calculations for $^{12}$C
\cite{KanadaEnyo:1998rf,KanadaEnyo:2006ze,Neff:2003ib,Chernykh:2007zz},
$3\alpha$-cluster structures are formed in the calculated results
without assuming the existence of  $\alpha$-clusters.

As mentioned previously, the model spaces of the AMD and FMD contain
the Bloch-Brink cluster wave functions and also cluster breaking configurations.
In  the $^{12}$C($0^+_1$), the cluster breaking component, i.e.,
the $p_{3/2}$ closed-shell component is significantly mixed
in the dominant  $3\alpha$ cluster structure
as seen in the compact intrinsic density distribution in Fig.~\ref{fig:exj}.
Due to the mixing of the cluster breaking component,
the band-head $^{12}$C($0^+_1$) gains extra energy of the spin-orbit attraction
resulting in stretching of the $0^+$-$2^+$ level spacing consistently with the experimental
energy spectra. It is also the case in the FMD calculation (see Fig.~\ref{fig:c12-comparison}).
The significant mixing of the cluster breaking component
in $^{12}$C($0^+_1$) is clearly indicated  in the AQCM calculation in Fig.~\ref{fig:ls-jj}
by the  finite value of the cluster breaking parameter $\Lambda$ at the energy minimum.

The cluster breaking component does not give drastic effects to excited $3\alpha$ cluster states.
However, excited $0^+$ structures are more or less affected by the cluster breaking component mixed
in the $^{12}$C($0^+_1$) through the orthogonality, and therefore,
quantitative differences can be seen between model calculations with and without the cluster breaking.
For example, in the calculated energy spectra shown in Fig.~\ref{fig:c12-comparison},
the AMD and FMD calculations show a trend of the larger $0^+_2$-$2^+_2$ level spacing than that obtained by
$3\alpha$ calculations without the cluster breaking, because the cluster breaking in the
$^{12}$C($0^+_1$) induces the global energy gain of excited $0^+$ states.
Moreover, the AMD calculation shows the larger $E2$ strength for $2^+_2\to 0^+_3$
than that for  $2^+_2\to 0^+_2$, differently from the $3\alpha$ calculations that give dominant
$E2$ strength for  $2^+_2\to 0^+_2$.
Suhara and Kanada-En'yo investigated the cluster breaking effects on $3\alpha$ cluster structures in $^{12}$C
by explicitly adding the $p_{3/2}$ closed-shell configuration (cluster breaking component) into the $3\alpha$ model
space \cite{Suhara:2014wua}.
Comparison of the results with and without the $p_{3/2}$ configuration is shown in the right two columns
in the upper panel of Fig.~\ref{fig:c12-comparison}.  The figure shows increasing of the
$2^+_1$ and $2^+_2$ energies relative to the $0^+_1$, $0^+_2$, and $0^+_3$ ones,
and also the inversion of the dominant $E2$ strengths between $2^+_2\to 0^+_2$ and $2^+_2\to 0^+_3$.
This may indicate that cluster breaking should not be ignored for detailed discussions of
band assignments in model calculations.

\subsubsection{Cluster and mean-field modes in monopole excitations in $^{12}$C}

In experimental and theoretical studies of nuclear clustering,
isoscalar monopole (ISM) and dipole (ISD) transitions are good probes to pin down cluster states
\cite{Kawabata:2005ta,Funaki:2006gt,Yamada:2011ri,KanadaEnyo:2015vkg,Chiba:2015zxa}.
Yamada {\it et al.} pointed out that two different modes of ISM excitations coexist
in $^{16}$O\cite{Yamada:2011ri}: one is the isoscalar giant monopole resonance
(ISGMR) known to be the collective breathing mode, and the other is
the low-energy ISM strengths for cluster states.
The low-energy ISM strengths were experimentally observed also
for $^{12}$C \cite{Youngblood:1998zz,John:2003ke}.
A hybrid model of the shifted basis AMD
(sAMD) and
$3\alpha$-GCM was applied to the ISM excitations in $^{12}$C
and described the low-energy ISM strengths for cluster modes separating
from high-energy ISGMR strengths \cite{KanadaEnyo:2015vkg}.
The separation of the low-energy and high-energy parts of the ISM strengths
qualitatively agrees with the experimental data (see Figs.~\ref{fig:c12-ewsr} (a) and (b)).
As explained in the previous section,
the sAMD bases describe coherent 1p-1h excitations for the GMR, whereas
the $3\alpha$-GCM bases are essential for the
large amplitude cluster modes which contribute to the low-energy strengths.
Figure \ref{fig:c12-ewsr} (e) shows the ISM strengths obtained only by the sAMD bases
without $3\alpha$ configurations, and
Figures \ref{fig:c12-ewsr} (c) and (d) show the ISM strengths calculated using specific
$3\alpha$ configurations in addition to the sAMD bases.
As clearly seen, the sAMD describes only the high-energy ISM strengths for the ISGMR
but fails to describe significant low-energy ISM strengths. As $3\alpha$ configurations
are added to the sAMD bases, a peak grows up and comes down to the low-energy
region (see Fig.~\ref{fig:c12-ewsr} (c)). Then, the low-energy peak finally splits into the $0^+_2$ and
$0^+_3$ in the full sAMD+$3\alpha$GCM calculation
because of the coupling of the radial motion with the rotational motion of clusters.
Namely, the large amplitude cluster motion is essential for the low-energy ISM strengths and
the fragmentation of the ISM strengths occurs by the coupling of the radial and rotational
motions in the $3\alpha$ dynamics.

\begin{figure}
\begin{center}
\includegraphics[width=0.35\textwidth]{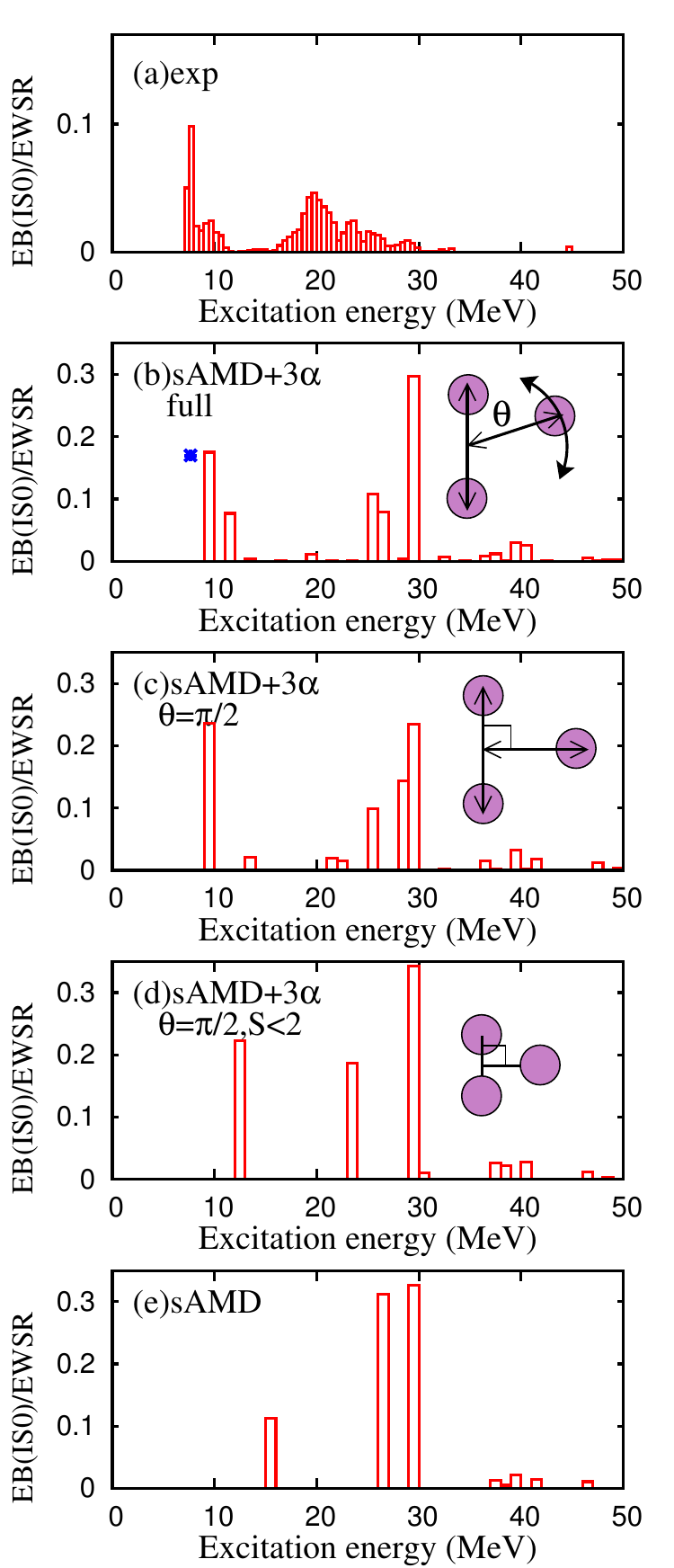}
\caption{\label{fig:c12-ewsr}(Color online) The energy weighted ISM
strength distributions
obtained by the sAMD+$3\alpha$GCM and those
measured by $(\alpha,\alpha')$ scattering \cite{John:2003ke}.
The experimental $E0$ strength for the $0^+_2$ measured by electron scattering
\cite{Chernykh:2010zu} is also shown in panel (b). (c) and (d) are
those calculated with a truncated model space of  $3\alpha$ configurations:
(c) calculation using $3\alpha$ configurations at $\theta=\pi/2$ and sAMD bases; (d)
same as (c) but only compact $3\alpha$ configurations with $|\bvec{S}_i|<2$ fm.
(e) shows strengths obtained by the sAMD bases without the $3\alpha$ configurations.
The figures (a) (b) (e) are from Ref.~\cite{KanadaEnyo:2015vkg}.}
 \end{center}
\end{figure}

\begin{figure}
\begin{center}
\includegraphics[width=0.45\textwidth]{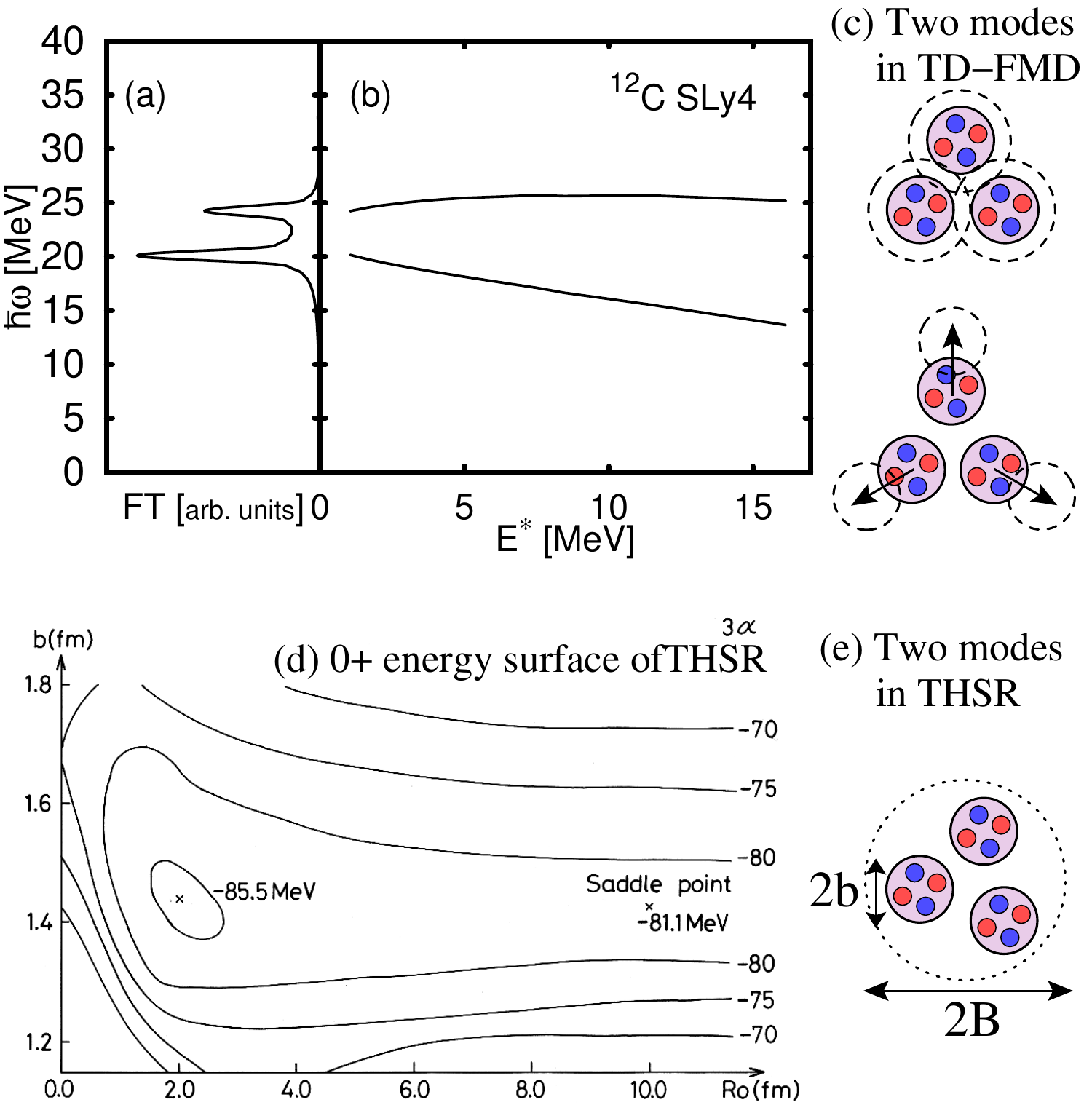}
\caption{\label{fig:Furuta}(Color online)
(a) (b) Monopole excitations in $^{12}$C calculated with the time-dependent FMD.
Oscillation frequencies of he root-mean-square radius are plotted against
the Fourier component in the small amplitude limit in (a), and those are plotted
as a function of the excitation energy given by the initial amplitude in (b).
(c) Sketches for width oscillation and inter-cluster modes in the time-dependent FMD.
(d) $0^+$ energy surface of $3\alpha$ TSHR wave function on the $B$-$b$ plane
($B$ is denoted by $R_0$). (e) Sketches for $b$ and $B$ modes
in the THSR. The figures (a) (b) are from Ref.~\cite{KanadaEnyo:2015vkg} and (d) from
Ref.~\cite{Tohsaki:2001an}.
}
 \end{center}
\end{figure}

The lowering mechanism of the ISM strengths by the large amplitude cluster motion
was also demonstrated in the time-dependent FMD calculation by Furuta {\it et al.}
\cite{Furuta:2010ad}.
In applications of time-dependent approaches to nuclear excitations, the response functions are
calculated by the Fourier transform (frequency) of the time evolution of the system. For the ISM excitations,
the initial state is prepared by imposing an external field (operator) $\sum_i \bvec{r}^2_i$ to the ground state, and starting from the initial state the time evolution of the system is solved
with time-dependent FMD.
In the FMD framework, the single-particle excitations are expressed by the time-dependent
width parameters of single-nucleon Gaussian wave packets, whereas the radial cluster motion is
described by the time-dependent Gaussian center positions. The Fourier transform of the root-mean-square
radius shows two modes with different frequencies corresponding to the width oscillation mode and
the radial cluster (inter-cluster) mode (see Fig.~\ref{fig:Furuta}).
They analyzed the
dependence of frequencies of two modes on the oscillation amplitude and found that
the higher frequency for the width mode, corresponding to the breathing mode,
does not depend on the amplitude. Rather, the lower peak frequency for the cluster mode moves down significantly to lower energy
as the amplitude becomes larger. This result is consistent with the sAMD+$3\alpha$GCM result discussed
previously, although the GMR mode is expressed by linear combinations of shifted Gaussian wave packets with
a fixed width in the sAMD instead of the variational width in the FMD wave function. Note that the quantization of excitation modes and spin-parity
projections are performed
in the sAMD+$3\alpha$GCM but they are not done in the TD-FMD.  It should be remarked that monopole vibrations in $^8$Be  have been investigated using
the TDCM and shows similar features for
the width oscillation and radial cluster modes \cite{DROZDZ1982145}.

It is also valuable to consider a link to the THSR wave function for two kinds of monopole modes.
In the THSR model, the width oscillation mode is expressed by the parameter $b$ for the $\alpha$-cluster size,
and the radial cluster mode is described by the parameter $B$ for the
$\alpha$ distribution size as shown in Fig.~\ref{fig:Furuta}(e). The $0^+$ energy surface on the $B$-$b$ plane
in  Fig.~\ref{fig:Furuta}(d) shows the coexistence of two modes.
The energy surface is very soft
along the $B$ mode and it is steep along the $b$ mode. It indicates that
the origin of the low-energy ISM strengths
is the large amplitude cluster motion decoupled from the width (coherent single-particle excitation) mode for
the ISGMR.
}

%%%%%%%%%%%%%%%%%%%%%%%%%%%%%%%%%%%%%%%%%%%%%%%%%%%%%%%%%%%%%%%%%%%%%%%
%
% \input{horiuchi_section.tex}
%

\section{Tohsaki-Horiuchi-Schuck-R\"opke wave function and container
model}
\label{THSR-section}
\subsection{Introduction}
\label{intro}

Cluster model studies in the 1970's showed that the Hoyle state of $^{12}$C has a gas-like structure of three $\alpha$
clusters which are weakly bound with predominantly $S$-wave correlations among the $\alpha$-particles~ \cite{HO74,uegaki1,Kamimura:1981oxj}. The gas-like structure of the
Hoyle state was reconsidered in a new light in Ref.~\cite{Tohsaki:2001an}.  In this paper it was proposed that the Hoyle state has a
3$\alpha$-condensate-like structure, and the Tohsaki-Horiuchi-Schuck-R\"opke (THSR) wave function was presented for the sake of expressing the $\alpha$-condensate-like structure.
It was soon discovered~\cite{Funaki:2003af} that the 3$\alpha$ THSR wave function was nearly identical to the 3$\alpha$ cluster-model
wave functions obtained in 1970's, namely the 3$\alpha$ Brink-GCM (generator coordinate method) wave function of Ref.~\cite{uegaki1} and
the 3$\alpha$ RGM (resonating group method) wave function of Ref.~\cite{Kamimura:1981oxj}.

About 10 years later it was found that the THSR wave
function for $^{16}{\rm O}+\alpha$ clustering in $^{20}$Ne was nearly identical to the Brink-GCM wave function for $^{16}{\rm O}+\alpha$ clustering ~\cite{zhou13}.
This finding was striking since the $^{16}{\rm O}+\alpha$ Brink-GCM wave functions with spatially-localized
$^{16}{\rm O}+\alpha$ structures describe accurately the states of the
so-called inversion-doublet bands of $^{20}$Ne where the even parity and odd parity levels are split into two separate bands.  The THSR wave function was found to
describe well the spatially-localized cluster structures even though it was originally
designed to describe gas-like delocalized cluster wave functions.  The fact that the THSR wave function can
describe both localized and delocalized clustering led to the introduction of the container model of cluster dynamics~\cite{zhou14}.

Here we discuss the THSR wave function and its history, starting from its initial introduction to the container model of cluster dynamics~\cite{Funaki:2015uya,SC16,TO17}.
We explain some  characteristics of the THSR wave function that might appear contradictory, such as the nucleon-density distribution
showing localized clustering despite the nonlocalized character of the THSR wave function and the equivalence of prolate
and oblate THSR wave functions after angular momentum projection. The container model is deeply connected to the evolution
of cluster structure, and  we demonstrate this in $^{12}$C and $^{16}$O.

\subsection{Alpha-condensate-like character of the Hoyle state}
\label{Alcond}

\subsubsection{$S$-wave dominance of $\alpha$-cluster motion in the Hoyle state}
\label{Sdominan}

The Hoyle state  is located slightly above the 3$\alpha$ and $^8$Be$(0^+_1)$ + $\alpha$ thresholds.
The small excitation energy of this state, 7.66~MeV, is very difficult to explain by the shell model.  The decay width $\Gamma$ of the Hoyle state is very small (8.7~eV) because the energy is well below the Coulomb barrier. Assuming for the moment two subsystems in a relative $S$ wave, the R-matrix calculation of the width yields
\begin{equation}
\Gamma = 2P_{L=0}(a) \gamma^2(a),
\end{equation}
where $\gamma^2(a)$ is the reduced width,
\begin{equation}
P_{L=0}(a) = ka/(F_{L=0}^2(ka) + G_{L=0}^2(ka))
\end{equation}
is the Coulomb barrier penetrability, $a$ is the channel radius, $F_L$ and $G_L$ are the regular and irregular Coulomb functions, respectively, and $k$ is the wave number.  See Ref. \cite{Descouvemont:2010cx} for a review.  The observed value $\gamma^2_{obs}(a)$ for the Hoyle state is very large.  It is comparable to or larger than the Wigner-limit value $\gamma^2_W(a)$ =
$3\hbar^2/(2\mu a^2)$ that corresponds to an $\alpha$ cluster with uniform density at radial distances less than $a$.  The very large value of $\gamma^2_{obs}(a)$ suggests that the structure of the Hoyle state is composed of an $^8$Be$(0^+_1)$ core and loosely attached $\alpha$ cluster in an $S$ wave.  This conclusion does not support the idea of the 3$\alpha$ linear-chain structure proposed by
Morinaga~\cite{morinaga66,Morinaga:1956zza},  since the 3$\alpha$ linear-chain structure would produce a reduced width $\gamma^2(a)$ that is significantly smaller than $\gamma^2_W(a)$ \cite{suzuki72}.

The $S$-wave dominance of the $^8$Be$(0^+_1)$-$\alpha$ relative wave function indicated by the observed $\alpha$-width was confirmed theoretically
by solving the 3$\alpha$ problem by the use of 3$\alpha$ OCM (orthogonality condition model)~\cite{HO74}.  Since $^8$Be$(0^+_1)$ consists of two
$\alpha$ clusters weakly coupled in a relative $S$ wave, the Hoyle state was concluded to have a weakly-coupled 3$\alpha$ structure in relative $S$
waves with large spatial extent.  It was therefore described as a gas-like state of $\alpha$ clusters.  A few years later, the results of the
3$\alpha$ OCM  study were confirmed by fully microscopic 3$\alpha$ calculations by two groups, namely the 3$\alpha$ GCM calculation of Ref.~\cite{uegaki1},
and 3$\alpha$ RGM calculation of Ref.~\cite{Kamimura:1981oxj}. These calculations nicely reproduced not only the excitation energy of the Hoyle state but
also other experimental properties including the $\alpha$-decay width, the inelastic electron-scattering charge form factor, and $E0$ and $E2$ transition properties.

\subsubsection{Equivalence of the 3$\alpha$ RGM/GCM wave function to a single 3$\alpha$ THSR wave function}
\label{AlconTHSR}

More than 20 years after the 3$\alpha$ OCM, GCM, and RGM  studies mentioned above, the Hoyle state was reconsidered in a new light in
Ref.~\cite{Tohsaki:2001an}. The authors of this paper proposed, for the description of the Hoyle state, the following new model wave function
$\Psi^{\rm THSR}_{3\alpha}$ called the THSR wave function.  Let $\Phi(3\alpha)$ be a simple product of three $\alpha$-cluster wave
functions,
\begin{equation}
\Phi(3\alpha) = \phi(\alpha_1) \phi(\alpha_2) \phi(\alpha_3).
\end{equation}
The THSR wave function has the form
\begin{eqnarray}
&& \Psi^{\rm THSR}_{3 \alpha}(B) \nonumber\\
&&={\cal A} \{ \exp [-\tfrac{2}{B^2} (\Vec{X}_1^2 + \Vec{X}_2^2 + \Vec{X}_3^2)] \Phi(3\alpha) \}      \label{eq:thsr}  \\
&& =\exp( -\tfrac{6\Vec{\xi}_3^2 }{B^2} ) {\cal A} \{ \exp( -\tfrac{4\Vec{\xi}_1^2}{3B^2}  - \tfrac{\Vec{\xi}_2^2}{B^2}  ) \Phi(3\alpha) \},
    \label{eq:thsr_be+a}
\end{eqnarray}
where $\Vec{X}_i$ is the center of mass of cluster $i$ and   $\Vec{\xi}_k$ are Jacobi coordinates defined as
\begin{eqnarray}
& \Vec{\xi}_1 = \Vec{X}_1 - \tfrac{1}{2} ( \Vec{X}_2 + \Vec{X}_3 ), \\
& \Vec{\xi}_2 = \Vec{X}_2 - \Vec{X}_3, \\
& \Vec{\xi}_3 = \tfrac{1}{3} ( \Vec{X}_1 + \Vec{X}_2 +
\Vec{X}_3 ).
\end{eqnarray}
With the center-of-mass dependence removed, the wave function has the form
\begin{equation}
\Phi (3\alpha {\rm THSR})  = C \Psi^{\rm THSR}_{3 \alpha}(B)/\exp(-\tfrac{6\xi_3^2}{B^2}),
\end{equation}
where $C$ is a normalization constant.  As shown in Eq.~(\ref{eq:thsr_be+a}), the THSR wave function can be regarded
as expressing the ${^8{\rm Be}(0^+_1)}+\alpha$ cluster structure,
where a ${^8{\rm Be}(0_1^+)}$-like cluster ${\cal A} \{ \exp(-\Vec{\xi}_2^2
/B^2) \phi(\alpha_2)\phi(\alpha_3) \}$ and the
$\alpha_1$  cluster couple via {\it S}-wave with inter-cluster wave function
$\exp[-4\Vec{\xi}_1^2/(3B^2)]$.
On the other hand, Eq.~(\ref{eq:thsr}) shows that the THSR wave function
represents the state where three $\alpha$-clusters occupy the same
single 0{\it S}-orbit $\exp(-2\Vec{X}^2/B^2)$, namely a 3$\alpha$ condensate
state which is a finite-size counterpart of the macroscopic
$\alpha$-particle condensation in infinite nuclear matter at low density \cite{Ro98}.
 What the authors of Ref.~\cite{Tohsaki:2001an} proposed was
that the ${^8{\rm Be}(0_1^+)} +\alpha$ structure of the Hoyle state can be
regarded as being a 3$\alpha$ condensate-like state. Furthermore
one can in general expect the existence of an $n \alpha$ condensate-like
state in the vicinity of the $n \alpha$ threshold in
$\alpha$-conjugate nuclei.  \\

An important and striking fact is that both the $3\alpha$ GCM wave function of Ref.~\cite{uegaki1} and the $3\alpha$ RGM wave function of
Ref.~\cite{Kamimura:1981oxj} are each nearly equivalent to a single $3\alpha$ THSR wave function \cite{Funaki:2003af}:  \begin{equation}
 | \langle \Phi (3\alpha {\rm THSR}) | \Phi (3\alpha {\rm GCM/RCM}) \rangle|^2 \approx 100\%.
\end{equation}
Hence the
3$\alpha$ THSR wave functions reproduce the same Hoyle state experimental data well described by the 3$\alpha$ RGM/GCM wave functions.
We refer the reader to the recent review in Ref. \cite{TO17} for applications of the THSR wave function to the Hoyle state and discussions of electric transitions, $\alpha$-condensation probabilities, and comparisons with quantum Monte Carlo calculations.

\subsection{Localized vs. nonlocalized clustering}
\label{locnonloc}

\subsubsection{Shell-model limit of THSR wave function}
\label{shellimit}

Let $b$ be the single-nucleon oscillator size parameter for the $0s$
H.O.
(harmonic oscillator) orbit,
\begin{equation}
\phi_{0s}(\Vec{r}) = (\pi b^2)^{-3/4} \exp(-\tfrac{\Vec{r}^2}{2b^2}).
\end{equation}
When $B = b$, we have
\begin{equation}
\exp(-\tfrac{2\Vec{X}^2}{B^2} ) \phi(\alpha) \propto
\det|(0s)^4|.
\end{equation}
Therefore
$\Psi^{\rm THSR}_{3 \alpha}(B=b)$ = 0 by the Pauli exclusion prin\-ciple.  By expressing the
normalization constant of $\Psi^{\rm THSR}_{3 \alpha}(B)$ with $B > b$ as
$n^{\rm THSR}_{3 \alpha}(B)$, one can prove the
relation
\begin{align}
\lim_{B \to b} n^{\rm THSR}_{3 \alpha}(B) & \Psi^{\rm THSR}_{3 \alpha}(B) \ \nonumber \\
& =|(0s)^4 (0p)^8; [444] L=0 \rangle,
\end{align}
where [444] refers to the spatial-symmetry Young diagram \cite{Hutzelmeyer1970}.
 This relation means that $\Psi^{\rm THSR}_{3 \alpha}(B)$ for $B$ close
to
$b$ is close to the shell-model wave function of the $^{12}$C
ground state.  So while the THSR wave function for large $B$ is a gas-like state of clusters, the THSR wave function with small
$B$ is a shell-model-like state.

In Ref.~\cite{Tohsaki:2001an}, the ground and Hoyle states of $^{12}$C were
obtained
as the lowest and second lowest energy states of the GCM equation
with the basis function $\Psi^{\rm THSR}_{3 \alpha}(B)$,
\begin{equation}
  \displaystyle\sum_B \langle \Psi^{\rm THSR}_{3 \alpha}(B') | ( H - E_k
) | \Psi^{\rm
THSR}_{3 \alpha}(B) \rangle f_k(B) = 0.   \label{eq:c12GCM}
\end{equation}
The GCM wave functions of the ground and Hoyle states were found to have
about 93\% and 98\% squared overlaps with single THSR wave functions.

\subsubsection{Inversion doublet bands of $^{20}$Ne and THSR wave function}
\label{invdoub}

In $^{20}$Ne the even-parity $K^\pi = 0^+$ rotational band upon the ground state and odd-parity $K^\pi = 0^-$ rotational band upon the
$J^\pi = 1^-$ state at 5.80~MeV  constitute
inversion-doublet bands having the same intrinsic $^{16}{\rm O}+\alpha$ cluster structure~\cite{hoik}.
The splitting between the even-parity and odd-parity bands can be understood as arising from tunneling of the $\alpha$ through the $^{16}{\rm O}$ core to form the corresponding mirror configuration.
The empirical success of this description constitutes evidence of spatial
localization of the clusters.  Much later it was discovered that the
GCM/RGM wave functions describing the inversion-doublet bands were found to be almost equivalent
to a single $^{16}{\rm O}+\alpha$ THSR wave
functions~\cite{zhou12,zhou13},
\begin{equation}
  |\langle \Phi(^{16}{\rm O}+\alpha\ {\rm THSR})|\Phi(^{16}{\rm O}+\alpha\
{\rm GCM/RGM}) \rangle|^2 \approx 100\%, \label{20NeEQUI}
\end{equation}
where the $^{16}$O + $\alpha$ THSR wave function has the form
\begin{eqnarray}
  && \Phi_L(^{16}{\rm O}+\alpha\ {\rm THSR}) \nonumber\\
  && \;= \lim_{|\vec{D}| \to 0} P_L {\cal A}
\{ e^{-\tfrac{8(\Vec{r}-\Vec{D})^2}{5B^2}} \phi(^{16}{\rm O})\phi(\alpha) \}, \label{20Ne}
\end{eqnarray}
where
\begin{equation}
\Vec{r} = \Vec{X}_{\rm CM}(^{16}{\rm O}) - \Vec{X}_{\rm CM}(\alpha),
\end{equation}
and $P_L$ is the projection operator for angular momentum $L$.

The relation in Eq.~(\ref{20Ne}) casts doubt on the use of the energy curve
for the Brink wave function in determining the spatial localization of
clusters. The $^{16}$O + $\alpha$ Brink wave function ${\cal A} \{ \exp[-8(\Vec{r}-\Vec{D})^2/(5b^2)]
\phi(^{16}{\rm O})\phi(\alpha)\}$ has
the inter-cluster separation parameter $\Vec{D}$.  The optimum value of $D=|\Vec{D}|$
is obtained from the minimum energy point of the
energy expectation value as a function of $D$. It was shown in Ref.~\cite{zhou13}
that this way of determining the inter-cluster distance is misleading because
if one uses as the size parameter of the inter-cluster motion
not $8/(5b^2)$ given by the Brink wave function, but the much smaller value
$8/(5B^2)$, the energy minimum point resides at $D$ = 0.

\subsubsection{Localized clustering from inter-cluster Pauli repulsion}
\label{linchain}

\begin{figure}[htbp]
\begin{center}
\includegraphics[scale=0.27]{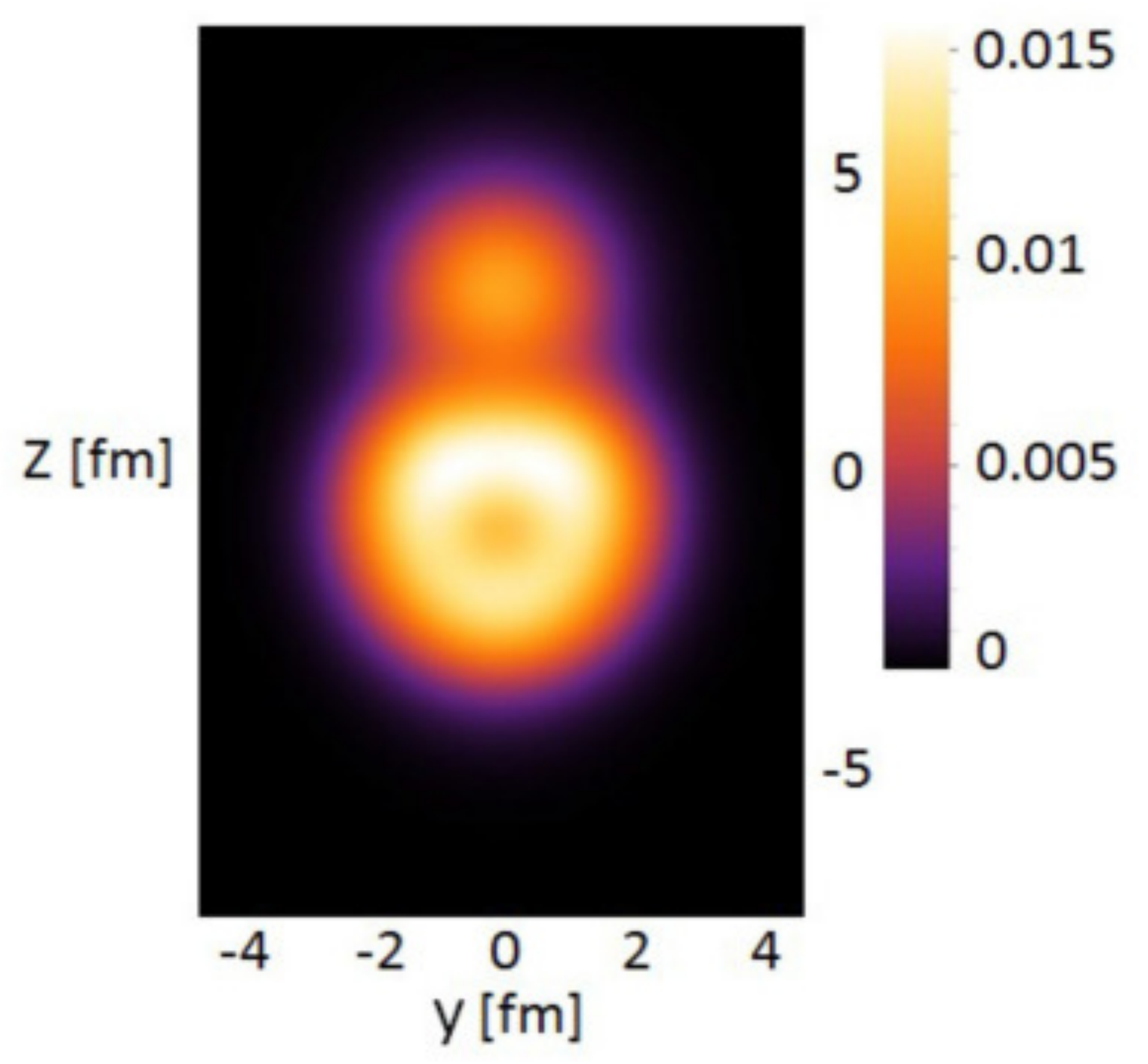}
\caption{(Color online) Nucleon density distribution of the $^{16}$O + $\alpha$ intrinsic THSR wave function of Eq.~(\ref{20Ne}) for 
$D$ = 0.6~fm.
A large distance of about 3.6~fm between $^{16}$O and $\alpha$ clusters is seen.}
\label{fig:AlpOdens}
\end{center}
\end{figure}

The limit
$D \to 0$ appearing in Eq.~(\ref{20Ne}) would seem to suggest that the distributions of the $^{16}$O and $\alpha$ clusters
are overlapping and therefore not localized.  The resolution of this apparent contradiction becomes clear after calculating the nucleon density distribution of
the $^{16}$O + $\alpha$ THSR wave function, as was performed in Ref.~\cite{zhou14}. In Fig.~\ref{fig:AlpOdens} we show the
nucleon density distribution of the $^{16}$O + $\alpha$ intrinsic THSR wave function of Eq.~(\ref{20Ne}) for the  small value $D=0.6\,{\rm fm}$. We do not directly set $D = 0$ as this would force the intrinsic THSR wave function to be symmetric under parity.  We see in Fig.~\ref{fig:AlpOdens} that there exists a large separation of about 3.6 ~fm between the
$^{16}$O and $\alpha$ clusters.
This spatial localization of clusters is due to the nucleon antisymmetrizer ${\cal A}$.
The antisymmetrizer ${\cal A}$ generates Pauli-forbidden states $\chi^F(\Vec{r})$ for the $^{16}{\rm O}-\alpha$ relative motion
which have the property
\begin{equation}
{\cal A} \{ \chi^F(\Vec{r}) \phi(^{16}{\rm O})\phi(\alpha)\} =0,
\end{equation}
where $\chi^F(\Vec{r})$ are H.O. eigenstates with
Gaussian factor $ \exp[-8\Vec{r}^2/(5b^2)]$ and oscillator quanta $2n + L$ smaller than 8.  Thus the probability for two clusters
$^{16}$O and $\alpha$ to be very close together is small.  This is nothing more than Pauli repulsion.

We can say that the dynamics favors non-localized clustering but the constraints of antisymmetrization make the system
exhibit localized clustering in the intrinsic frame.  While this conclusion holds for two-cluster systems in general, the pairwise Pauli repulsion between clusters generally does not produce static localization in the intrinsic frame for more than two clusters.
This is why a nonlocalized gas-like structure of three $\alpha$ clusters arises in the Hoyle state even though a localized
dumbbell-like structure appears in the $2\alpha$ system.

\begin{figure}[h]
\includegraphics[scale=0.27]{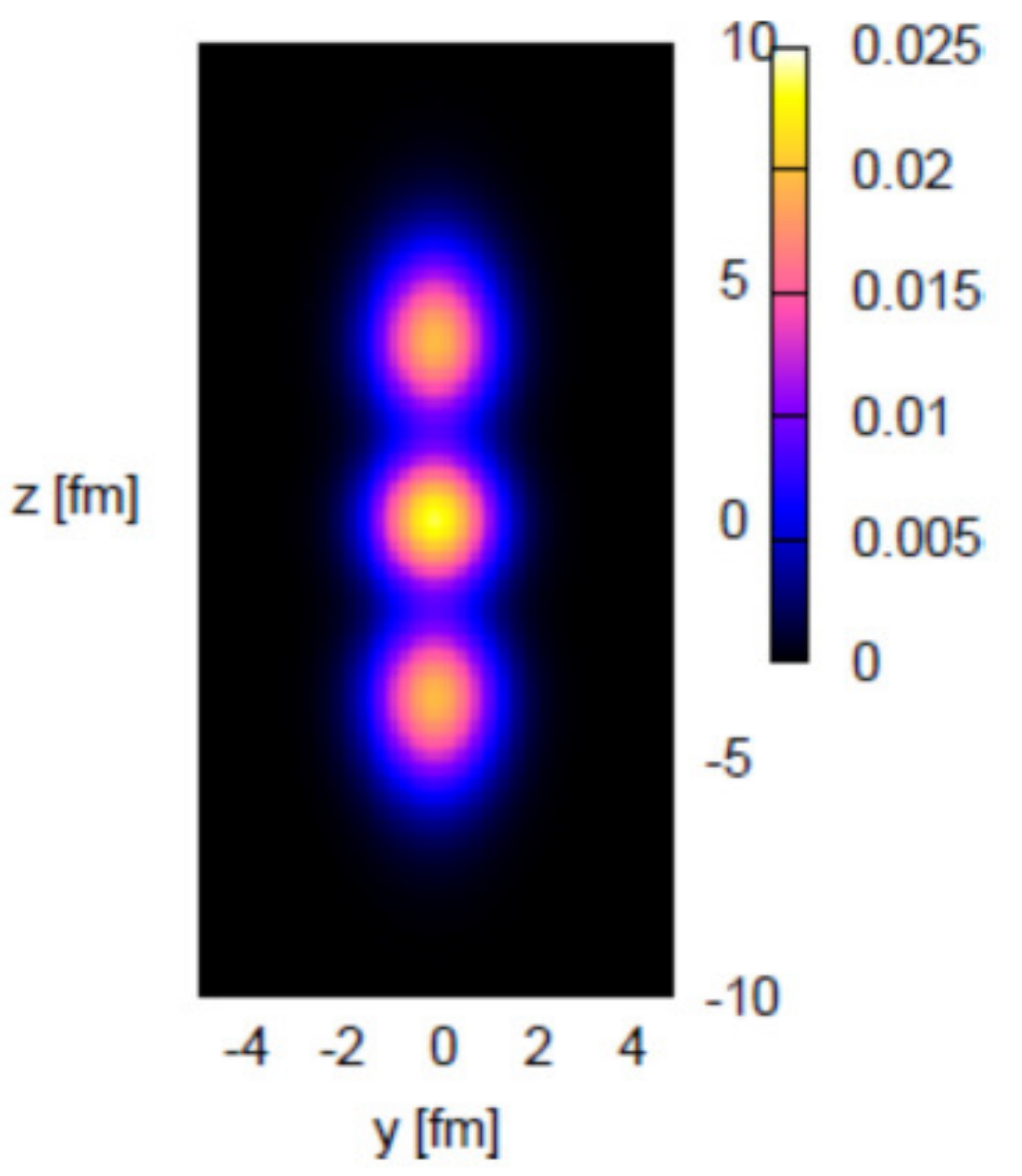}\hspace{2pc}%
\caption{(Color online) Nucleon density distribution for a strongly prolate THSR wave function of $3\alpha$s.}
\label{fig:thrlinal}
\end{figure}

If there are additional constraints, however, there can be localized
clustering even in systems with three or more clusters.  An excited state can be viewed as the minimum energy state under the requirement of orthogonality
to all energy eigenstates at lower energies.  This requirement of orthogonality can constrain the possible deformations of the
excited state.  Consider, for example, an excited state of the $3\alpha$ system that is orthogonal to the ground state, the Hoyle state,
and also the next excited state above the Hoyle state.  Suppose furthermore that these constraints energetically favor a strongly prolate THSR wave function of the form, 
\begin{equation}
 {\cal A} \left\{ \exp\left[-\sum_{k=1}^3[\tfrac{2(X_k^2+Y_k^2)}{b^2} + \tfrac{2Z_k^2}{B^2}]\right] \Phi(3\alpha)\right\},
 \end{equation}
 where with $B \gg b$.  The nucleon density distribution for this THSR wave function is shown~in Fig.~\ref{fig:thrlinal} \cite{zhou14}.
We see three localized $\alpha$ clusters forming a linear-chain structure.

\subsubsection{Equivalence of prolate and oblate THSR wave functions after angular momentum projection}
\begin{figure}[htbp]
\begin{center}
\includegraphics[scale=0.35]{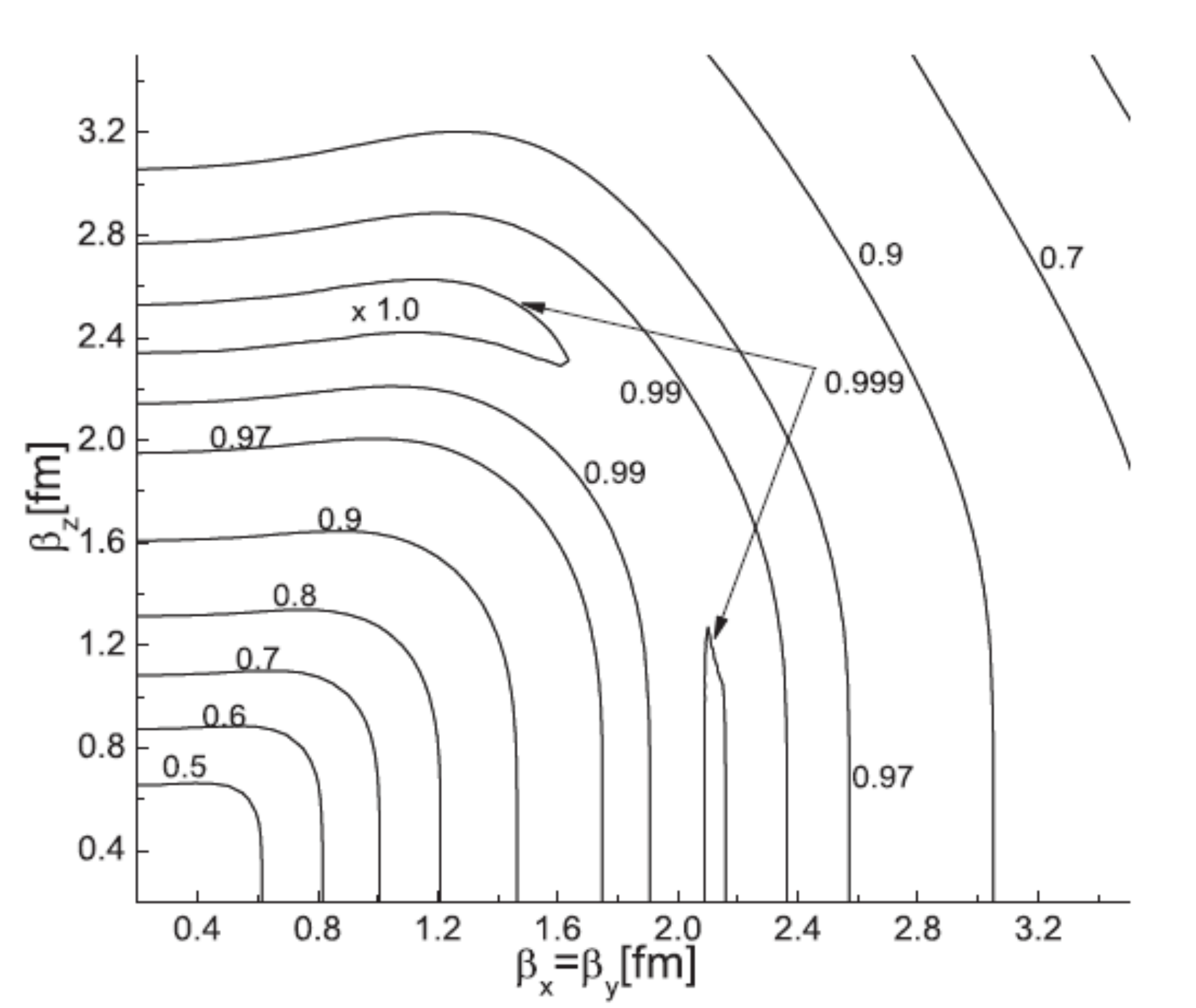}
\caption{Contour map of the squared overlap between a $0^+$ wave function with $\beta_x$ = $\beta_y$ = 0.9~fm, $\beta_z$ = 2.5~fm and
$0^+$ wave functions with various deformations $\beta_x$ = $\beta_y$ and $\beta_z$~\cite{zhou14}.  Numbers attached to the contour curves are
squared overlap values.}
\label{fig:overproobl}
\end{center}
\end{figure}

One unusual feature of THSR wave functions is that prolate and oblate wave functions can become equivalent after angular
momentum projection.  Fig.~\ref{fig:overproobl} shows a contour
map of the squared overlaps between a prolate 0$^+$ THSR wave function with 0$^+$ THSR wave functions
with various deformations in $^{20}$Ne~\cite{zhou14}.  The deformed THSR wave function $\Phi_{\rm Ne}$ of $^{20}$Ne has the form
${\cal A}[ \chi(\Vec{r}) \phi(\alpha) \phi(^{16}{\rm O})]$ where $\chi(\Vec{r})$ is $\exp[-\sum_{k=x,y,z}(8/5B_k^2) r_k^2]$ and
$B_k^2 = b^2 + 2\beta_k^2$. We see in this figure that the prolate THSR wave function with $\beta_x$ = $\beta_y$ = 0.9~fm, 
$\beta_z$ = 2.5~fm is
almost 100\% equivalent to oblate THSR wave functions with $\beta_x$ = $\beta_y$ $\approx$ 2.1 fm and $\beta_z$ between 0 and 1.2~fm after
angular momentum projection onto $0^+$. The equivalence of prolate and oblate THSR wave functions after angular momentum projection is true for
all the spin-parity states of $^{20}$Ne.

Despite this equivalence of prolate and oblate wave functions after angular momentum projection, we can say that the $^{20}$Ne states
expressed by the THSR wave functions all have prolate deformation as the actual deformation. This conclusion is obtained from the fact that the
expectation values of the quadrupole moments of all the $^{20}$Ne states expressed by THSR wave functions have negative sign.  From the
well-known formula
\begin{equation}
Q(J) = -\frac{J}{2J+3}\cdot Q({\rm intrinsic}),
\end{equation}
we know that when the expectation value $Q(J)$ of the quadrupole moment
of the wave function with good spin $J$ is negative, the quadrupole moment of the intrinsic state, $Q({\rm intrinsic}),$ is positive and therefore prolate.  The THSR wave function after angular momentum projection has the form
$\Phi^J_{\rm Ne}$ = ${\cal A}[ \chi_J(\Vec{r}) \phi(\alpha) \phi(^{16}{\rm O})]$ and we can prove
that this type of wave function $\Phi^J_{\rm Ne}$ gives us the following formula for $Q(J)$~\cite{zhou14}:
\begin{equation}
  Q(J) = - \frac{J}{2J+3} \frac{16}{5} \langle r^2 \rangle,
\end{equation}
where
\begin{equation}
  \frac{16}{5} \langle r^2 \rangle = \langle \Phi^J_{\rm Ne} | \sum_{j=1}^{20} (\Vec{r}_j - \Vec{X}_{\rm CM})^2 |\Phi^J_{\rm Ne} \rangle
 -  R^2(^{16}{\rm O}) - R^2(\alpha),
 \end{equation}
 and
 \begin{equation}
R^2(C_k) = \langle \phi(C_k) | \sum_{j \in C_k} (\Vec{r}_j - \Vec{X}_{\rm CM}(C_k) )^2  | \phi(C_k) \rangle.
\end{equation}
This shows that $Q(J)$ has negative value and explains why the calculated values of $Q(J)$ by THSR wave functions have all negative sign.
Of course the negative sign of $Q(J)$ by THSR wave functions is in accordance with the prolate distribution of nucleon density shown in
Fig.~\ref{fig:AlpOdens}.

\begin{figure}[htbp]
\begin{center}
\includegraphics[scale=0.30]{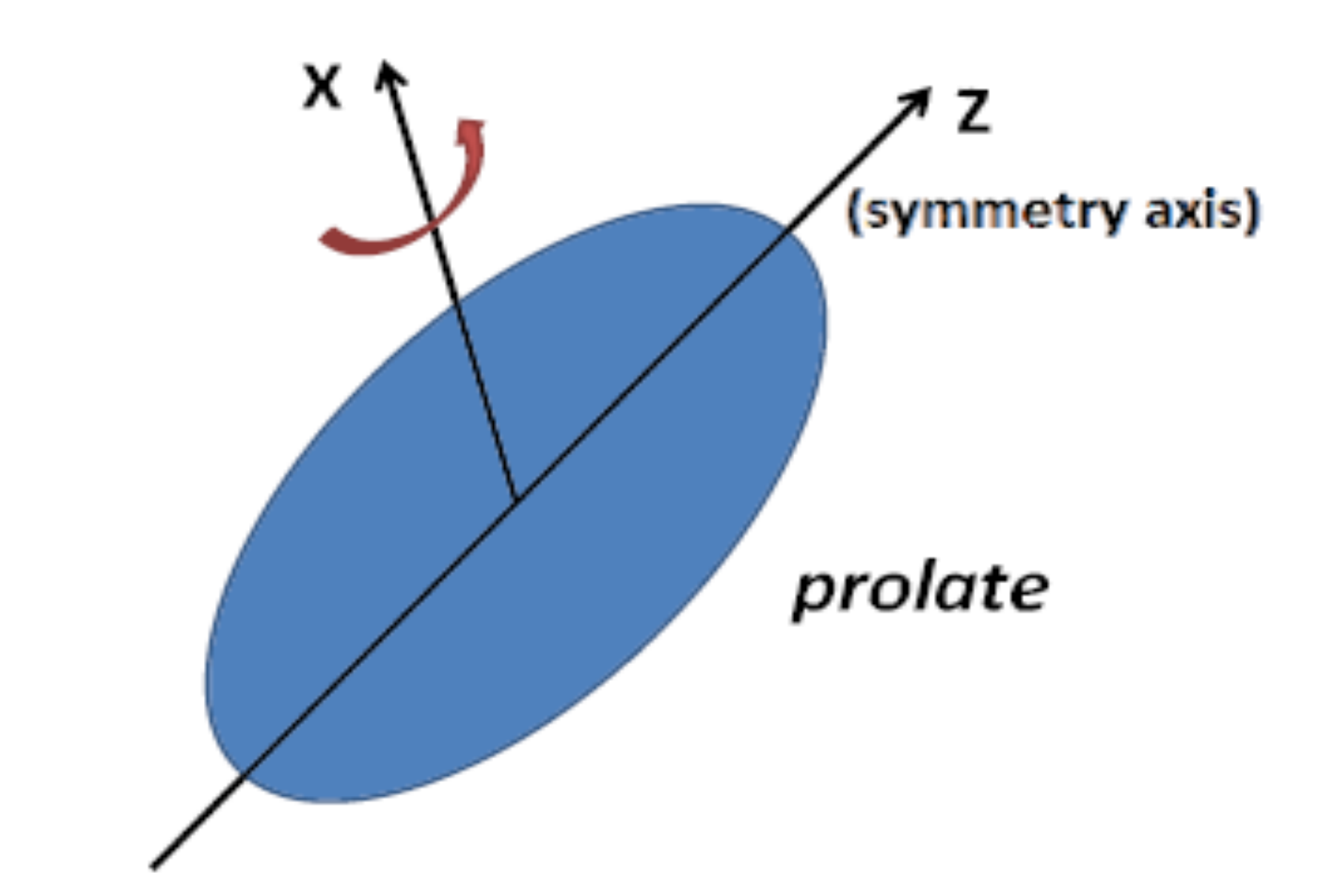}
\caption{(Color online) Rotation average of a prolate THSR wave function around an axis ($x$ axis) perpendicular to the symmetry axis ($z$ axis) of
the prolate deformation.}
\label{fig:rotaverage}
\end{center}
\end{figure}

The reason why prolate and oblate THSR wave functions are almost equivalent after angular momentum projection is explained by the fact
that the rotation-average of a prolate THSR wave function is almost equivalent to an oblate THSR wave function.  The rotation-averaged wave function
$\Phi^{\rm ave}(\beta_x=\beta_y,\beta_z)$ generated from a prolate THSR wave function $\Phi^{\rm prolate}(\beta_x=\beta_y,\beta_z)$ is defined as
\begin{eqnarray}
  &&\Phi^{ave}(\beta_x=\beta_y,\beta_z)   \nonumber  \\
  &&= \left( \frac{1}{2\pi} \int_0^{2\pi} d\theta e^{i\theta J_x} \right) \Phi^{\rm prolate}(\beta_x=\beta_y,\beta_z).
\end{eqnarray}
If we rotate a prolate THSR wave function around an axis ($x$ axis) perpendicular to the symmetry axis of the prolate deformation ($z$ axis)
and construct a wave function by taking an average over this rotation, the density distribution of the rotation-average wave function will be
oblate (see Fig.~\ref{fig:rotaverage}).  In the case of the $0^+$ state, when we construct the rotation-average wave function
from the prolate THSR wave function with $(\beta_x,\beta_y,\beta_z)$ =
(0.9, 0.9, 2.5~fm) which gives the minimum energy for $0^+$,  it is almost 100\% equivalent to the oblate THSR wave function with
$(\beta_x,\beta_y,\beta_z)$ = (0.9, 2.1, 2.1~fm).
\begin{eqnarray}
  &&\Phi^{\rm oblate}(\beta_x=0.9\,{\rm fm}, \beta_y=\beta_z=2.1\,{\rm fm})    \nonumber \\
  &&\approx \Phi^{\rm ave}(\beta_x=\beta_y=0.9\,{\rm fm},\beta_z=2.5\,{\rm fm}).
\end{eqnarray}

\subsection{Container picture of cluster dynamics}
\label{container}

The ground and Hoyle states in $^{12}$C are obtained as the eigenstates with lowest and
second lowest energies of Eq.~(\ref{eq:c12GCM}).  This equation is the GCM equation with respect to the size parameter $B$ of the THSR wave
function. The excitation of the system is described by the dynamics of the system size.  This description is very different
from the traditional description of the system excitation by RGM/GCM equation, which treats the dynamics of inter-cluster motion.
In Ref.~\cite{zhou14}, this new description of the cluster dynamics is called the container model of cluster dynamics.  The container
refers to the self-consistent field with size $B$ in which clusters are accomodated and make nonlocalized motion.

The GCM equation with respect to the container size parameter $B$ was also solved in the $4\alpha$ system~\cite{4alpTHSR}.
Table~\ref{tab:4alpTHSROCM} shows the energy spectra of $0^+$ states obtained by $4\alpha$ THSR-GCM. In this table we show the
energy spectra of $0^+$ states obtained by $4\alpha$ OCM \cite{4alpOCM} together with experimental spectra.
This $4\alpha$ OCM study confirmed the assignments by former $^{12}$C + $\alpha$ cluster model studies that the observed $0_2^+$ and
$0_3^+$ states are dominantly $^{12}$C$(0_1^+)$ + $\alpha$ ($S$ wave) and $^{12}$C$(2_1^+)$ + $\alpha$ ($D$ wave) configurations respectively. On the other hand, the $4\alpha$ OCM study newly assigned the dominant
configurations for the $0_4^+$ and $0_5^+$ states to be  $^{12}$C$(0_1^+)$ + $\alpha$ (higher-nodal $S$ wave) and $^{12}$C$(1_1^-)$ + $\alpha$ ($P$ wave) respectively.
What is interesting about this $4\alpha$ OCM study is the assignment of the $4\alpha$ condensate-like structure to the observed $0_6^+$ state
at 15.1~MeV excitation.  The reason for this assignment is that the reduced-width amplitude of the calculated $0_6^+$ state
is large only in the  $^{12}$C(Hoyle) + $\alpha$ ($S$ wave) channel~\cite{4alpOCM}.   The good accordance of the calculated decay
width $\Gamma_{\rm cal}$ = 136 keV with the observed width $\Gamma_{\rm exp}$ = 166~keV for this  $0_6^+$ state gives high reliability to
this OCM assignment.  In Ref.~\cite{Yamada:2011ri}, it is reported that the fine structures of the observed isoscalar monopole strength function
up to about 16~MeV in $^{16}$O are well reproduced by this $4\alpha$ OCM.
Compared to the OCM calculation, the THSR-GCM calculation gives us only four $0^+$ states in the excitation-energy region
of observed six $0^+$ states.  However the fourth $0^+$ state of the THSR-GCM calculation can be considered to correspond to the $0_6^+$ state
of the $4\alpha$ OCM and hence to the observed $0_6^+$ state at 15.1~MeV excitation.  This is because the reduced-width amplitude of the $0_4^+$ state
of the THSR-GCM calculation is markedly large only in the channel  $^{12}$C(Hoyle) + $\alpha$ ($S$ wave)~\cite{4alpTHSR}.
Thus also in the case of the $4\alpha$ system, the THSR-GCM calculation describes the excitation of the system from the ground state
to the $4\alpha$ gas-like excited state, although the description of the excitation to other states is incomplete.
In order to remedy the incompleteness of the description of Ref.~\cite{4alpTHSR} we have to extend the $4\alpha$ THSR wave function
so that the THSR wave function includes not the single size parameter $B$ but two or more $B$ parameters.  In the case of two $B$ parameters,
one $B$ is for the container containing three $\alpha$ clusters and the other $B$ is for the container for the relative motion
between $3\alpha$ system and fourth $\alpha$ cluster.  In the next subsection we discuss the extension of the THSR wave function so that it includes
two or more $B$ parameters.

\begin{table}
\caption
{Comparison of the $0^+$ energy spectra from experiments, $4\alpha$ OCM calculation~\cite{4alpOCM}, and $4\alpha$ THSR
calculation \cite{4alpTHSR}. Energies $E$ are measured in MeV from the $4\alpha$ threshold, and  RMS radii $R_{\rm rms}$ are in fm.}
\label{tab:4alpTHSROCM}
\begin{center}
\begin{tabular}{ccc|ccc|c}
\hline
\hspace{0.1mm} & \hspace{0.1mm} 4$\alpha$ THSR \hspace{0.1mm} & \hspace{0.1mm} &  \hspace{0.1mm} & \hspace{0.1mm} 4$\alpha$ OCM  \hspace{0.1mm} &
\hspace{0.1mm} & \hspace{0.1mm} Exp.  \hspace{0.1mm}    \\
 \hspace{0.1mm} & \hspace{0.1mm} $E$ \hspace{0.1mm} & \hspace{0.1mm} $R_{\rm rms}$ \hspace{0.1mm} &
 \hspace{0.1mm} & \hspace{0.1mm} $E$ \hspace{0.1mm} & \hspace{0.1mm} $R_{\rm rms}$ \hspace{0.1mm} &
 \hspace{0.1mm} $E$ \hspace{0.1mm}     \\
\hline
$0^+_1$ \hspace{0.1mm} & \hspace{0.1mm} $-$15.05 \hspace{0.1mm} & \hspace{0.1mm} 2.5 \hspace{0.1mm} & \hspace{0.1mm} $0^+_1$  \hspace{0.1mm} &
  \hspace{0.1mm} $-$14.37 \hspace{0.1mm} & 2.7  \hspace{0.1mm} & \hspace{0.1mm} $-$14.44 \hspace{0.1mm}   \\
$0^+_2$ \hspace{0.1mm} & \hspace{0.1mm} $-$4.7 \hspace{0.1mm} & \hspace{0.1mm} 3.1 \hspace{0.1mm} & \hspace{0.1mm} $0^+_2$  \hspace{0.1mm} &
  \hspace{0.1mm} $-$8.0 \hspace{0.1mm} & 3.0  \hspace{0.1mm} & \hspace{0.1mm} $-$8.39 \hspace{0.1mm}   \\
        \hspace{0.1mm} & \hspace{0.1mm}       \hspace{0.1mm} & \hspace{0.1mm}     \hspace{0.1mm} & \hspace{0.1mm} $0^+_3$  \hspace{0.1mm} &
  \hspace{0.1mm} $-$4.41 \hspace{0.1mm} & 3.1  \hspace{0.1mm} & \hspace{0.1mm} $-$2.39 \hspace{0.1mm}   \\
$0^+_3$ \hspace{0.1mm} & \hspace{0.1mm} 1.03 \hspace{0.1mm} & \hspace{0.1mm} 4.2 \hspace{0.1mm} & \hspace{0.1mm} $0^+_4$  \hspace{0.1mm} &
  \hspace{0.1mm} $-$1.81 \hspace{0.1mm} & 4.0  \hspace{0.1mm} & \hspace{0.1mm} $-$0.84 \hspace{0.1mm}   \\
       \hspace{0.1mm} & \hspace{0.1mm}       \hspace{0.1mm} & \hspace{0.1mm}     \hspace{0.1mm} & \hspace{0.1mm} $0^+_5$  \hspace{0.1mm} &
  \hspace{0.1mm} $-$0.25 \hspace{0.1mm} & 3.1  \hspace{0.1mm} & \hspace{0.1mm} $-$0.43 \hspace{0.1mm}   \\
$0^+_4$ \hspace{0.1mm} & \hspace{0.1mm} 3.04 \hspace{0.1mm} & \hspace{0.1mm} 6.1 \hspace{0.1mm} & \hspace{0.1mm} $0^+_6$  \hspace{0.1mm} &
  \hspace{0.1mm} 2.08  \hspace{0.1mm} &  5.6 \hspace{0.1mm} & \hspace{0.1mm} 0.66 \hspace{0.1mm}   \\
\hline
\end{tabular}
\end{center}
\end{table}

The GCM wave functions $\Phi_{\lambda}^{\rm
THSRGCM}$ of the obtained four $0^+$ states
with $\lambda = 1,2,3,4$ are found to have almost 100\% squared overlaps
with
single orthogonalized THSR wave functions $\widehat{\Phi}_{\lambda}(\beta_0)$
for a certain value of $\beta_0$ satisfying $B^2 = b^2 + 2\beta_0^2$.  $\widehat{\Phi}_{\lambda}(\beta_0)$
is
defined by
\begin{equation}
\widehat{\Phi}_{\lambda}(\beta_0) = N_{\lambda} P_{\lambda-1}
\Phi_{4\alpha}^{\rm THSR}(\beta_0)
\end{equation}
where
\begin{equation}
P_{\lambda-1} = 1 - \sum_{k=1}^{\lambda-1} |\Phi_k^{\rm THSRGCM}
\rangle \langle \Phi_k^{\rm THSRGCM}|,\end{equation}
for $\lambda = 2 ,3, 4$.
Here, $N_{\lambda}$ is a normalization constant and $P_0 = 1$.  Since the
orthogonalization operator $P_{\lambda-1}$ expresses the
necessary property which any excited state should satisfy, the
essential character of $\Phi_{\lambda}^{\rm THSRGCM}$ is
expressed by $\Phi_{4\alpha}^{\rm THSR}(\beta_0)$.  Thus, although $\Phi_{\lambda}^{\rm
THSRGCM}$ is constructed by a linear combination of
many $4\alpha$ THSR wave functions, its essential character is described
by only a single $4\alpha$ THSR wave function.
The optimum values of $\beta_0$ for four $0_{\lambda}^+$ states are
1.2~fm, 2.5~fm, 4.0~fm, 6.5~fm, for $\lambda = 1,2,3,4,$ respectively, which
means that the system size becomes larger with increasing $\lambda$.

\subsection{Extended THSR wave function and examples of its application}
\label{extendTHSR}

\subsubsection{Breathing-like excitation of the Hoyle state}    \label{3aextend}

\begin{figure}[h]
\includegraphics[scale=0.20]{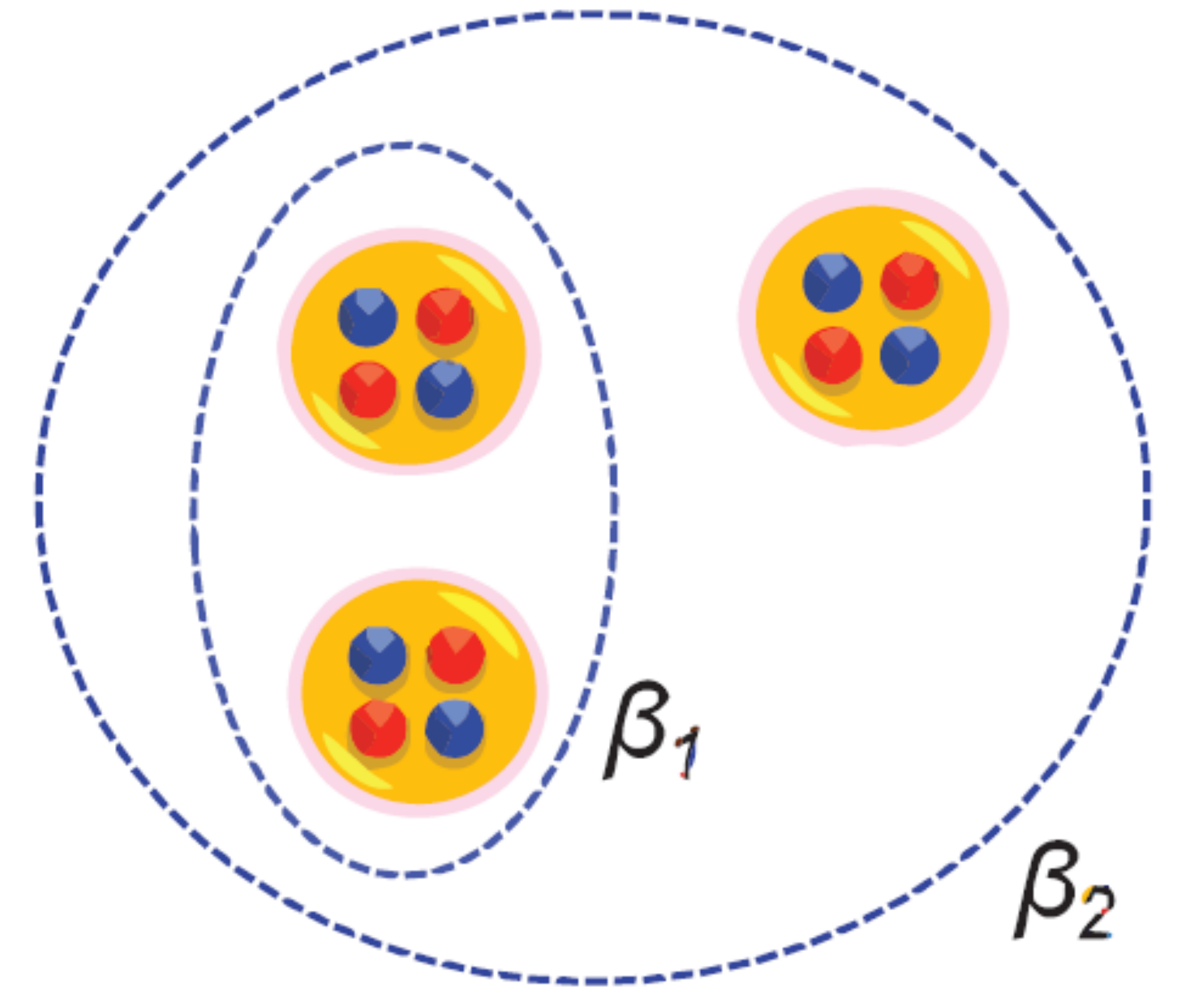}\hspace{2pc}%
\caption{(Color online) Size parameters $B_1$ and $B_2$ for $2\alpha$ and $3\alpha$ containers, respectively.
$B_1^2$ = $b^2$ + $\beta_1^2$ and $B_2^2$ =(3/4) $b^2$ + $\beta_2^2$.}
\label{fig:twobeta}
\end{figure}

The container model of cluster dynamics uses the system size parameter as the generator coordinate for clustering motion.
In the case of $3\alpha$ system, we can introduce size parameters $B_1$ and $B_2$ for $2\alpha$ and $3\alpha$ containers as shown in
Fig.\ref{fig:twobeta}.  The extended THSR wave function for this double container system is given as~\cite{zhou14B}
\begin{equation}
  \Phi^{\rm exTHSR}_{3 \alpha}(B_1,B_2) =  {\cal A} \left\{ \exp\left( -\tfrac{\Vec{\xi}_1^2}{B_2^2}  - \tfrac{\Vec{\xi}_2^2}{B_1^2}  \right) \Phi(3\alpha)
 \right\}
    \label{eq:exthsr}.
\end{equation}
When $B_2^2 = (3/4)B_1^2$, $\Phi^{\rm exTHSR}_{3 \alpha}(B_1,B_2)$ = $\Phi^{\rm THSR}_{3 \alpha}(B_1).$   As has been done in the traditional THSR
wave function, we can use deformed containers.  In this case, we replace $\Vec{\xi}_1^2/B_2^2$ by $\sum_{i=x}^z \xi_{1i}^2/B_{2i}^2$ and also
$\Vec{\xi}_2^2/B_1^2$ by $\sum_{i=x}^z \xi_{2i}^2/B_{1i}^2$.

The extended THSR wave function for $3\alpha$ system has been applied to the studies of the ground state~\cite{zhou14B} and the positive-parity
excited states in $^{12}$C~\cite{funaki15,zhou16}.  Refs.~\cite{funaki15} and ~\cite{zhou16} supported the existence of two $0^+$ states
($0_3^+$ and $0_4^+$) around 10~MeV excitation energy above the Hoyle state ($0_2^+$) that was proposed by the $3\alpha$ OCM studies
combined with the CSM (complex scaling method) in Refs.~\cite{Kurokawa:2005ax,ohtsubo13}.   The $0_4^+$ state is the state whose existence had been long
known since 1970's, and this state is considered to have a bent-chain structure of $3\alpha$ with a large component of $^8$Be($2_1^+$) +
$\alpha$($D$~wave)~\cite{uegaki1,enyo,Neff:2003ib}.  On the other hand the
$0_3^+$ state is the state whose existence was newly proposed and which was suggested to be a breathing excitation of the Hoyle state
in Ref.~\cite{Kurokawa:2005ax,KuKaH}.  The theoretical proposal of the existence of two $0^+$ states ($0_3^+$ and $0_4^+$) around 10 MeV excitation energy was
soon supported experimentally by Itoh et al.~\cite{Itoh}.  In Ref.~\cite{Itoh}, it is reported that the observed broad $0^+$ state at 10 MeV consists of
two components. The lower $0^+$ state do $\alpha$-decay to the ground state of $^8$Be only while the higher $0^+$ state has a distinct peak at
10.8~MeV with a width of 0.4~MeV in the coincidence spectrum for the first excited state of $^8$Be channel.  These two $0^+$ states were considered
to have consistent properties predicted for a higher nodal state of the Hoyle state and a linear-like 3$\alpha$ state, respectively.

\begin{table}
\caption
{Energies and rms radii of $0^+_2$, $0^+_3$, and $0^+_4$ states of $^{12}$C and monopole transition values $M(E0)$ between
three $0^+$ states calculated by extended THSR wave functions.
Energies (MeV) are measured from the $3\alpha$ threshold, radii are in fm, and $M(E0)$ values are in $e$~fm$^2$.  Cal. A and Cal. B denote the results
of Refs.~\cite{funaki15} and \cite{zhou16}, respectively.}
\label{tab:3aextd}
\begin{center}
\begin{tabular}{c|c|c}
\hline
 & \hspace{0.1mm}  Cal. A  \hspace{0.1mm} & \hspace{0.1mm} Cal. B \hspace{0.1mm}     \\
\hline
 $E(0^+_2)$ , \hspace{0.1mm} $R_{\rm rms}(0^+_2)$ \hspace{0.1mm} & \hspace{0.1mm} 0.23, \hspace{0.1mm} 3.7 \hspace{0.1mm}
    & \hspace{0.1mm} 0,22, \hspace{0.1mm} 3.9 \hspace{0.1mm}   \\
 $E(0^+_3)$ , \hspace{0.1mm} $R_{\rm rms}(0^+_3)$ \hspace{0.1mm} & \hspace{0.1mm} 2.6, \hspace{0.1mm} 4.7 \hspace{0.1mm}
    & \hspace{0.1mm} 1.7, \hspace{0.1mm} 5.2 \hspace{0.1mm}   \\
 $E(0^+_4)$ , \hspace{0.1mm} $R_{\rm rms}(0^+_4)$ \hspace{0.1mm} & \hspace{0.1mm} 3.9, \hspace{0.1mm} 4.2 \hspace{0.1mm}
    & \hspace{0.1mm} 2.7, \hspace{0.1mm} 4.0 \hspace{0.1mm}   \\
 $M(E0, 0^+_2 \rightarrow 0^+_1)$ \hspace{0.1mm} & \hspace{0.1mm} $6.3-6.4$ \hspace{0.1mm} & \hspace{0.1mm} 6.2 \hspace{0.1mm}  \\
 $M(E0, 0^+_2 \rightarrow 0^+_3)$ \hspace{0.1mm} & \hspace{0.1mm} $34-37$ \hspace{0.1mm} & \hspace{0.1mm} 47 \hspace{0.1mm}  \\
 $M(E0, 0^+_2 \rightarrow 0^+_4)$ \hspace{0.1mm} & \hspace{0.1mm} $0.5-1.4$ \hspace{0.1mm} & \hspace{0.1mm} 7.7 \hspace{0.1mm}  \\
\hline
\end{tabular}
\end{center}
\end{table}

Table~\ref{tab:3aextd}  shows energies and rms radii of $0^+_2$, $0^+_3$, and $0^+_4$ states of $^{12}$C and monopole transition values
$M(E0)$ between three $0^+$ states which are calculated by extended $3\alpha$ THSR wave functions~\cite{funaki15,zhou16}.
We see that the Hoyle state ($0^+_2$) and other two  $0^+$ states have very large rms radii.
The very large value of the calculated $E0$ strength $M(E0; 0_3^+ \to 0_2^+)$ = 35~$e$~fm$^2$ or 47~$e$~fm$^2$ supports the idea to regard the $0^+_3$ state as a breathing-like excited state of the Hoyle state $0^+_2$.
In Ref.~\cite{funaki15}, the total width of the $0^+_4$ state is calculated to be 0.7~MeV which is to be compared with the observed width 1.42~MeV.
As for the $0^+_3$ state the calculated $\alpha$ width is 1.1~MeV which is rather close to the observed width 1.45~MeV~\cite{Itoh:2011zz}.

In Ref.~\cite{zhou16}, two kinds of inter-cluster relative wave functions were analysed which are contained in the four $0^+$ states
($0^+_1 \sim 0^+_4$) obtained by the extended $3\alpha$ container model.  The first kind is an $S$-wave relative wave function between
$^8$Be($0^+_1$) and the remaining $\alpha$ cluster, and the second kind is an $S$-wave relative wave function between two $\alpha$ clusters after the
integration over the $\Vec{\xi}_1$ Jacobi coordinate by using a single Gaussian weight.  It was found that both kinds of relative wave
functions have one more node in the $0^+_3$ state than in the Hoyle state.  This result implies that $0^+_3$ state is the breathing-like
excited state of the Hoyle state.  It is because the generating operator for the breathing excitation
\begin{equation}
O_B = \sum_{i=1}^{12}(\Vec{r}_i - \Vec{r}_{\rm CM})^2
\end{equation}
can be rewritten as
\begin{equation}
O_B = \sum_{k=1}^3\sum_{i \in \alpha_k}(\Vec{r}_i - \Vec{X}_k)^2 + \tfrac{8}{3}\Vec{\xi}_1^2 + 2\Vec{\xi}_2^2.
\end{equation}

\subsubsection{Container evolution in $^{16}$O}
\begin{figure}[htbp]
\begin{center}
\includegraphics[scale=0.70]{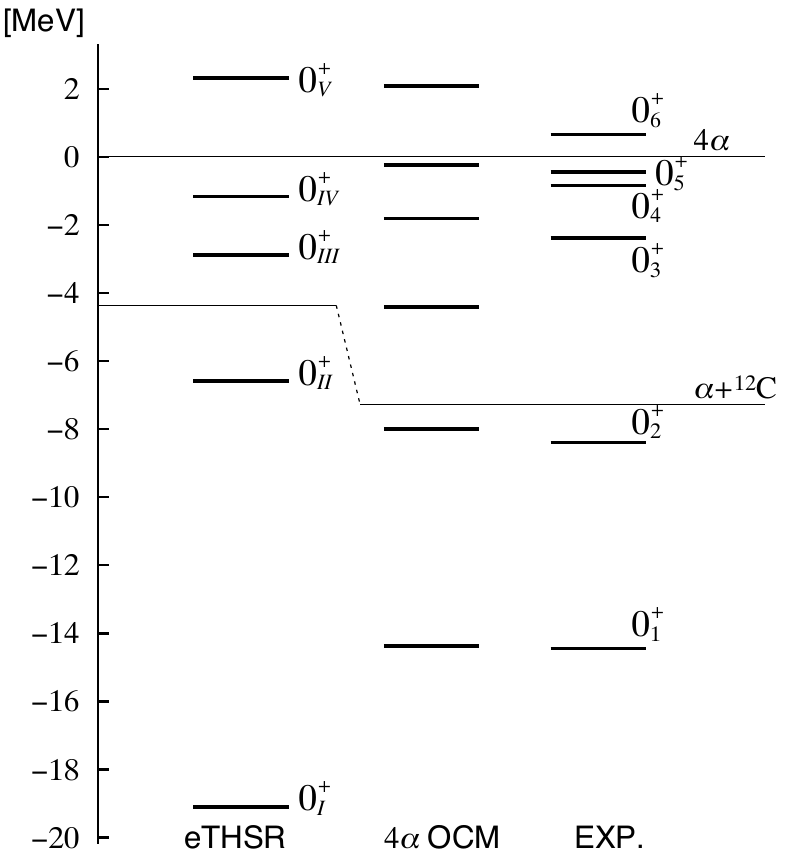}
\caption{{Energy spectrum obtained by extended $4\alpha$ THSR of Ref.~\cite{Funaki:2017tia} which is compared with those by $4\alpha$ OCM
and experiment.}}
\label{fig:4aextend}
\end{center}
\end{figure}

The extended THSR wave function was applied recently to study the evolution of cluster structure in $^{16}$O \cite{Funaki:2017tia}.
As in the $3\alpha$ system in subsection \ref{3aextend}, two deformed containers are adopted where the first container is for the $3\alpha$
subsystem and the second container is for the relative motion between the $3\alpha$ subsystem and the fourth $\alpha$ cluster.
Fig.~\ref{fig:4aextend} shows the energy spectrum obtained by the extended $4\alpha$ THSR  (denoted as eTHSR) compared with the  $4\alpha$ OCM
and experiment~\cite{Funaki:2017tia}. The fifth $0^+$ state ($0^+_5$) is just above the $4\alpha$ threshold and the two size parameters of
its eTHSR wave function are nearly the same.  This means that $4\alpha$ clusters in this state are accommodated approximately in a single
container. Since the size parameters of this state are large, $\beta_x = \beta_y \approx 5.6\,{\rm fm}$, $\beta_z \approx 2.0\, {\rm fm}$, the $0^+_5$ state
represents a Hoyle-analogue state in $^{16}$O.

The ground state ($0^+_1$) has the smallest radius and its intrinsic shape is reported to be terahedral. It is based on the calculated result
that while the $3\alpha$ sub-container has oblate shape the container describing the relative motion between the $3\alpha$ subsystem and the 4th
$\alpha$ cluster is of prolate shape.  The calculated second $0^+$ state ($0^+_2$) is reported to have $^{12}$C$(0^+_1)$ + $\alpha$ ($S$ wave)
structure.  The reason for this identification is that the size parameters of the $3\alpha$ sub-container are close to those of
the $3\alpha$ container of the  $^{12}$C ground state and are nearly spherical. Also, the container describing the relative motion between
the $3\alpha$ subsystem and the 4th $\alpha$ cluster is also nearly spherical but with a much larger radius.  This structure of the second $0^+$ state
is in good accordance with previous cluster model studies~\cite{suzuki72,4alpOCM}.   Similarly, the structure of the calculated third $0^+$ state
($0^+_3$) is in good accordance with previous cluster model studies~\cite{suzuki72,4alpOCM}; namely the $0^+_3$ state is reported to have
$^{12}$C$(2^+_1)$ + $\alpha$($D$~wave) structure.  The structure of the calculated fourth $0^+$ state ($0^+_{IV}$) is reported to have
$^{12}$C$(0^+_1)$ + $\alpha$($S$~wave), where the container describing the relative motion between the $3\alpha$ subsystem and the 4th
$\alpha$ cluster is much larger than that of the second $0^+$ state.

The container model and the extended THSR wave function describes the evolution of the cluster structure from the
ground state up to the $4\alpha$ condensate-like state (Hoyle-analogue state) through the various $^{12}$C + $\alpha$ structures and is therefore well-suited for studying the evolution of cluster structure.
In describing this evolution, we go from a single container to several containers, a process called container evolution \cite{Funaki:2017tia}.

\subsubsection{Neutron-rich Be isotopes}

Extended THSR wave functions have also been applied to neutron-rich nuclei.  Since the container describes valence
neutrons with size parameters different from the container for the core part of the system, the use of the
extended THSR wave function for neutron-rich nuclei is quite natural. Here we report the works of Refs.~\cite{LyuBe9} and \cite{LyuBe10} which
treat $^9$Be and $^{10}$Be, respectively.

In the case of $^9$Be, the valence-neutron wave function $F(\Vec{r})$ in the extended THSR wave function should have negative parity, and it is given by
\begin{equation}
F_n(\Vec{r}) = \int d\Vec{R} \exp\left(-\sum_{k=x}^z \tfrac{R_k^2}{\beta_k^2}\right) \exp(i\phi_R) \exp\left[-\tfrac{(\Vec{r} - \Vec{R})^2}{2b^2}\right].
\end{equation}
The phase factor $\exp(i\phi_R)$ makes the parity of $F_n(\Vec{r})$ negative.  In Ref.~\cite{LyuBe9},
the ground rotational-band levels, $3/2^-$, $5/2^-$, $7/2^-$, were treated, and it was found that the extended THSR wave functions of
these levels have about 95\% squared overlaps with the wave functions obtained by GCM calculation by using $2\alpha + n$ three-body Brink
wave functions.

In the case of $^{10}$Be, the energy spectra of two rotational bands upon the ground state and the $0_2^+$ state were calculated using single
extended THSR wave functions and were compared with those obtained by AMD calculations~\cite{AMDDC,AMDGCM}.  For the ground band, the extended
THSR wave functions where two valence neutrons occupy the orbit $F_n(\Vec{r})$ were used.  The modification of these extended THSR wave
functions were also made by introducing the distance parameter $\Vec{R}_{\rm pair}$ between the center of mass of the $2\alpha$ system and center of mass of the $2n$ system.
It was reported that both kinds of extended THSR wave functions give very similar energy spectra to that of the AMD calculations \cite{AMDDC}.
For the excited band, the extended THSR wave functions were constructed by accommodating two valence neutrons into the $\sigma$-type
single-neutron orbit.
The obtained energy spectrum is very similar to but a little higher than the AMD energy spectra in Refs.~\cite{AMDDC,AMDGCM}.
The extended THSR wave function of the $0_2^+$ state is not orthogonalized to that of the ground state, but the squared overlap between them is
as small as 1.4\%.  We see thus that the wave functions as simple as the single extended THSR wave functions give good results that are quite similar to AMD calculations.

%%%%%%%%%%%%%%%%%%%%%%%%%%%%%%%%%%%%%%%%%%%%%%%%%%%%%%%%%%%%%%%%%%%%%%%
%
% \input{no_core_section.tex}
%

\section{No-core shell model}

In contrast with the traditional shell model approach which
starts with an inert core of nucleons filling a closed shell,
the no-core shell model treats all nucleons as active. The many-body basis states are the energy eigenstates
of the spherical
harmonic oscillator,
\begin{align}
H_{\rm osc} & =  \sum_{i=1}^A H_i,\\
H_i & = -\frac{\hbar^2}{2m} \nabla^2_i + \frac{1}{2}m\Omega^2r^2_i,
\label{simpleHO}
\end{align}
with some finite truncation imposed in the total
oscillator excitation energy
\cite{Navratil:2000ww,Navratil:2000gs}. Here $m$ is the nucleon mass and
$\Omega$ is the oscillator frequency.
The truncation of the basis in terms of the total
sum of oscillator excitation energies allows for an exact factorization of
the wave function into separated center-of-mass and relative-coordinate degrees
of freedom.

In these no-core shell model calculations the interactions among nucleons include a nucleon-nucleon potential fitted to experimental nucleon-nucleon scattering data as well as higher-nucleon interactions fitted to few-nucleon observables. Some take the approach of using a high-quality phenomenological potential \cite{Wiringa:1994wb}, while others apply the organizational principles of chiral effective field theory to produce effective chiral interactions for nucleons
\cite{Epelbaum:2008ga,Machleidt:2011zz}.

The method has had many remarkable successes in recent years in describing
nuclear structure from first principles, e.g., \cite{Navratil:2007we,Maris:2008ax,
Roth:2011ar,Barrett:2013nh}.  For the study of nuclear clustering, however,
the no-core shell model in its basic form is typically not efficient in describing
spatial correlations among nucleons forming localized clusters.   
\subsection{Symmetry-adapted no-core shell model approaches}

The symmetry-adapted no-core shell model overcomes the problem of efficiently
describing clustering by making use of exact and dynamical symmetries of the spherical harmonic
oscillator Hamiltonian associated with collective mode excitations
\cite{Dytrych:2007sv,Draayer:2011a,Dreyfuss:2012us,Dytrych:2013cca,Dreyfuss:2016ezg}.
We can rewrite the single-particle spherical harmonic oscillator Hamiltonian
in terms of the usual ladder operators,
\begin{equation}
H_i = \hbar\Omega \left[ c^{\dagger}_{x,i}c_{x,i}^{}+c^{\dagger}_{y,i}c_{y,i}^{}+c^{\dagger}_{z,i}c_{z,i}^{}
+\frac{3}{2}\right].
\end{equation}
We see there is a U(3) symmetry group associated with unitary $3\times3$
rotations of the $x,y,z$ quanta, and the component continuously connected
to the identity forms an SU(3) symmetry group \cite{Elliott:1958zj,Elliott:1958yc}.
 But the symmetry group can be expanded further by also allowing SU(1,1) transformations
of the form
\begin{align}
c_{x,i}^{} & \rightarrow \alpha c_{x,i}^{} + \beta c^{\dagger}_{x,i}, \\
c^{\dagger}_{x,i} & \rightarrow \alpha^* c^{\dagger}_{x,i} + \beta^* c^{}_{x,i},
\\
& |\alpha|^2-|\beta|^2 = 1.
\end{align}
The transformation can also be applied to $c_{y,i}$, $c_{z,i}$, and any set
of real orthogonal linear combinations of $c_{x,i}$, $c_{y,i}$, and $c_{z,i}$,
and thus we also have an ${\rm SU}(1,1)\otimes{\rm O}(3)$ symmetry.  It can
be shown that the full dynamical group of the spherical harmonic oscillator is the real symplectic group
${\rm Sp}(6,{\bf R})$ for  $6\times 6$ matrices \cite{Rowe:2010a}.

The symmetry-adapted no-core shell model uses the real symplectic group
${\rm Sp}(6,{\bf R})$ and its subgroup ${\rm SU}(3)$ to generate
linear combinations of spherical harmonic basis states which form complete
representations of the $\rm{SU}(3)$ subgroup for some selected quantum numbers
$(\lambda,\mu)$ of the Cartan subalgebra of $\rm{SU}(3)$.  As the quantum numbers
$(\lambda,\mu)$ correspond to different deformation geometries, the problem
of capturing the collective behavior induced by clustering can be considerably
more efficient in the symplectic basis.
One of the future challenges for the symmetry-adapted no-core shell model
approach is to 
handle realistic nuclear forces with significant terms breaking symplectic or SU(3) symmetry.

A specific version of the symmetry-adapated no-core shell model called the
no-core symplectic model (NCSpM) was used to compute the low-lying even parity
states of $^{12}$C \cite{Dreyfuss:2012us}.  The results for the rms matter
radii and electric quadrupole moments are shown in Table \ref{C12observable}
 \cite{Dreyfuss:2012us}.
The NCSpM calculation gives a  point matter rms radius for the ground state
in agreement with experiment.  The calculation yields a point matter radius
of $r_{\rm rms}=2.93$ fm for the
Hoyle state, which is slightly larger than that of the ground state.
While this result is smaller than the results typically obtained in cluster
model calculations, it is close to a recent value deduced from experiment,
 $2.89(4)$~fm \cite{Danilov:2009zz},
and is similar to {\it ab initio} lattice EFT results at leading
order, 2.4(2)~fm  \cite{Epelbaum:2011md}.

The NCSpM calculations yield an electric quadrupole moment for the $2_1^+$
state in agreement with the experimental
value.  Similarly, a positive quadrupole moment is found for the $4_1^+$ state,
and the $0_1^+$, $2_1^+$, and $4_1$ are consistent with a rotational band
with an oblate structure.  On the other hand, a large negative result is
found for the $2^+_2$ state above the Hoyle state and the same for the $4^+$
state above the Hoyle state.  The results are consistent with a rotational
band associated with the Hoyle state with a substantial prolate deformation.
Such a prolate deformation has also been found in {\it ab initio} lattice
EFT results \cite{,Epelbaum:2011md,Epelbaum:2012qn}.

We note that SU(3)-symmetry has been also used to study clustering in shell model calculations with a core.  The cluster-nucleon configuration interaction model is one such approach \cite{Volya:2015a}.  This method has recently been used to probe the cluster structure of $^{20}$Ne resonances in elastic $^{16}\rm{O}+\alpha$ scattering \cite{Nauruzbayev:2017chb}.

\begin{table}[t]
\caption{ NCSpM point rms matter radii and electric quadrupole moments for
$^{12}$C compared to experimental data.  $^a$Ref. \cite{Tanihata:1986kh};  $^b$Ref. \cite{Danilov:2009zz};
$^c$Ref.
\cite{Ogloblin:2013rda}; and $^d$Ref. \cite{AjzenbergSelove:1990zh}. *Experimentally
deduced, based on model-dependent analyses of diffraction
scattering.}
{\footnotesize
\begin{tabular}{l|llll}
\hline
 & \multicolumn{2}{c}{matter radius (fm)} & \multicolumn{2}{c}{$Q$
($e\,$fm$^2$)} \\
&  Expt. & NCSpM         &  Expt. & NCSpM \\
        \hline
$0^+_{gs}$  & $2.43(2)^{a}$
 & $2.43(1)$ & & \\
$0^+_{2}$  {\scriptsize (Hoyle)}  & $2.89(4)^{b*}$  & $2.93(5)$ & & \\
$0^+_{3}$ &  N/A & $2.78(4)$  & & \\
$2_1^+$  & $2.36(4)^{b*}$ & $2.42(1)$  & $+6(3)^{d}$ & $+5.9(1)$\\
 $2^+$ above $0^+_{2}$  & $3.07(13)^{c*}$ & $2.93(5)$  &  N/A  & $-21(1)$\\
$4_1^+$  & N/A & $2.41(1)$  & N/A & $+8.0(3)$\\
$4^+$ above $0^+_{2}$  & N/A & $2.93(5)$  &  N/A  & $-26(1)$\\
 \hline
 \end{tabular}
}
\label{C12observable}
\end{table}%

\subsection{Continuum no-core shell model approaches}
\label{ncsmc}
Another way to incorporate clustering in the no-core shell model is to consider
spherical harmonic oscillator states corresponding to more than one center.
This is done by combining the no-core shell model formalism with the resonating
group method (RGM).  A review article summarizing recent developments can be found in Ref.~\cite{Navratil:2016ycn}.  In the following we discuss the case with two clusters.

Let the  binary-cluster state of interest have total angular momentum $J$,
parity $\pi$, and isospin $T$.  We start with binary-channel basis states
of the form \cite{Quaglioni:2010xj}
\begin{eqnarray}
|\Phi^{J^\pi T}_{\nu r}\rangle &=& \Big [ \big ( \left|A{-}a\, \alpha_1 I_1^{\,\pi_1}
T_1\right\rangle \left |a\,\alpha_2 I_2^{\,\pi_2} T_2\right\rangle\big )
^{(s T)}\nonumber\\
&&\times\,Y_{\ell}\left(\hat r_{A-a,a}\right)\Big ]^{(J^\pi T)}\,\frac{\delta(r-r_{A-a,a})}{rr_{A-a,a}}\,.\label{basis}
\end{eqnarray}
Here $\left|A{-}a\, \alpha_1 I_1^{\,\pi_1} T_1\right\rangle$ and $\left |a\,\alpha_2
I_2^{\,\pi_2} T_2\right\rangle$ are the internal wave functions of the first
and second clusters, containing $A{-}a$ and $a$ nucleons respectively. They
carry angular momentum quantum numbers $I_1$ and $I_2$ which are coupled
together to form spin $s$, and the clusters have orbital angular momentum
$\ell$. Their parity, isospin and additional quantum numbers are written
as $\pi_i, T_i$, and $\alpha_i$, respectively, with $i=1,2$. The separation
vector between the cluster centers is
\begin{equation}
\vec r_{A-a,a} = r_{A-a,a}\hat r_{A-a,a}= \frac{1}{A - a}\sum_{i = 1}^{A
- a} \vec r_i - \frac{1}{a}\sum_{j = A - a + 1}^{A} \vec r_j\,,
\end{equation}
where $\vec{r}_i$ are the single-particle coordinates for $i=1,\cdots A$.
It is convenient to group all relevant quantum numbers into a collective
index $\nu=\{A{-}a\,\alpha_1I_1^{\,\pi_1} T_1;\, a\, \alpha_2 I_2^{\,\pi_2}
T_2;$ $\, s\ell\}$.
In order to enforce the correct fermionic statistics, one uses the inter-cluster
antisymmetrizer,
\begin{equation}
\hat{\mathcal A}_{\nu}=\sqrt{\frac{(A{-}a)!a!}{A!}}\sum_{P}{\rm sgn}(P)P\,,
\end{equation}
where the sum runs over all possible permutations $P$
%$P$ are permutations
that can be carried out
among nucleons, and ${\rm sgn}(P)$ is the sign of the permutation.

The antisymmetrized basis states can be used to expand the many-body
wave function as
\begin{equation}
|\Psi^{J^\pi T}\rangle = \sum_{\nu} \int dr \,r^2\frac{g^{J^\pi T}_\nu(r)}{r}\,\hat{\mathcal
A}_{\nu}\,|\Phi^{J^\pi T}_{\nu r}\rangle\,. \label{trial}
\end{equation}The coefficient functions  $g^{J^\pi T}_\nu(r)$ correspond
to the relative-motion radial
wave functions between the clusters. These unknown coefficient functions
are solved by the non-local integral-differential coupled-channel equations
\begin{equation}
\sum_{\nu}\int dr \,r^2\left[{\mathcal H}^{J^\pi T}_{\nu^\prime\nu}(r^\prime,
r)-E\,{\mathcal N}^{J^\pi T}_{\nu^\prime\nu}(r^\prime,r)\right] \frac{g^{J^\pi
T}_\nu(r)}{r} = 0\,,\label{RGMeq}
\end{equation}
where $E$ is the total energy in the center-of-mass frame,
and the two integration kernels are the Hamiltonian kernel,
\begin{equation}
{\mathcal H}^{J^\pi T}_{\nu^\prime\nu}(r^\prime, r) = \left\langle\Phi^{J^\pi
T}_{\nu^\prime r^\prime}\right|\hat{\mathcal A}_{\nu^\prime}H\hat{\mathcal
A}_{\nu}\left|\Phi^{J^\pi T}_{\nu r}\right\rangle\,,\label{H-kernel}
\end {equation}
and the norm kernel,
\begin{equation}
{\mathcal N}^{J^\pi T}_{\nu^\prime\nu}(r^\prime, r) = \left\langle\Phi^{J^\pi
T}_{\nu^\prime r^\prime}\right|\hat{\mathcal A}_{\nu^\prime}\hat{\mathcal
A}_{\nu}\left|\Phi^{J^\pi T}_{\nu r}\right\rangle.\label{N-kernel}
\end{equation}
The nontrivial norm kernel is the result of the non-orthogonality of the
basis states~(\ref{basis}).  Furthermore, the exchange terms in the antisymmetrizer
give rise to non-local terms in the two kernels.

This no-core shell model with resonating group formalism has been used very
successfully to calculate many elastic scattering processes and inelastic
reactions involving light nuclei
\cite{Quaglioni:2008sm,Navratil:2010jn,Navratil:2011zs}.  The method has
recently been improved further by also including basis states corresponding
to the regular no-core shell basis with the full $A$-body space
in one cluster.  This has the advantage of encoding the short-range interactions
between clusters more efficiently than the resonating group method would otherwise.
 This approach, known as the no-core shell model with continuum approach, has been
used to describe two-body reactions \cite{Dohet-Eraly:2015ooa,Raimondi:2016njp},
unbound states \cite{Baroni:2013fe}, and even three-body reactions \cite{Quaglioni:2013kma}.
 Quite recently there have also been no-core shell model with continuum studies
of the cluster structure of $^6$Li \cite{Hupin:2014iqa} as an $\alpha$-cluster
and deuteron and also of $^6$He \cite{Romero-Redondo:2014fya,Romero-Redondo:2016qmc}
in terms of an $\alpha$-cluster and two neutrons.

\begin{figure}[ht]
\includegraphics[width=8cm]{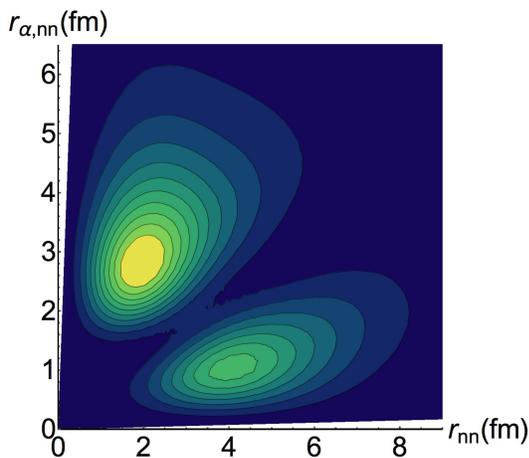}\hspace{2pc}%
\caption{(Color online) Results for the $^6$He wave function using
no-core shell model with continuum \cite{Romero-Redondo:2016qmc}.  The horizontal
axis is the separation between the two halo neutrons, $r_{\rm nn}$, and the vertical
axis is the separation between the $\alpha$-particle core and the center of
mass of the two  halo neutrons, $r_{\alpha,{\rm nn}}$.  Adapted with permission from Ref.~\cite{Romero-Redondo:2016qmc}.  Copyrighted by the American Physical Society. }
\label{fig:He6}
\end{figure}

In Fig.~\ref{fig:He6} we show results for the $^6$He wave function using no-core shell model with continuum \cite{Romero-Redondo:2016qmc}.  The horizontal axis is the separation between the two halo neutrons, $r_{\rm nn}$, and the vertical
axis is the separation between the alpha-particle core and the center of mass of the two halo neutrons, $r_{\alpha,{\rm nn}}$. The plots shows the dominance of a di-neutron configuration where the two neutrons are about 2 fm apart and the $\alpha$-particle about 3 fm away.  There is also a smaller contribution from a much smaller contribution from a split configuration where the two neutrons are far from each other with the $\alpha$-particle situated in between.

The no-core shell model with continuum can be viewed as one of several continuum
shell model methods with a long history  \cite{Mahaux:1969}.  Some
other recent developments are the shell model embedded in the continuum \cite{Okolowicz:2003a},
continuum shell model \cite{Volya:2005a}, and 
no-core Gamow shell model \cite{Papadimitriou:2013a}.
 
One recent work with particular relevance for nuclear clustering is Ref.~\cite{Kravvaris:2017a},
which uses the no-core shell model and resonating
group method for clusters, but also applies the harmonic oscillator expansion
for the relative separation between clusters.  In this work they compute spectroscopic amplitudes for the low-lying even parity states of $^8$Be, $^{10}$Be, $^{12}$C into open $\alpha$-separation thresholds.

%%%%%%%%%%%%%%%%%%%%%%%%%%%%%%%%%%%%%%%%%%%%%%%%%%%%%%%%%%%%%%%%%%%%%%%
%
% \input{continuum_qmc_section.tex}
%

\section{Continuum quantum Monte Carlo}

A recent review on continuum Quantum Monte Carlo methods in nuclear physics
has been recently been published \cite{Carlson:2014vla}.  Here we give an
overview of the methods and studies which have been used to investigate clustering
in nuclei.

\subsection{Variational Monte Carlo}

Variational Monte Carlo (VMC) relies on the variational principle that the
energy of any trial wave function will be greater than or equal to the ground
state energy.  We are of course assuming only physical states antisymmetrized
with respect to the exchange of all identical fermions. The strategy is to
start with some general functional form for the trial wave function $\Psi_T^{\{
\alpha_i \}}$ which depends on some set of unknown parameters $\{ \alpha_i
\}$.  One then computes the energy expectation
$E_T^{\{ \alpha_i \}}$ for the trial state\begin{equation}
E_T^{\{ \alpha_i \}} = \frac{ \langle \Psi_T^{\{ \alpha_i \}} |  H |\Psi_T^{\{
\alpha_i \}} \rangle }{\langle \Psi_T^{\{ \alpha_i \}} | \Psi_T^{\{ \alpha_i
\}}
\rangle},
\label{energy1}
\end{equation} and minimizes with respect to $\{ \alpha_i \}$. Instead of
minimizing the energy, one can also minimize the expectation value of the
variance operator $(H-\lambda I)^2$, which vanishes only when $\lambda$ is
an exact energy eigenvalue.

Since the trial wave function is typically a function with many degrees of
freedom, the inner products in Eq.~(\ref{energy1}) are computed using Monte
Carlo integration.  If the interactions in $H$ have a local structure in
position space, then the required integration can be performed quite simply
by selecting points in the space of the particle coordinates, ${\bf r}_1,{\bf
r}_2,\cdots$, chosen according to the squared absolute value of the trial
wave function, $|\Psi_T^{\{ \alpha_i
\}}({\bf r}_1,{\bf r}_2,\cdots)|^2$ \cite{McMillan:1965a,Ceperley:1977a}.
 For each set of points, the expectation of $H$ correspond to the value of
the function $\Psi_T^{\{ \alpha_i
\}*} H\Psi_T^{\{ \alpha_i
\}}({\bf r}_1,{\bf r}_2,\cdots)$. If one divides by the relative probability
of selecting the points ${\bf r}_1,{\bf r}_2,\cdots$, then the value one records
in the Monte Carlo integration of this observable is $H\Psi_T^{\{ \alpha_i
\}}({\bf r}_1,{\bf r}_2,\cdots)/\Psi_T^{\{ \alpha_i
\}}({\bf r}_1,{\bf r}_2,\cdots).$

The quality of the variational Monte Carlo result depends entirely on the
functional form used for the trial wave function.  Therefore it is important
to incorporate particle correlations into $\Psi_T^{\{
\alpha_i \}}$.  In variational Monte Carlo calculations
of the structure of $^{16}$O \cite{Pieper:1992gr}, the trial wave function included non-central two-body and three-body correlations acting on Slater determinants of $S$-wave and $P$-wave one-body wave functions.  The expectation values of operators were calculated using a cluster expansion for the spin- and isospin-dependent terms up to four-body order.  In many cases
variational Monte Carlo is also used to optimize the trial wave function serving as a starting point for other Monte Carlo calculations such as diffusion or Green's function
Monte Carlo.

\subsection{Diffusion or Green's function Monte Carlo}
\label{GFMC}
Diffusion or Green's function Monte Carlo (GFMC)  starts with a trial wave
function $| \Psi_T \rangle$ and uses Euclidean time evolution to extract
the ground state wave function  \cite{Kalos:1962}.  Originally diffusion
Monte Carlo and Green's function Monte Carlo referred to slightly different
algorithms.  However in today's usage, they refer to the same method. The
ground state wave function is obtained in the large time limit as
\begin{equation}
| \Psi_0 \rangle \propto \lim_{\tau \rightarrow \infty} \exp [ - (H - \lambda)
\tau ] | \Psi_T \rangle.
\end{equation}
The parameter $\lambda$ is used to stabilize the normalization of the wave
function and gives an estimate of the ground state energy $E_0$.
  A more direct calculation of the ground state energy is given by the ratio
\begin{equation}E_0 =
\lim_{\tau \rightarrow \infty}\frac{\langle \Psi_T |  H\exp [ - (H - E_0)
\tau ] | \Psi_T \rangle}{\langle \Psi_T |   \exp [ - (H - E_0)
\tau ] | \Psi_T \rangle}~.
\label{E_estimator}
\end{equation}

When exponentiated over a short time step $\Delta \tau$, the kinetic energy
term in $H$ gives rise to a diffusion process which is modeled as a random
walk in the space of all possible particle coordinates. Meanwhile, the particle
interactions result in an exponential growth or decay for each possible spin
and isospin channel.

One of the main computational challenges in GFMC\ is the sign oscillation
problem associated with the exchange of identical fermions.  These sign oscillations
will render the numerator and denominator to be vanishingly small in the
limit of large time $\tau$. For real-valued wave functions the fixed-node
approximation gives a remedy for this problem by restricting the random walk
in the space of particle coordinates to a region where the trial wave function
remains positive.  For complex-valued wave functions as one finds in nuclear
physics, a generalization of the approach called the constrained path approximation
is used \cite{Wiringa:2000gb}. In the constrained
path approximation one restricts the random walk
to a region where the overlap of the propagated state with the trial wave
function is positive \cite{Carlson:2014vla}.

GFMC has been used to compute the spectra of many light nuclei \cite{Wiringa:2000gb,Pieper:2001mp,Pieper:2002ne}.
This includes a well-known study of the $\alpha$-cluster structure of the $^8$Be
ground state \cite{Wiringa:2000gb}.  There have also been also recent studies
of the Hoyle state of $^{12}$C \cite{Carlson:2014vla} and its transitions
to the ground state.  These calculations find a radius for the Hoyle state of more than 3.1~fm, which is much larger than the ground state radius 2.43~fm. Fig. 20 shows the density distributions $r^2\rho(r)$ of the ground
state ($0^+_1$) and the Hoyle state ($0^+_2$) of
$^{12}$C \cite{Carlson:2014vla}.   Similar results have been obtained using the THSR wave function \cite{TO17}.
\begin{figure}[ht]
\includegraphics[scale=0.30]{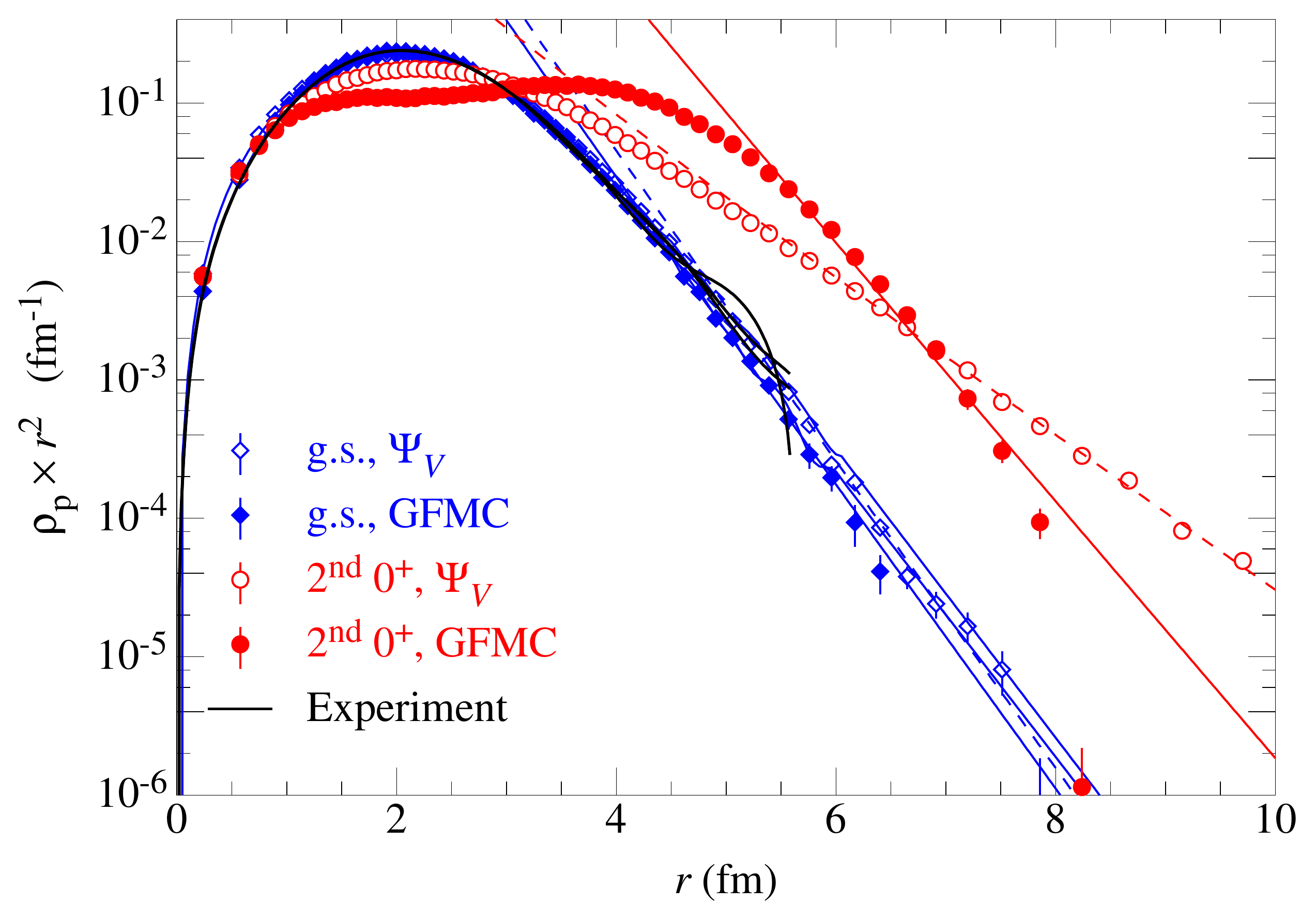}%
\caption{(Color online)  The density distributions $r^2\rho(r)$ of the ground state ($0^+_1$) and the Hoyle state ($0^+_2$) of
$^{12}$C \cite{Carlson:2014vla}.  The variational Monte Carlo results are indicated by $\Psi_V$ while Green's function Monte Carlo results are labelled as GFMC. Adapted with permission
from Ref.~\cite{Carlson:2014vla}.  Copyrighted by the American Physical
Society. }
\label{fig:HoyleQMC}
\end{figure}

In order to the improve the computational scaling of the diffusion Monte
Carlo simulations with the number of particles, one approach being pursued
is introducing an auxiliary field to rewrite the spin-dependent interactions
in terms of one-body spin operators.  This method is called auxiliary-field
diffusion Monte Carlo \cite{Gandolfi:2007hs,Gezerlis:2013ipa}.

\subsection{Monte Carlo shell model}
The Monte Carlo Shell Model (MCSM) approach is a variational method which uses auxiliary-field Monte Carlo simulations to determine a set of low-energy
basis states $|\phi_n\rangle $ \cite{Abe:2012wp}.  In this discussion we focus on the no-core version of MCSM where all nucleons are active.  Each $|\phi_n\rangle$
is a Slater determinant of deformed single-particle shell model states. The
resulting states are given good angular momentum and parity quantum
quantum numbers by explicit projection. In order to remove residual errors
due to the basis truncation, extrapolations are performed as a function of
the energy variance \cite{Shimizu:2010mp,Shimizu:2012rt}.  This method has
been used to study the alpha-two-neutron cluster structure of $^6$He, two-alpha
structure of $^8$Be, and two-alpha-two-neutron
structure of $^{10}$Be \cite{Shimizu:2012mv,Yoshida:2013dwa}.

For the case with total angular momentum $J=0$, the projected wave function is
\begin{equation}
  |\Psi \rangle = P^{J=0} | \Phi\rangle,
  \ \ \ \ \   | \Phi\rangle = \sum_n  f_n |\phi_n\rangle.
\end{equation}
The linear combination of the unprojected basis states, $|\Phi\rangle$,
% ,  which is referred to as ``before rotation'' (BR),
cannot be considered  as an intrinsic state
since the principal axis of each basis state,
$|\phi_n\rangle$, are not all aligned in the same direction.
This is fixed by performing a rotation $R(\Omega_n)$ so that the quadrupole moment is diagonalized, and $Q_{zz} \ge Q_{yy} \ge Q_{xx}$ so that the principal axis is aligned with the $z$-axis.
The intrinsic wave function $|\Phi^{{\rm intr}} \rangle$ is then defined as
\begin{eqnarray}
  |\Phi^{{\rm intr}} \rangle &\equiv& \sum_{n} f_{n} R(\Omega_n)|\phi_n \rangle.
  \label{DENS}
\end{eqnarray}

\begin{figure}[ht]
\includegraphics[scale=0.35]{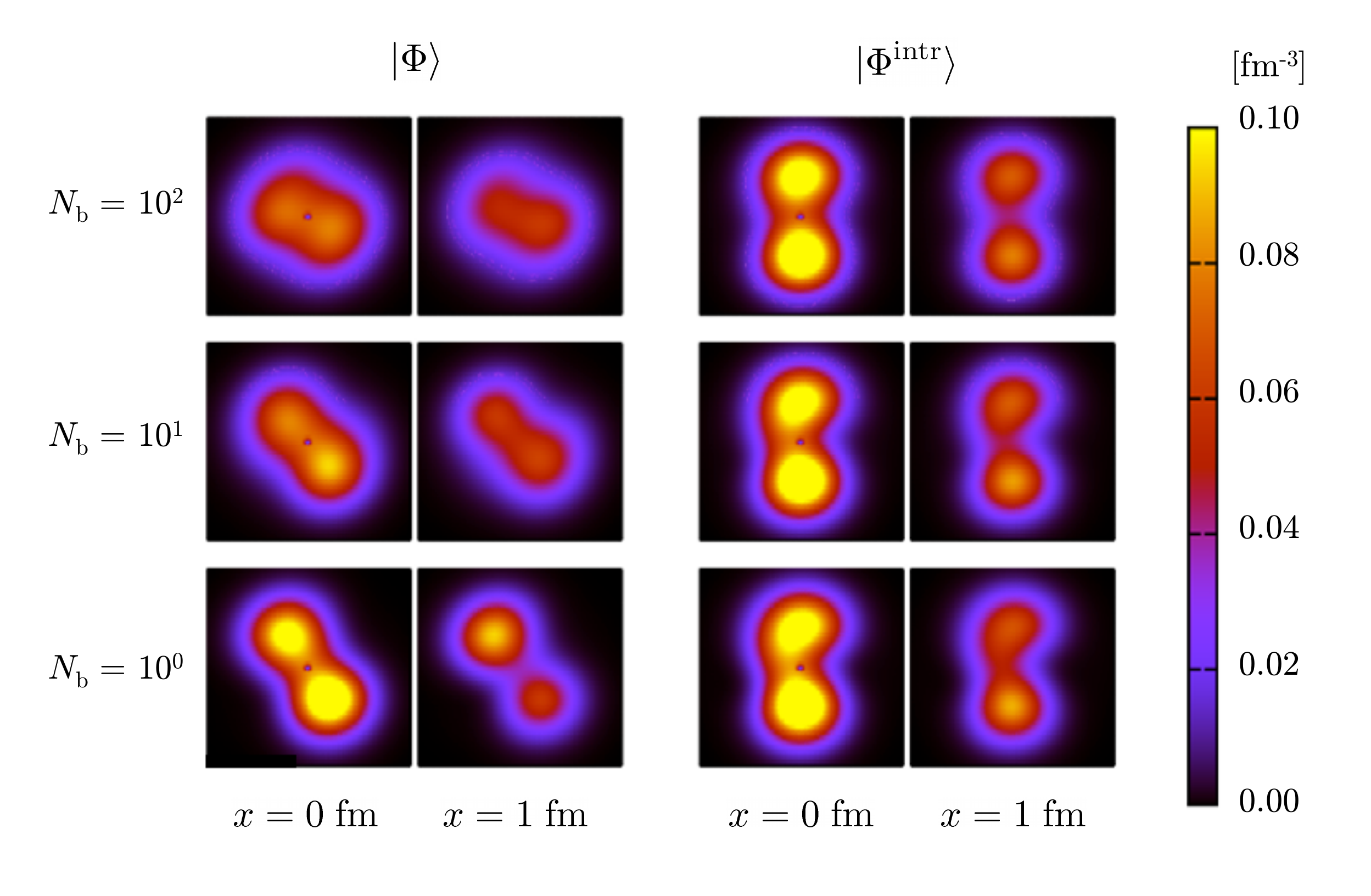}%
\caption{(Color online) the $^{8}$Be proton densities for $|\Phi\rangle$
and $|\Phi^{{\rm intr}} \rangle$ \cite{Shimizu:2012mv,Yoshida:2013dwa}.
Results are shown for $N_b=10^0, 10^1, 10^2$ basis states. Each density distribution shows the $yz$ plane for intercepts  $x=0$ fm and $x=1$~fm.  Adapted with permission
from Ref.~\cite{Shimizu:2012mv}.  Copyrighted by The Physical Society of Japan.}
\label{fig:Be8_MCSM}
\end{figure}

In Fig.~\ref{fig:Be8_MCSM} we show the $^{8}$Be proton densities for $|\Phi\rangle$
and $|\Phi^{{\rm intr}} \rangle$
\cite{Shimizu:2012mv,Yoshida:2013dwa}.
We show results for $N_b=10^0, 10^1, 10^2$ basis states.
Each density distribution shows the the $yz$ plane for intercepts $x=0$ fm and $x=1$~fm.

%%%%%%%%%%%%%%%%%%%%%%%%%%%%%%%%%%%%%%%%%%%%%%%%%%%%%%%%%%%%%%%%%%%%%%%
%
% \input{ulfnl_section.tex}
%
%

\section{Nuclear lattice effective field theory }
\label{sec:nleft}

\subsection{Chiral effective field theory on a lattice}
\label{sec:basics}

The basic idea of Nuclear Lattice Effective Field Theory (NLEFT) is to merge
the successful
chiral EFT for nuclear forces pioneered by Weinberg~\cite{Weinberg:1990rz,Weinberg:1991um}
with lattice Monte Carlo methods, that allow for numerically exact solutions
of the nuclear $A$-body problem.
First,  the ingredients to construct the chiral nuclear EFT are briefly discussed.
The EFT is formulated in terms of the asymptotically
observed states, the nucleons and the pions, the latter being the Goldstone
bosons of the
spontaneously broken chiral symmetry of QCD. The basic idea of the Weinberg
approach is
to use chiral perturbation theory to construct the potential between two,
three and four nucleons.
The various contributions are organized according to the power counting based
on the small
parameter $Q$, with $Q \in \{ p/\Lambda, M_\pi/\Lambda\}$. Here, $p$ denotes
some soft external momentum,
$M_\pi$ the pion mass and $\Lambda$ the hard scale that accounts for all
physics integrated out.
Usually, this scale is set by the appearance of the first resonance, like
the $f_0(500)$ in pion-pion scattering
or the $\Delta(1232)$ in pion-nucleon scattering. For the nuclear force problem,
the leading order (LO) contributions
are of order ${\mathcal O}(Q^0)$, comprising the leading one-pion exchange
(OPE) and two local four-nucleon
contact interactions without derivatives. At next-to-leading order (NLO),
${\mathcal O}(Q^2)$,
one has the leading two-pion exchange (TPE)  interactions and seven further
four-nucleon terms with two
derivatives (for on-shell scattering) as well as two isospin symmetry-breaking
terms that account for the
dominant strong interaction difference between the proton-proton, proton-neutron
and neutron-neutron
systems. Finally, at next-to-next-to-leading order (N$^2$LO), ${\mathcal
O}(Q^3)$,
that is the accuracy to which most NLEFT calculations have been carried out
so far, one has further
TPE corrections proportional to the dimension-two low-energy constants (LECs)
$c_i$ of the effective pion-nucleon
Lagrangian that can be precisely determined from the dispersive Roy-Steiner
equation analysis of pion-nucleon scattering~\cite{Hoferichter:2015tha}.
At this order, three-nucleon forces start to contribute. These fall in three
topologies. The two-pion exchange diagram
is entirely given in terms of the LECs $c_{1,2,4}$. The one-pion exchange
coupling to a four-nucleon term and the
local six nucleon contact term are parametrized by the LECs $D$ and $E$,
respectively. These are commonly
determined from the triton binding energy and the axial-vector current contribution
to triton decay~\cite{Gazit:2008ma}.
For further details, we refer the reader to the reviews~\cite{Epelbaum:2008ga,Machleidt:2011zz}.

\begin{figure}[!tb]
\centerline{\includegraphics*[width=4.5cm]{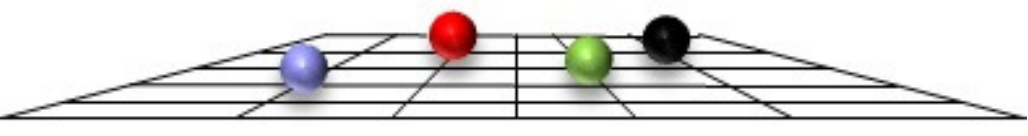}~{\includegraphics*[width=4.5cm]{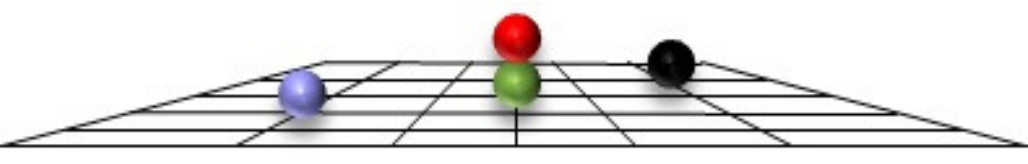}}}
\centerline{\includegraphics*[width=4.5cm]{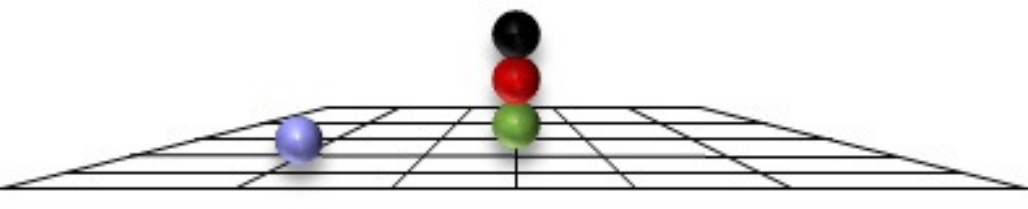}~{\includegraphics*[width=4.5cm]{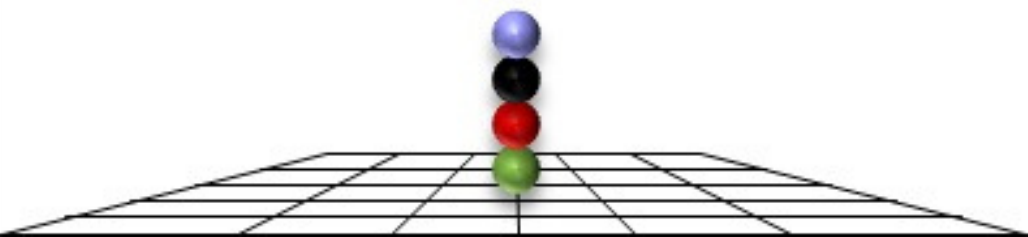}}}
\caption{\label{fig:configs}
(Color online) Possible configurations of four nucleons on the lattice (here shown in a two-dimensional sketch).}
\end{figure}

In the lattice formulation, Euclidean space-time is given by a finite hypercubic
volume, with $L$ the length in
any of the spatial directions and $L_t$ the extension in the temporal direction. Further, the lattice is defined by
a minimal spatial distance $a$, the lattice spacing, and similarly by $a_t$
in the temporal direction. In most calculations
discussed in what follows,  a coarse spatial lattice with $a = 1/(100~{\rm
MeV}) = 1.97$~fm was used, while $a_t$ is
chosen to be $a_t = 1/(150~{\rm MeV}) = 1.32$~fm. One important feature of
the finite lattice spacing is the UV finiteness of the theory,
as the largest possible momentum is given by $p_{\rm max} = \pi/a \simeq
314\,$MeV.   Thus, the interaction is
very soft and therefore most higher order corrections, including also the
Coulomb effects, can be treated in perturbation theory.
Another advantage of this approach is the fact that all possible configurations
of nucleons are sampled, as
depicted in Fig.~\ref{fig:configs}.
%%%%%%%%%%%%%%%%%%%%%%%%%%%%%%%%%%%%%%%%%%%%%%%%%%%%%%%%%%%%%%%%%%%%%%%
%%%%%%%%%%%%%%%%%%%%%%%%%%%%%%%%%%%%%%%%%%%%%%%%%%%%%%%%%%%%%%%%%%%%%%%%
This gives a first hint that the phenomenon of  clustering indeed will arise
quite naturally
in this approach. In the actual calculations, the interactions between the
nucleons are described in terms of auxiliary
fields, which makes the approach particularly suited for highly parallel
computation. In essence, each nucleon evolves in time
from the starting at $t=t_i$ up to the final time $t_f$. The value of $t_f$
has to be large enough so that the asymptotic
behavior of any observable for  the $A$-nucleon state can be extracted. For
further details, we refer to the detailed
description of the LO chiral EFT interactions on the lattice in Ref.~\cite{Borasoy:2006qn}.
See also the review in Ref.~\cite{Lee:2008fa}.

Another important fact is the approximate Wigner SU(4) symmetry of the nuclear
interactions~\cite{Wigner:1937zz}.  This is the observation that combined spin-isospin rotations of
the nucleon four-vector $(p\uparrow, p\downarrow, n\uparrow, n\downarrow)
$
leave the nuclear forces in the $S$-wave  approximately invariant. This symmetry
is broken by the OPE and the Coulomb interaction,
but rather well respected by the four-nucleon short-range operators~\cite{Mehen:1999qs}.
Most importantly, in case of an
exact Wigner symmetry, nuclei with spin and isospin zero do not show any
sign oscillations~\cite{Chen:2004rq}, which make finite density
lattice simulations so difficult. This approximate symmetry can therefore
be used as an inexpensive filter in the actual
simulations. It can also be proven that for the case of attractive SU(4) interactions, the resulting nuclear binding energies must satisfy spectral convexity bounds that correspond to an alpha clustering phase \cite{Lee:2007eu}.  For further work on understanding  Wigner  symmetry within
QCD  and its consequences,
see e.g. Refs.~\cite{Lee:2004ze,CalleCordon:2008cz,Beane:2013br}.  We will
come back to this topic in subsection~\ref{sec:tech}.

Before continuing, let us define what we mean by {\em ab initio} calculation
in this context. The various parameters appearing in the lattice
approach, like the LECs and the smearing parameters as defined below, are
determined in fits to properties of few-particle
systems like phase shifts and binding energies. Here, few means less or equal
four. The properties of nuclei with larger atomic
number can then be predicted without further parameter tuning to a precision that is  given by the
accuracy of the underlying chiral EFT Hamiltonian. Note that recently it has
been found that  fitting also to the low-energy  $\alpha$-$\alpha$ $S$-wave phase shifts
for determining the pertinent LECs
provides some advantage in controlling higher-body interactions in larger systems \cite{Elhatisari:2016owd}.

\subsection{Lattice formalism}
\label{sec:tech}

To calculate the energy or any other static observable, we need an initial
wave function for the nucleus under consideration,
$|\Psi_{A}^\mathrm{in} \rangle$.  Such a state can on one hand be chosen
as a Slater-determinant states composed of delocalized standing
waves in the periodic cube with $A$ nucleons, and on the other hand as localized
$\alpha$-cluster trial states (or any other type of cluster state).
Such localized states have been used in the investigations of $^{12}$C and
$^{16}$O~\cite{Epelbaum:2011md,Epelbaum:2012qn,Epelbaum:2013paa}.
These can be used to check the calculations with the delocalized initial
states, but also  allow to assess the spatial structure of the nuclei.
It has to be understood that these states are always prepared with a given
total angular momentum $J$ and parity $\pi$, that is a fixed $J^\pi$. Rotational symmetry breaking due to the lattice is an issue that will be discussed
later.
The central object of NLEFT is the Euclidean-time projection amplitude
\begin{equation}
Z_A^{}(t) \equiv \langle\Psi_A^{}(t^\prime_{})| \exp(-H_\mathrm{LO}^{} t)
|\Psi_A^{}(t^\prime_{})\rangle,
\end{equation}
that allows to compute the ``transient energy'' $E_A^{}(t) = -\partial[\ln
Z_A^{}(t)]/\partial t$. Here $H_\mathrm{LO}$ is the leading-order Hamiltonian. In the infinite time limit, this gives
the ground-state energy, as all excited states have a larger energy and thus
 fall off faster. The initial and final states  have been prepared using
\begin{equation}
|\Psi_A^{}(t^\prime_{})\rangle = \exp(-H_{\rm SU(4)}t')|\Psi_A^{\rm in}\rangle
\end{equation}
where $H_{\rm SU(4)}$ is a lattice Hamiltonian that approximates $H_\mathrm{LO}$ but maintains an exact Wigner SU(4) symmetry, as detailed in Ref.~\cite{Borasoy:2006qn}.
%
%For a coarse lattice spacing, higher order corrections can be calculated
%in perturbation theory.
This method has been considerably improved by the
so-called ``triangulation'' procedure introduced in Ref.~\cite{Lahde:2013uqa}
using several initial states, which allows significant reduction in the
error due to Euclidean time extrapolation. For example, using this
method, the ground state energies of $^{12}$C and $^{16}$O
can now be calculated with an absolute uncertainty of $\pm 200\,$keV. A detailed discussion of this method and the associated
uncertainties is given in Ref.~\cite{Lahde:2014sla}.

\begin{figure}[!t]
\includegraphics[trim=0.5cm 3cm 0.5cm 13cm,width=8cm]{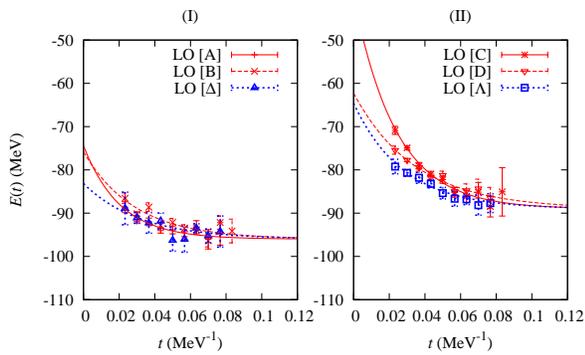}
\caption{(Color online) Lattice results for the $^{12}$C spectrum at leading order~(LO).
Panel~I shows the results using initial states $A$, $B$ and $\Delta$,
each of which approaches the ground state energy. Panel~II shows the results
using initial states $C$, $D$ and $\Lambda$. These trace out an
intermediate plateau at an energy $\simeq 7$~MeV above the ground state.
Here, $A-D$ are configurations that start with delocalized nucleons, where
as
$\Delta$ refers to a compact triangle, while $\Lambda$ denotes an obstuse
triangular
configuration. For details, see Ref.~\cite{Epelbaum:2012qn}.
\label{fig:C12LO}}
\vspace{-5mm}
\end{figure}

In order to compute the low-lying excited states of a given nucleus,
the Euclidean time projection method is extended to a multi-channel calculation.
Take the $^{12}$C nucleus as an example~\cite{Epelbaum:2011md}.
 Using the auxiliary-field formalism, one applies the exponential operator $\exp({-Ht})$ to a set of different single-nucleon
standing waves
in the periodic cube. From these standing waves one then builds initial states
consisting of Slater determinants of $6$ protons and $6$ neutrons each and extracts orthogonalized
energy eigenstates with the desired quantum properties.  In Ref.~\cite{Epelbaum:2011md} four states were found with even parity
and total momentum equal to zero.  As is well known, the lattice discretization of space and periodic boundaries
reduce the full rotational group to a cubic subgroup.  This complicates the
identification of spin states. However, the number
of
energy levels seen for each value of $J_{z}$ allows one to identify
different spin states. This method can be refined by not only using
delocalized standing waves but also initial cluster states with nucleons grouped into  Gaussian wave packets arranged
in certain geometrical configurations, see Ref.~\cite{Epelbaum:2012qn}.
As shown in Fig.~\ref{fig:C12LO}, various configurations
that corresponds to $J^\pi = 0^+$ indeed leads to the ground state (left panel)
or the first
excited $0^+$ Hoyle state about 7~MeV above the ground state.

We have already noted that the proximity of the Hoyle state energy to the triple-$\alpha$ threshold is important for the production of carbon in the universe.  As with any near-threshold state, this very low-energy scale is well-separated from other energy scales, and this separation of scales forms the basis for halo effective field theory \cite{Bertulani:2002sz,Bedaque:2003wa,Higa:2008dn}.
One interesting theoretical question is whether this proximity of the Hoyle state to the triple-$\alpha$ threshold is a generic feature of quantum chromodynamics or is something needs to be fine-tuned.
The quark mass dependence of the Hoyle state energy has been studied
using lattice simulations \cite{Epelbaum:2012iu,Epelbaum:2013wla} in connection
with the anthropic principle, the production of carbon and oxygen, and the
fine-tuning of the parameters of nature \cite{Beane:2002vs,Epelbaum:2002gb,Beane:2002xf,
Epelbaum:2002gk,Berengut:2013nh,Bedaque:2010hr,Meissner:2014pma}.  

%%%%%%%%%%%%%%%%%%%%%%%%%%%%%%%%%%%%%%%%%%%%%%%
%%%%%%%%%%%%%%%%%%%%%%%%%%%%%%%%%%%%%%%%%%%%%%%

We now discuss two-body scattering on the lattice. This is not only an important
ingredient to fix the LECs of the effective Lagrangian but
 can also be used to investigate the nuclear dynamics encoded in nuclear
reactions.
The best known and most frequently used method is due to L\"uscher, who showed that the energy
of an interacting two-particle
system in a finite volume can be related to the infinite-volume phase shift at
the same energy \cite{Luscher:1986pf,Luscher:1990ux}. This
method has by now been extended to cope with higher partial waves, partial-wave
mixing,  multi-channel scattering,
boosted frames and all possible types of boundary conditions, see e.g. Refs.~\cite{He:2005ey,Liu:2005kr,Bernard:2008ax,Lage:2009zv,Luu:2011ep,Konig:2011nz,Bour:2011ef,Konig:2011ti,Gockeler:2012yj,Li:2012bi,Briceno:2013bda,Li:2014wga,Li:2015npa}.
However, for the case of nucleon-nucleon
scattering, which involves higher energies, spin-orbit coupling and partial-wave
mixing, the method has more limited accuracy.
A more robust approach that makes use of the non-relativistic character of
the nuclear problem is the so-called spherical wall approach,
used for numerical calculations as early as in Ref.~\cite{Carlson:1984zz}, but reinvented for
the lattice formulation used here in Ref.~\cite{Borasoy:2007vy}.
Here, one imposes a hard spherical wall boundary on the relative separation
between the two particles at some radius $R_{\rm W}$.
In that way, copies of the interactions produced by the periodic lattice
are removed and, from the solution of the Schr\"odinger
equation for spherical standing waves at $r= R_{\rm W}$, one can easily recover
the phase shift for a given partial wave. Mixing of the
partial waves caused by spin-orbit coupling is also easily dealt with. This
method has been improved significantly in Ref.~\cite{Lu:2015riz}.
First, so-called radial position states for a given partial wave are constructed
according to
\begin{equation}\label{eq:radial}
|{r}\rangle^{\ell,\ell_z} = \sum_{\vec{r}^\prime} Y_{\ell,\ell_z}(\hat{r}^\prime)
\, \delta_{r,|\vec{r}^{\,\prime}|} \, |\vec{r}^{\,\prime}\rangle~,
\end{equation}
with $ Y_{\ell,\ell_z}$ spherical harmonics with angular momentum quantum
numbers $\ell, \ell_z$, and $r$ is to be restricted to be
less than half the box size $L/2$. Angular momentum is not conserved on the lattice. However the amount rotational invariance breaking decreases with increasing radial distance, and we can use spherical harmonics to dial the corresponding partial waves. This projection allows one to construct the so-called radial
lattice Hamiltonian. Second, one introduces auxiliary
potentials in the region immediately in front of the spherical wall. By tuning the depth of
this potential, one can dial the scattering energy. In case of
partial-wave mixing, this potential is chosen such that time-reversal
symmetry is broken. This allows to extract phase shifts and scattering
angles from the real and imaginary parts of the wave function. For details,
see Ref.~\cite{Lu:2015riz}.

The aforementioned SU(4) symmetry can be further utilized to suppress the
sign oscillations in auxiliary-field Monte Carlo calculations.
The underlying idea is to smoothly connect the LO lattice Hamiltonian with
an SU(4)-symmetric counterpart that does not suffer from
any sign oscillations. In that way, one can construct a one-parameter family
of Hamiltonians, $H(d) = d H_{\rm LO} + (1-d)H_{\rm SU(4)}$.
For $d=1$, one obviously recovers the microscopic chiral Hamiltonian. One
can then perform simulations for various values of $d$ and extrapolate to the limit $d=1$. This method is called symmetry-sign
extrapolation (SSE) \cite{Lahde:2015ona}.  In contrast to techniques introduced earlier
in shell model Monte Carlo
calculations~\cite{Alhassid:1993yd,Koonin:1996xj}, the magnitude of the sign oscillations
in this approach are typically quadratic in $d$ and are thus milder.

Another important issue in the lattice simulations is interaction smearing. The interactions are not strictly point-like but
distributed over neighboring lattice sites. This method is very common in lattice
QCD to enhance the strength of a given quark source,
see e.g. Refs.~\cite{Gusken:1989qx,Daniel:1992ek,Allton:1993wc,Morningstar:2003gk,Hasenfratz:2007rf,Edwards:2008ja}
for some groundbreaking work. In NLEFT, smearing is done for various reasons.
First, the nucleon-nucleon interaction terms
are smeared with a Gaussian-type function, whose depth and width is fixed from
the averaged nucleon-nucleon $S$-wave scattering lengths and effective ranges. As discussed in
detail in Ref.~\cite{Borasoy:2006qn}, this type of smearing is required to
avoid overbinding due to the configurations with four
nucleons on one lattice site, cf. Fig.~\ref{fig:configs}.  This has the added
value that important effective range corrections are treated
non-perturbatively rather than perturbatively, i.e. some important higher-order
corrections are being resummed.
Second, a novel type of non-local smearing was introduced in Ref.~\cite{Elhatisari:2016owd}.
For that, one considers non-local nucleon annihilation (creation) operators
and two-nucleon densities, such as
\begin{eqnarray}
a^{(\dagger)}_{\rm NL}({\bf n}) &=& a^{(\dagger)}({\bf n})+s_{\rm NL}\displaystyle\sum_{\langle{\bf
n'\, n}\rangle}a^{(\dagger)}({\bf n'})~, \nonumber\\
\rho_{\rm NL}({\bf n}) &=& a^\dagger_{\rm NL}({\bf n}) a_{\rm NL}^{}({\bf
n})~,
\end{eqnarray}
where the three-vector ${\bf n}$ denotes a lattice site, and $\sum_{\langle{\bf
n'\, n}\rangle}$ denotes the sum over
nearest-neighbor lattice sites of ${\bf n}$, so that $|{\bf n'}-{\bf n}|
= a$.
%, as shown in Fig.~\ref{fig:smear}.
%%%%%%%%%%%%%%%%%%%%%%%%%%%%%%%%%%%%%%%%%%%%%%%%%%%%%%%%%%%%%%%%%%%%%%%
%\begin{figure}[!tb]
%\centerline{\includegraphics[width=5.5cm,angle=0]{Ulf/smearing.pdf}}
%\caption{\label{fig:smear}
%Non-local smearing.
%}
%\end{figure}
%%%%%%%%%%%%%%%%%%%%%%%%%%%%%%%%%%%%%%%%%%%%%%%%%%%%%%%%%%%%%%%%%%%%%%%%
The smearing parameter $s_{\rm NL}$ is to be determined together with the
other parameters and LECs as discussed later.
This non-local smearing offers the possibility of considering non-local short-range interactions
 on the lattice,
as discussed in subsection~\ref{sec:alphagas}.

Another issue to be addressed is the lattice spacing dependence. In contrast
to lattice QCD, in NLEFT one does not perform
the continuum limit $a\to 0$ as we are dealing with an effective field theory
that  only makes sense below some hard (breakdown)
momentum scale $\Lambda$. Physically, one can understand this very intuitively, the
EFT is not appropriate to resolve the inner structure of the
nucleon at distances less than the proton charge radius of about 0.85~fm. Therefore,
one expects that the calculations within NLEFT are invariant
under variations of $a$ between  1 and 2~fm, provided that the LECs are properly
readjusted. This expectation is indeed borne out
by explicit calculations. In Ref.~\cite{Klein:2015vna} it was shown within
the pionless as well as the pionful LO EFT that the $S$-wave
phase shifts and the deuteron binding energy can be reproduced for $0.5 \lesssim
a \lesssim 2.0\,$fm. This has recently been sharpened by studying the neutron-proton
interactions to NNLO for lattice spacings from $1\ldots 2$~fm \cite{Alarcon:2017zcv}.
Presently, larger systems are
being systematically investigated to establish this $a$-independence in general.

 Finally, we mention that simple $\alpha$-cluster models have been used in
Refs.~\cite{Lu:2014xfa,Lu:2015gfa}
 to gain a deeper understanding of the effects of the rotational symmetry
breaking on the lattice and to develop
 methods to overcome this. It was demonstrated in  Ref.~\cite{Lu:2014xfa}
that  lattice spacing errors are closely related to the commensurability
of the lattice with
 the intrinsic length scales of the system and that rotational symmetry breaking
effects can be significantly reduced by using improved lattice actions.
 In particular,  the physical energy levels are accurately reproduced by
the weighted average of a given spin multiplets. Further, in Ref.~\cite{Lu:2015gfa}
 the matrix elements of multipole moment operators were studied. It could
be shown that the physical reduced matrix element is well reproduced
 by averaging over all possible orientations of the quantum state, and this
is expressed as a sum of matrix elements weighted by the corresponding
 Clebsch-Gordan coefficients. These methods will become important when further investigations of the electromagnetic structure of
 nuclei using NLEFT are performed.

\subsection{Adiabatic projection method}
\label{sec:APM}

To study reactions  and inelastic processes on the lattice, one makes use
of the adiabatic projection method (APM).  The APM  has been developed in Refs.~\cite{Rupak:2013aue,Pine:2013zja}
and further refined in  Refs.~\cite{Elhatisari:2014lka,Rokash:2015hra,Elhatisari:2016hby}.
From the set-up, it is similar
to the recent studies combining the resonating group method with the no-core-shell-model
\cite{Quaglioni:2008sm,Navratil:2010jn,Navratil:2011zs,Romero-Redondo:2014fya}.
  Within the APM,   the cluster-cluster scattering problem on the lattice
is evaluated in a two-step procedure. First, one uses Euclidean time projection
 to
determine an  adiabatic Hamiltonian for the participating clusters.
Strictly speaking, for finite temporal lattice spacing, an adiabatic transfer
matrix rather than the Hamiltonian is constructed, but the method is
essentially the same, and for simplicity,  the Hamiltonian formulation will
be discussed here.
In the second step, this adiabatic Hamiltonian is then used to calculate
the pertinent
phase shifts. The biggest advantage of the APM is that the computational
time appears to
scale with the number of interacting constituents, $t_{\rm CPU}\sim (A_1+A_2)^2$,
with $A_i$
the number of nucleons in cluster $i$, while more conventional approaches
exhibit factorial
scaling with increasing atomic number.

%%%%%%%%%%%%%%%%%%%%%%%%%%%%%%%%%%%%%%%%%%%%%%%%%%%%%%%
\begin{figure}[!tb]
%\begin{center}
\includegraphics[scale=0.3]{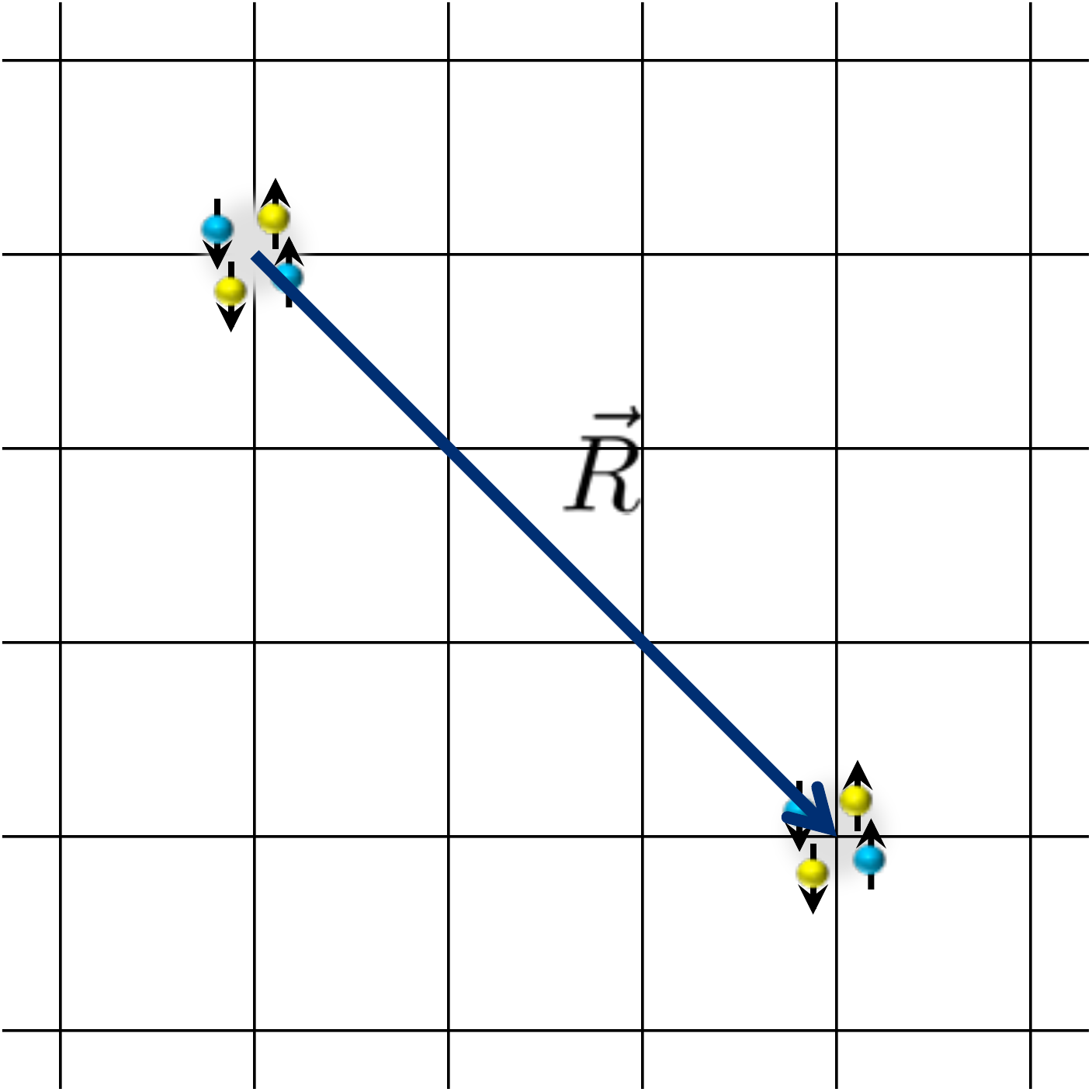}
\caption{(Color online) An initial state formed of two clusters $|\vec{R}\rangle$. The two
clusters are separated by
  the displacement vector $\vec{R}$.}
\label{fig:clusters}
%\end{center}
\vspace{-5mm}
\end{figure}
%%%%%%%%%%%%%%%%%%%%%%%%%%%%%%%%%%%%%%%%%%%%%%%%%%%%%%%%%

Consider  an $L^3$ periodic  lattice and a set of two-cluster states $|\vec{R}\rangle$
labeled by their separation vector $\vec{R}$, as illustrated in
Fig.~\ref{fig:clusters}. In general, there are spin and flavor indices
for these states, but we suppress writing the indices for notational
simplicity.  Also, it is favorable to perform a radial projection as given
in Eq.~(\ref{eq:radial}).
However, the exact form of these two-cluster states is not
important except that they are localized so that for large separations
they factorize as a tensor product of two individual clusters,
%\begin{equation}
$|\vec{R}\rangle=\sum_{\vec{r}} |\vec{r}+\vec{R}\rangle_1\otimes|\vec{r}\rangle_2.
\label{eqn:single_clusters}$
%\end{equation}
These states are propagated in Euclidean time to form dressed cluster states,
% \begin{align}
$\vert \vec{R}\rangle_\tau   =\exp(-H\tau)\vert \vec{R}\rangle$.
%\end{align}
An important consequence of this  evolution in Euclidean time with the microscopic
Hamiltonian is the
fact that deformations and polarizations of the interacting clusters are
incorporated automatically. Also, in this
way one projects onto the space of low-energy scattering states in the finite
volume, so that in
the limit of large Euclidean time, these dressed cluster states span the
 low-energy subspace of   two-cluster continuum states.
Next, matrix elements of the microscopic Hamiltonian
with respect to the dressed cluster states are formed,
%\begin{align}
$\left[H_{\tau}\right]_{\vec{R},\vec{R}'} =\ _{\tau}\langle\vec{R}\vert H
\vert\vec{R}'\rangle_{\tau}$.
%\end{align}
However, since the dressed cluster states  $|\vec{R}\rangle_\tau$
are, in general, not orthogonal, one needs to construct the  norm matrix
$N_{\tau}$,
%\begin{align}
$\left[N_{\tau}\right]_{\vec{R},\vec{R}'} =\ _{\tau}\langle\vec{R}\vert\vec{R}'\rangle_{\tau}$,
%\end{align}
so that the  Hermitian adiabatic Hamiltonian can be readily calculated
\begin{equation}
\left[ {H^a_{\tau}} \right]_{\vec{R},\vec{R}'} = \!\!\!\! \ \sum_{\vec{R}'',\vec{R}'''}
\left[ N_{\tau}^{-1/2} \right]_{\vec{R},\vec{R}''} \left[H_{\tau}^{\phantom{1/2}}\!\!\!\!\!\right]_{\vec{R}'',\vec{R}'''}
\left[ N_{\tau}^{-1/2} \right]_{\vec{R}''',\vec{R}'}.
\label{eqn:Adiabatic-Hamiltonian}
\end{equation}
%%%%%%%%%%%%%%%%%%%%%%%%%%%%%%%%%%%%%%%%%%%%%%%%%%%%
\begin{figure}[!t]
\includegraphics[scale=0.3]{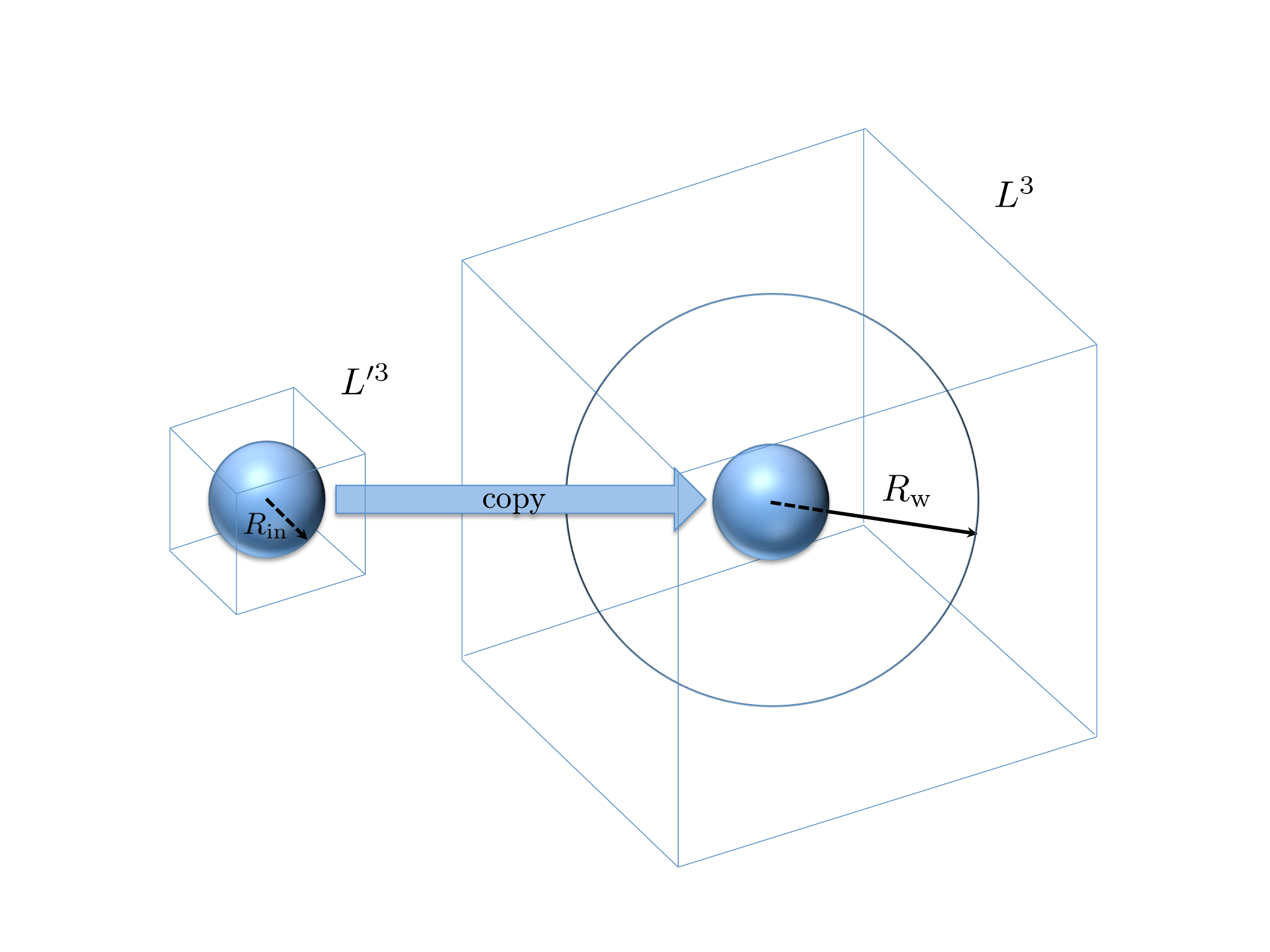}
\vspace{0.2cm}
\caption{(Color online) A sketch of the lattices for the cluster-cluster calculations in
the overlapping
and the non-interacting regions. $R_{\text{in}}$ is the largest radial distance
that is free of
systematic errors due to the periodic boundary in the cubic box with volume
${L'} ^{3}$.
 $R_{\text{w}}$ indicates the radius of the spherical wall discussed in section~\ref{sec:tech}.}
\label{fig:two_cluster_simulation}
\end{figure}
%%%%%%%%%%%%%%%%%%%%%%%%%%%%%%%%%%%%%%%%%%%%%%%%%%%
In the limit of large $\tau$, the spectrum of $H^a_{\tau}$ exactly
reproduces the low-energy finite volume spectrum of the microscopic
Hamiltonian $H$. From this adiabatic Hamiltonian, elastic phase shifts can
be calculated
using the methods discussed above. Inelastic processes can also be dealt
within this
scheme by including additional channels, see Ref.~\cite{Pine:2013zja} for
details.
One remark is in order. Since one is working in Euclidean time,  the time
evolution operator
acts indeed as a diffusion operator. The precise definition of the asymptotic
states must therefore
account for this, and in fact one can define an asymptotic radius $R_{\epsilon}$
as the
radius such that for $|\vec{R}|>R_{\epsilon}$ the amount of overlap
between the cluster wave packets is less than $\epsilon$~\cite{Rokash:2015hra}.
Consequently,
in  the asymptotic region $|\vec{R}|>R_{\epsilon}$, the dressed
clusters are widely separated and interact only through long range
forces such as the Coulomb interaction.
For cases where there are no long range interactions, the scattering
states of the adiabatic Hamiltonian are given by a superposition of Bessel
functions in the asymptotic region.  For the case with Coulomb
interactions, the scattering  states of the adiabatic Hamiltonian in the
asymptotic
region correspond to a superposition of Coulomb  wave functions. The latter
case is schematically shown in Fig.~\ref{fig:two_cluster_simulation}. A much
refined version of the adiabatic Hamiltonian based on an improved radial
``binning''
was given in Ref.~\cite{Elhatisari:2016hby}. Thus, large-scale numerical
computations of
nucleus-nucleus scattering and reactions using Monte Carlo methods are possible.
We discuss the archetypical process of elastic $\alpha$-$\alpha$ scattering
in
subsection~\ref{sec:alphaalpha}.

\subsection{Results}
\label{sec:resNL}

\subsubsection{Alpha-cluster nuclei}
\label{sec:alphachain}

\begin{comment}  %removed
%%%%%%%%%%%%%%%%%%%%%%%%%%%%%%%%%%%%%%%%%%%%%%%%%%%%%%%%%%%%%%%%%%%%%%%
\begin{figure}[!tb]
\centerline{\includegraphics[width=6.95cm,angle=0]{Ulf/c12spec.pdf}}
\vspace{-7mm}

\caption{\label{fig:12C}
(Color online) Lowest lying even-parity states of $^{12}$C from NLEFT at NNLO (right) in
comparison to
the experimental values (left). Note that these are absolute energies, not
just excitation energies.
The corresponding theoretical uncertainties are discussed in the text.
}
\vspace{-5mm}
\end{figure}
%%%%%%%%%%%%%%%%%%%%%%%%%%%%%%%%%%%%%%%%%%%%%%%%%%%%%%%%%%%%%%%%%%%%%%%%
\end{comment}

In Refs.~\cite{Epelbaum:2011md,Epelbaum:2012qn}   the even-parity spectrum  and
structure of $^{12}$C was calculated.
The underlying Hamiltonian was given at NNLO precision, which includes the
first contributions of the three-nucleon force (3NF). The 11 LECs related
to the nucleon-nucleon interactions were fixed from the $S$- and $P$-wave np phase shifts
as well from the pp and nn scattering lengths. The two
LECs related to the 3NF were fixed from the triton binding energy and the
weak axial-vector current. With that, the binding energy of $^4$He
is $-28.3(6)\,$MeV, in agreement with the empirical value. The next $\alpha$-type
nucleus, $^8$Be, is bound with $-55(2)\,$MeV,
compared to the empirical value of $-56.5\,$MeV, which is above the $2\alpha$
threshold, i.e. $^8$Be is unbound in nature. Nevertheless, $^8$Be is long-lived,
so given the accuracy of the NNLO calculation, this agreement is satisfactory.
The resulting even-parity spectrum  of $^{12}$C
is shown in the NLEFT results presented in Fig.~\ref{fig:c12-comparison}. The uncertainties on the energy levels have been considerably reduced to what was quoted in the original papers~\cite{Epelbaum:2011md,Epelbaum:2012qn}, the ground state can now
be calculated with an uncertainty of about 200~keV, and similar errors are
expected for the excited states. Most importantly, the
clustering arises very naturally, as already discussed in subsection~\ref{sec:tech}.
Also, by using initial cluster-type states, one
can map out the most important contributions for a state of given energy,
spin and parity. We find that the ground and the first
excited $2^+$ state of $^{12}$C are mostly given by a compact triangular
configuration of three alphas, while the Hoyle state
and the second $2^+$ receive a large contribution from the so-called ``bent-arm''
configuration (obtuse triangle).  This is an
indication that the second $2^+$ state is indeed a rotational excitation
of the Hoyle state. However, one has to be aware that
such ``pictorials'' of the wave function are resolution-dependent, that means
for a finer lattice spacing one will be able to resolve
these structures in more detail. The charge radii, quadrupole moments and
electromagnetic transitions among the low-lying
even-parity states of $^{12}$C have also been calculated at LO. These results
tend to be on the low side of the experimental
values. This can be traced back to the fact that at LO, the charge radius
comes out about 10\% too small. If one scales the corresponding moments
and transition elements with appropriate powers of $r(0^+_1)^{\rm exp}/r(0^+_1)^{\rm
LO}$, the agreement is quite satisfactory. Of course, this needs
to be backed up in the future by higher order calculations of these observables.

%%%%%%%%%%%% table %%%%%%%%%%%%%%%%%%%%%%%%%%%%%%%%%%%%%%%%%%%%%%%%%%
\begin{table}[t]
\begin{tabular}{|c|cccc|}
\hline
           &   $^{16}$O & $^{20}$Ne & $^{24}$Mg & $^{28}$Si\\
\hline
 LO         & $-147.3(5)$ & $-199.7(9)$ & $-255(2)$ & $-330(3)$ \\
 NNLO   & $-138.8(5)$ & $-184.3(9)$ & $-232(2)$ & $-308(3)$ \\
 NNLO* & $-131.3(5)$ & $-165.9(9)$ & $-198(2)$ & $-233(3)$ \\
\hline
 Exp.      &  $-127.62$ & $-160.64$ & $-198.26$ & $-236.54$ \\
\hline
\end{tabular}
\caption{Ground state energies for $\alpha$-cluster nuclei above $^{12}$C.
Shown are
the results at LO and NNLO. NNLO* denotes the force supplied with a four-nucleon
interaction. The experimental values are also given. Units are MeV.
\label{tab:4N}}
\vspace{-5mm}
\end{table}
%%%%%%%%%%%%%%%%%%%%%%%%%%%%%%%%%%%%%%%%%%%%%%%%%%%%%%%%%%%%%%%%%%%
Before elaborating on the structure of heavier nuclei, it is important to
scrutinize the NNLO forces. This was done in Ref.~\cite{Lahde:2013uqa},
where it was shown that for $\alpha$-cluster nuclei beyond $A=12$ an overbinding
appears, that grows with atomic number, as shown in
Table~\ref{tab:4N}. This has also been observed in other {\em ab initio} approaches
using soft interactions, see e.g.
Refs.~\cite{Hagen:2012fb,Jurgenson:2013yya,Roth:2011ar}\footnote{We note that the NNLO$_{\rm sat}$ interaction in Ref.~\cite{Ekstrom:2015rta} is a soft interaction that does not overbind medium-mass nuclei, and thus there are other aspects of the interactions that also come into play.}. In Ref.~\cite{Lahde:2013uqa}
this problem was overcome by adding an effective
repulsive four-nucleon force, whose strength was determined from the ground
state energy of $^{24}$Mg. As one can see from
Table~\ref{tab:4N}, including this, one achieves a very good description of
the ground state energies of all $\alpha$-cluster nuclei up to $^{28}$Si.
Another method to overcome this deficiency will be discussed in subsection~\ref{sec:alphagas}.

\begin{figure}[!t]
\centering
\includegraphics[width=0.42\columnwidth]{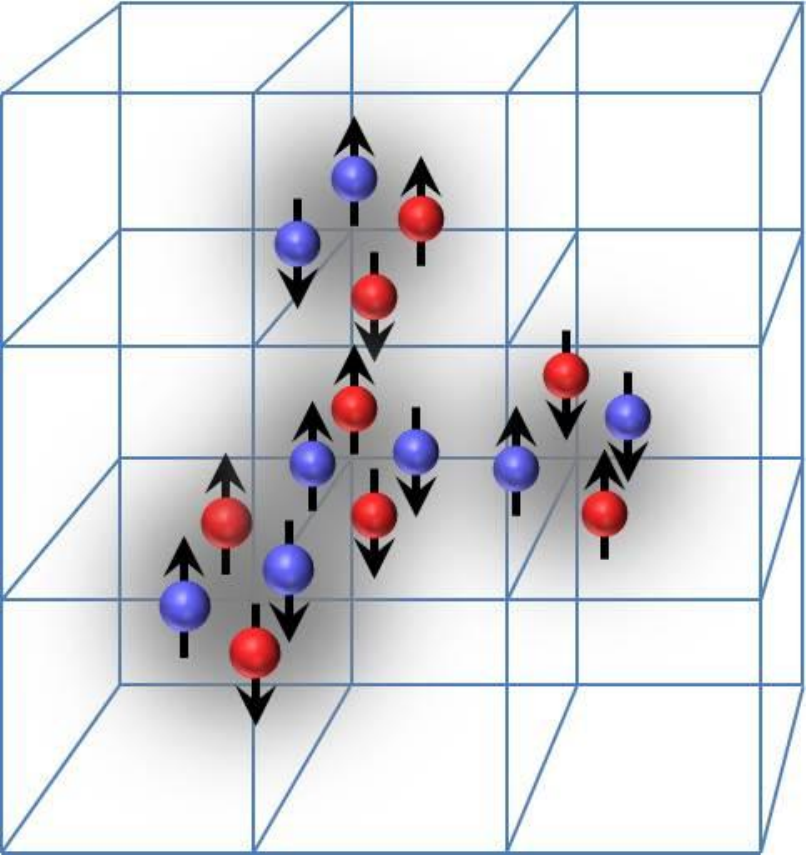}
\hspace{.5cm}
\includegraphics[width=0.42\columnwidth]{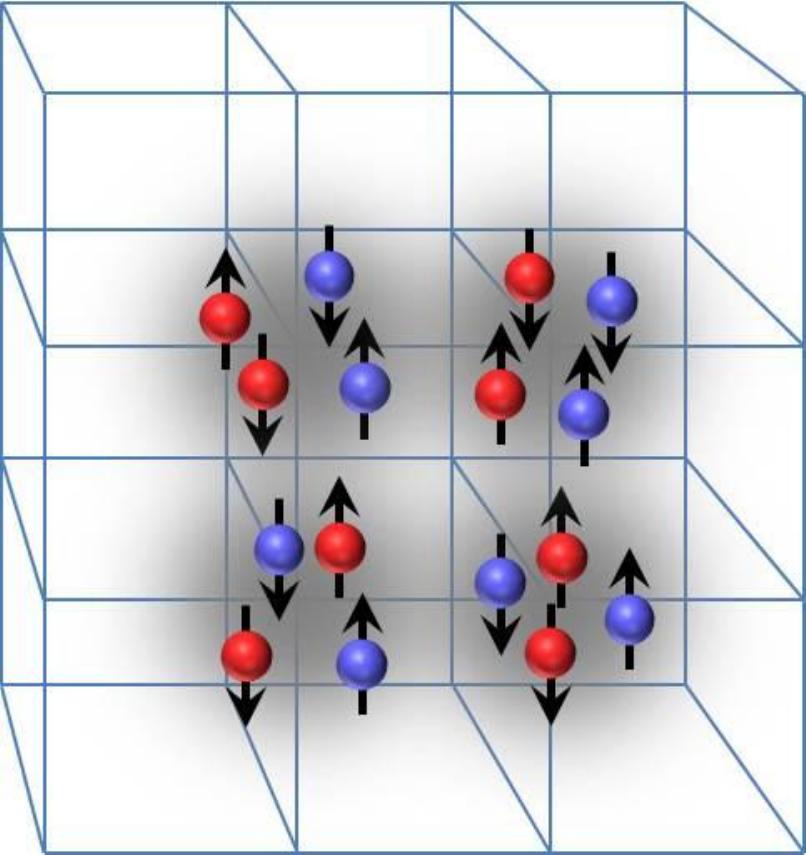}
\caption{(Color online) Schematic illustration of the $\alpha$-cluster initial states with
tetrahedral and planar configurations.
\label{fig:16O_1}}
\vspace{-5mm}
\end{figure}

The even-parity spectrum and structure of $^{16}$O has discussed in Ref.~\cite{Epelbaum:2013paa}.
The ground state has $J^\pi=0^+$ and
its energy is within 3\% of the empirical value, cf. Table~\ref{tab:4N}. One
finds a second $0^+$ state at $-123(2)\,$MeV
and the first $2^+$ state at the same energy. This is consistent with the
empirical values, $E(0^+_2) = -121.57\,$MeV and $E(2^+_1) = -120.70\,$MeV.
By measuring  four-nucleon correlations, one finds that the dominant cluster
configuration on the lattice is the tetrahedron,
see Fig.~\ref{fig:16O_1} (left), while the excited states have a strong overlap
with the planar-type configurations also shown in Fig.~\ref{fig:16O_1} (right).

This implies that the first $2^+$ state is a rotational excitation of the
first excited $0^+$. As in the case of $^{12}$C, the charge
radius of the ground state comes out too small, we get $r(0^+_1)^{\rm LO}
= 2.3(1)\,$fm, while the empirical value is $2.710(15)\,$fm.  This again
is due
to the overbinding at LO. If one rescales as described above,  one find that
the predictions for the $E2$ and
$E0$ transitions are in good agreement with the experimental values.
In particular, NLEFT is able to explain the empirical value of $B(E2,2^+_1
\to 0^+_2)$, which is $\simeq 30$ times larger than the Weisskopf
single-particle shell model estimate. This provides further confirmation
of the interpretation of the $2^+_1$ state as a rotational excitation of
the
$0^+_2$ state. Again, more detailed higher calculations of the electromagnetic
response of $^{16}$O within NLEFT are needed.

%%%%%%%%%%%%%%%%%%%%%%%%%%%%%%%%%%%%%%%%%%%%%%%%%%%%%%%%%%%%%%%%%%%%%%%%
%%%%%%%%%%%%%%%%%%%%%%%%%%%%%%%%%%%%%%%%%%%%%%%%%%%%%%%%%%%%%%%%%%%%%%%%%

\subsubsection{{\it Ab initio} alpha-alpha scattering}
\label{sec:alphaalpha}

\begin{figure}[!t]
\centering
\includegraphics[width=0.45\columnwidth]{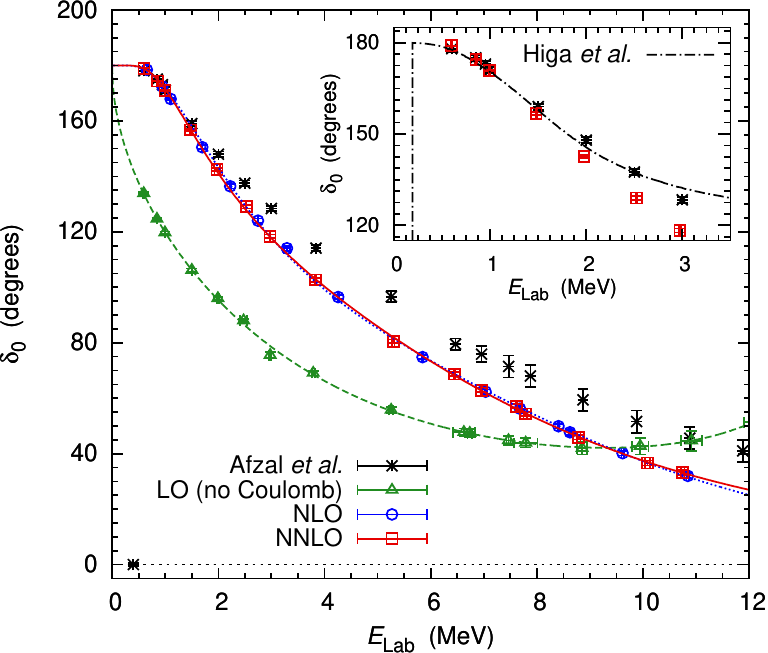}
\hspace{.2cm}
\includegraphics[width=0.45\columnwidth]{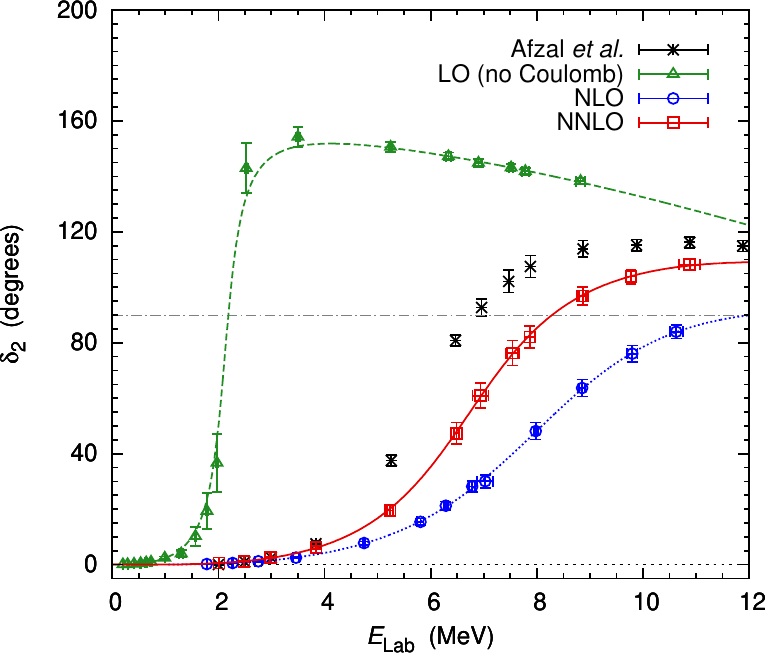}
\caption{(Color online) Left panel: $S$-wave phase shifts  for $\alpha$-$\alpha$ scattering
at LO, NLO and NNLO.
In the inset, the calculation based on an EFT with point-like $\alpha$-particles
is shown~\cite{Higa:2008dn}.
Right panel: $D$-wave phase shifts at LO, NLO and NNLO . The  experimental
data are from Refs.~ \cite{Heydenburg:1956zza,Nilson:1958zz,Afzal:1969zz}.
\label{fig:alalSD}}
\vspace{-5mm}
\end{figure}

Although there has been impressive progress in {\em ab initio} calculations
of nuclear scattering and reactions in the recent years,
see e.g. Refs.~\cite{Nollett:2006su,Navratil:2010jn,Navratil:2011zs,Hagen:2012rq,Orlandini:2013eya},
the aforementioned computational limits did so far not allow to consider
 astrophysically relevant reactions like elastic $\alpha$-$\alpha$,
$\alpha$-$^{12}$C or $^{12}$C-$^{12}$C scattering. A major step forward done
in these directions was reported in Ref.~\cite{Elhatisari:2015iga}.
There, the first {\em ab initio} calculation of $\alpha$-$\alpha$ scattering
based on chiral EFT and using the lattice formulation was
discussed. It is based on the same NNLO chiral Hamiltonian that was used
for the analysis of $^{12}$C and $^{16}$O, and so all
parameters had been determined before. Using the APM, the $S$- and $D$-wave
scattering phase shifts could be calculated as shown
in Fig.~\ref{fig:alalSD}. For more details on the actual computations, see
Ref.~\cite{Elhatisari:2015iga}.
In the chiral counting employed, the Coulomb interactions only appear at
NLO, therefore the LO curves
deviate significantly from the data. However, already at NLO one finds a
good description of the $S$-wave and a fair description for the $D$-wave.
While the NNLO corrections in the $S$-wave are very small, these corrections
bring the $D$-wave close to the data, although there is
still some room for improvement. The observed energy of the $S$-wave resonance
 is 0.09184~MeV
above threshold. For the lattice results, the ground state is found at  0.79(9)~MeV
below threshold at LO, and 0.11(1)~MeV below
threshold at both NLO and NNLO. The $D$-wave resonance is located at  $E_R
= 2.92(18)\,$MeV and $\Gamma = 1.34(50)\,$MeV~ \cite{Afzal:1969zz},
but there is some model-dependence as discussed in Ref.~\cite{Elhatisari:2015iga}.
In NLEFT, one finds at NNLO
$E_R = 3.27(12)\,$MeV and $\Gamma = 2.09(16)\,$MeV. This calculation can
be considered a benchmark for {\em ab initio} calculations of
nuclear scattering processes. Clearly, it needs to be refined by going to
higher orders and also working with finer lattices. However, arguably
the most significant finding of this investigation is the fact that the computing
time scales approximately quadratically with the number
of nucleons involved. Therefore, the computation of the ``holy grail'' of
nuclear astrophysics~\cite{Fowler:1984zz},
namely the reaction $\alpha + ^{12}{\rm C} \to ^{16}{\rm O} + \gamma$ at
 stellar  energies,  is in reach.

%%%%%%%%%%%%%%%%%%%%%%%%%%%%%%%%%%%%%%%%%%%%%%%%%%%%%%%%%%%%%%%%%%%%%%%%
%%%%%%%%%%%%%%%%%%%%%%%%%%%%%%%%%%%%%%%%%%%%%%%%%%%%%%%%%%%%%%%%%%%%%%%%%

\subsubsection{Nuclear binding near a quantum phase transition}
\label{sec:alphagas}

\begin{figure}[!t]
\centering
\includegraphics[width=0.99\columnwidth]{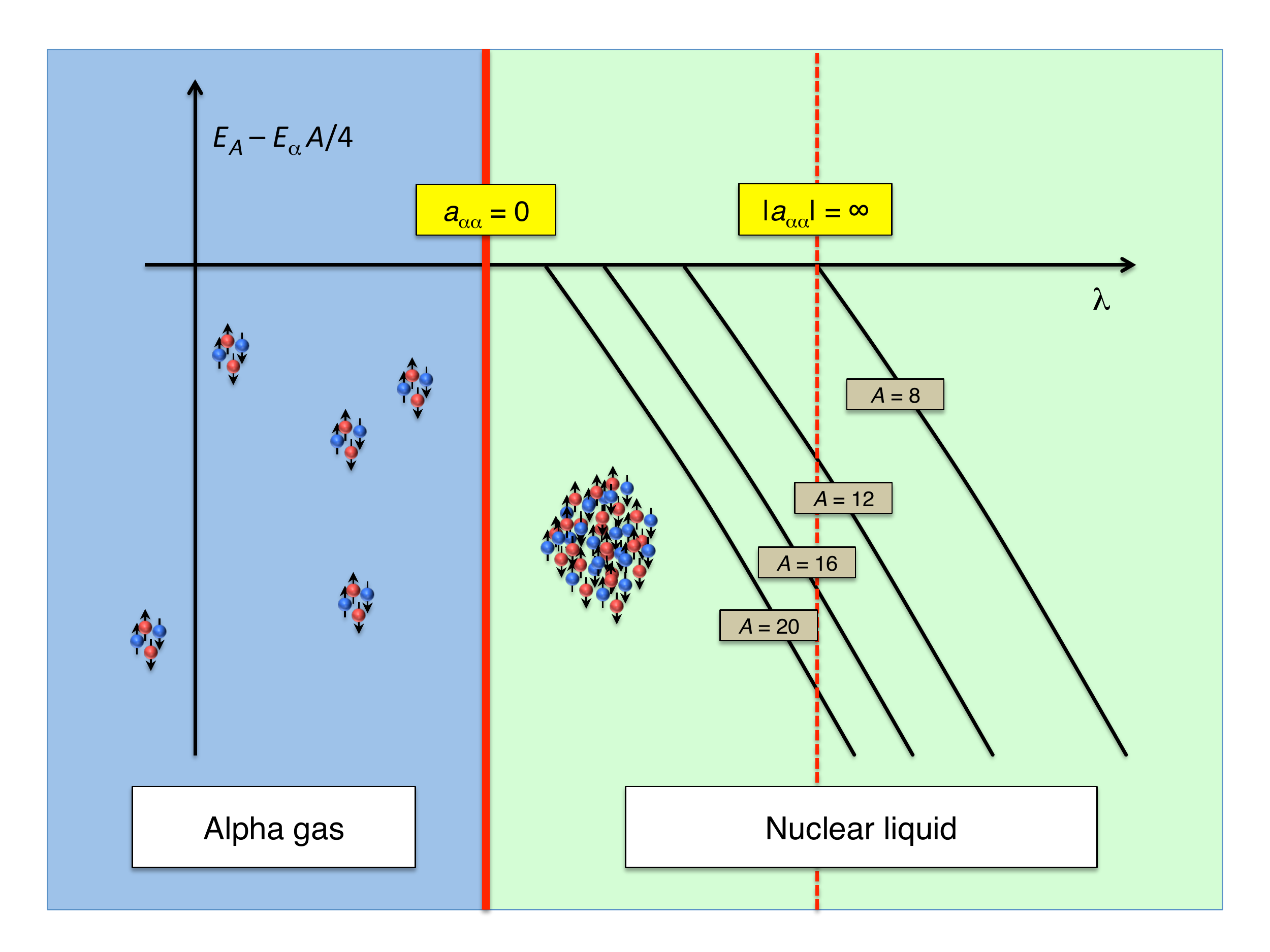}
\caption{(Color online) Zero-temperature phase diagram as a function of the parameter $\lambda$
in the strong interaction $V_{\lambda}=(1-\lambda)V_{\rm
A} + \lambda V_{\rm B}$.   A first-order quantum phase transition from a
Bose gas to nuclear liquid at the point
appears where the scattering length $a_{\alpha\alpha}$ crosses zero.  This is very close to the value $\lambda = 0$.  Also
shown are the $\alpha$-like nuclear ground state energies
$E_A$ for $A$ nucleons up to $A=20$ relative to the corresponding multi-alpha
threshold $E_{\alpha}A/4$. The last
$\alpha$-like nucleus to be bound is $^8$Be at the unitarity point where $|a_{\alpha\alpha}|=\infty$.  This
unitarity point is very close to the value $\lambda = 1$.  
\label{fig:phases}}
\vspace{-5mm}
\end{figure}

We had already seen that the NNLO forces overbind in larger nuclei, so higher-order calculations will be needed and eventually higher-body
forces might be required.  In Ref.~\cite{Elhatisari:2016owd} two ideas were
combined to give further insight into how nuclei are
formed and what role $\alpha$-clustering plays. First, the non-local smearing
 already discussed in subsection~\ref{sec:tech} was utilized to construct
two new LO interactions, motivated by the hope that the smearing would further
suppress the sign oscillations. Second, it was speculated that
determining  the LECs from fitting also to data from nucleus-nucleus scattering
might make the troublesome higher order corrections small.
To quantify these ideas, two different  LO interactions were constructed.
More  precisely, interaction~A consists of non-local short-range interactions
and one-pion exchange, supplemented by the
Coulomb interaction. Interaction~B has in addition local short-distance interactions.
Second, while interaction~A was entirely determined
by a fit to np scattering data and the deuteron binding energy, interaction~B
was in addition tuned to the $S$-wave $\alpha$-$\alpha$ phase shifts.
The resulting ground state energies for $^3$H, $^3$He, $^4$He and $\alpha$-cluster
nuclei are given in Table~\ref{tab:alpha_nuclei}. While
the results up to $^8$Be are similar, interaction~A fails to describe the
heavier nuclei, quite in contrast to interaction~B, which gives an
amazingly good description. From this one concludes that  $\alpha$-$\alpha$
scattering is quite sensitive to the degree of locality of the
nucleon-nucleon lattice interactions.  This can be understood from  the compactness of
the $\alpha$-particle wave function, as explained in more detail
in Ref.~\cite{Elhatisari:2016owd}.
%%%%%%%%%%%%%%%%%%%%%%%%%%%%%%%%%%%%%%%%%%%%%%%%%%%%%%%%%%%%%%%%%%%%%%%
\begin{table}[t!]
\centering{}%
\begin{tabular}{|c|c|c|c|c|}
\hline
Nucleus & A (LO) & B (LO) & A (LO + C) & B (LO + C)
\tabularnewline
\hline
$^{3}$H & $-7.82(5)$ & $-7.78(12)$ & $-7.82(5)$ & $-7.78(12)$
\tabularnewline
$^{3}$He & $-7.82(5)$ & $-7.78(12)$ & $-7.08(5)$ & $-7.09(12)$
\tabularnewline
$^{4}$He & $-29.36(4)$ & $-29.19(6)$ & $-28.62(4)$ & $-28.45(6)$
\tabularnewline
$^{8}$Be & $-58.61(14)$ & $-59.73(6)$ & $-56.51(14)$ & $-57.29(7)$
\tabularnewline
$^{12}$C & $-88.2(3)$ & $-95.0(5)$ & $-84.0(3)$ & $-89.9(5)$
\tabularnewline
$^{16}$O & $-117.5(6)$ & $-135.4(7)$ & $-110.5(6)$ & $-126.0(7)$
\tabularnewline
$^{20}$Ne & $-148(1)$ & $-178(1)$ & $-137(1)$ & $-164(1)$
\tabularnewline
\hline
\end{tabular}
\caption{Ground state energies of various nuclei for interactions A and B.
 Shown are  results for  LO
and LO + C(oulomb).  All energies are in units of MeV.}
\label{tab:alpha_nuclei}
\end{table}
%%%%%%%%%%%%%%%%%%%%%%%%%%%%%%%%%%%%%%%%%%%%%%%%%%%%%%%%%%%%%%%%%%%%%%%%
From Table~\ref{tab:alpha_nuclei} one further reads off that in the absence
of Coulomb interactions, the binding energy for a nucleus made of $N$
$\alpha$-particles is exactly $N$ times the $\alpha$ energy for interaction~A,
that is it describes a Bose-condensed gas of particles. These observations
allows one to draw interesting conclusions about the many-body limit. As
usual, the Coulomb interactions are switched off in order to take the many-body limit.  One can then define
 a one-parameter family of interactions via $V_{\lambda}=(1-\lambda)V_{\rm
A} + \lambda V_{\rm B}$. While the properties of the two, three, and four
nucleon systems vary only slightly with $\lambda$, the many-body ground state
of $V_{\lambda}$ undergoes a quantum phase transition from a
Bose-condensed gas to a nuclear liquid. The corresponding  zero temperature
phase diagram is sketched  in Fig.~\ref{fig:phases}. The phase transition
occurs when the $\alpha$-$\alpha$ $S$-wave scattering length $a_{\alpha\alpha}$
crosses zero, and the Bose gas collapses due to the attractive interactions
\cite{Stoof:1994zza,Kagan:1998a}. At slightly larger $\lambda$,
finite $\alpha$-type nuclei also become bound, starting with the largest
nuclei first. The last $\alpha$-like nucleus to be bound is $^8$Be
in the  so-called unitarity limit $|a_{\alpha\alpha}|=\infty$.  Superimposed
on the phase diagram,  the $\alpha$-like nuclear ground state
energies $E_A$ for $A$ nucleons up to $A=20$ relative to the corresponding
multi-alpha threshold $E_{\alpha}A/4$ are also depicted.  This shows that
by varying
$\lambda$, one can move  any  $\alpha$-cluster state up or down with respect
to the $\alpha$ separation thresholds. This can be used as a
new window to view the structure of these exotic nuclear states. In particular,
this allows  one to  continuously connect the Hoyle state wave function
without Coulomb interactions to a universal Efimov trimer \cite{Efimov:1971zz,Braaten:2004rn,Kraemer:2006nat}.

Another interesting system is  the second $0^+$ state of $^{16}$O , which
should be continuously connected to a universal Efimov
tetramer \cite{Hammer:2006ct,Kraemer:2006nat,vonStecher:2009a}.  In summary,
the main findings of this work are that the
$\alpha$-$\alpha$ interaction is a key control parameter which determines
whether the ground state  of a many-nucleon system is a
Bose-condensed gas of $\alpha$-particles or a nuclear liquid. The proximity
of this first-order quantum phase transition may explain
why seemingly similar nuclear interactions can produce very different results
in {\em ab initio} nuclear structure calculations. These conclusions
need to be solidified by more detailed higher order calculations. Similar results have been found in Ref.~\cite{Ebran:2012ww,Ebran:2012hi,Ebran:2014boa,Ebran:2014pda,Ebran:2015kba} using density functional methods.

One might ask what the dependence on $\lambda$ means for future nuclear structure calculations for heavier systems using chiral effective field theory.  It suggests that the order-by-order convergence of chiral effective field theory might benefit from some optimization of the forces and regulators used in the chiral interactions.  This need for optimization may not be visible in few-nucleon observables until very high-orders in chiral effective field theory.  But the dependence on $\lambda$ appears as a leading-order effect in the framework of cluster effective field theory for two low-energy $\alpha$-particles.  This suggests that some acceleration of the convergence of chiral effective field theory in heavier systems might be possible by making links to cluster effective field theory. \\  

%%%%%%%%%%%%%%%%%%%%%%%%%%%%%%%%%%%%%%%%%%%%%%%%%%%%%%%%%%%%%%%%%%%%%%%%

%\begin{figure}[!t]
%\centering
%\includegraphics[width=0.44\columnwidth]{Ulf/corelDeltaBEv5.pdf}
%\hspace{.2cm}
%\includegraphics[width=0.47\columnwidth]{Ulf/hoyle_band_new.pdf}
%\caption{Left panel:  Sensitivities of $\Delta E_h^{}$, $\Delta E_b^{}$
% and $\varepsilon$
%to changes in $M_\pi^{}$, as a function of $K_{E_4^{}}^\pi$ under independent

%variation of $\bar A_s^{}$ and $\bar A_t^{}$ over the range $\{-1 \ldots % 1\}$.
%The bands correspond to $\Delta E_b^{}$,
%$\varepsilon$ and  $\Delta E_h^{}$ in clockwise order.
%Right panel: ``Survivability bands'' for carbon-oxygen based life from
 % Eq.~(\ref{final_res}), due to  $0.5\%$ (broad outer band), $1\%$ (medium
 % band) and $5\%$ (narrow inner band) changes in $m_q^{}$ in terms of the
  %input parameters $\bar A_s^{}$ and $\bar A_t^{}$.
%\label{fig:fate}}
%\vspace{-5mm}
%\end{figure}

\subsubsection{Clustering in neutron-rich nuclei}
\label{sec:neutron-rich}

In addition to the discussion of the quantum phase transition, another development in Ref.~\cite{Elhatisari:2016owd} was the use of non-local interactions to reduce sign oscillations in the lattice Monte Carlo simulations.  This idea was utilized in Ref.~\cite{Elhatisari:2017eno} to perform lattice simulations of neutron-rich nuclei.  While this work only considered interactions at leading order in chiral effective field theory, the  ground state energies of the hydrogen, helium, beryllium, carbon, and oxygen isotopes could be reproduced with an error of 0.7~MeV per nucleon or less with only three adjustable parameters.

In Ref.~\cite{Elhatisari:2017eno} a new model-independent method was also introduced for measuring clustering in nuclei using localized three- and four-nucleon operators.  Let $\rho(\bf
n)$ be the total nucleon density operator on lattice site $\bf
n$. $\rho_3$ is defined as the expectation value of $:\rho^3({\bf n})/3!:$
summed over $\bf n$, where the $::$ symbols denote normal-ordering where all annihilation operators
are moved to the right and all creation operators are moved to the left. Similarly $\rho_4$ is defined
as the expectation value of $:\rho^4({\bf n})/4!:$
summed over $\bf n$.

Although the expectation values $\rho_3$ and $\rho_4$ depend
on the manner in which short-distance physics is regularized, the leading part of this dependence is an overall factor which does not depend on the nucleus being considered.  So if $\rho_{3,\alpha}$
and $\rho_{4,\alpha}$ are the corresponding values for the $\alpha$-particle,
then the ratios $\rho_3/\rho_{3,\alpha}$ and $\rho_4/\rho_{4,\alpha}$ are
free from short-distance divergences and are model-independent quantities
up to contributions from higher-dimensional operators in an operator product
expansion. In Ref.~\cite{Elhatisari:2017eno} the quantities
$\rho_3/\rho_{3,\alpha}$
and $\rho_4/\rho_{4,\alpha}$ were computed and used the quantify the amount of $\alpha$-clustering in the helium, beryllium, carbon, and
oxygen isotopes in  a model-independent manner.  It was observed that these ratios $\rho_3/\rho_{3,\alpha}$
and $\rho_4/\rho_{4,\alpha}$ could be used to probe the shape of the $\alpha$-clusters as well as the amount of quantum entanglement of nucleons from different $\alpha$-clusters.

Another development in Ref.~\cite{Elhatisari:2017eno}
was the determination of $\alpha$-cluster correlations in the carbon isotopes $^{12}$C, $^{14}$C, and $^{16}$C by measuring density correlations among the three spin-up protons.  This approach relies on the fact that, on average, there is only one spin-up proton within each $\alpha$-cluster.  The similarities among the $^{12}$C, $^{14}$C, and $^{16}$C $\alpha$-cluster geometries suggest that there should be $\alpha$-cluster states in $^{14}$C and $^{16}$C that are analogs of the $\alpha$-cluster states in $^{12}$C.  For example, the bound $0^+_2$ state at 6.59~MeV above the ground state of $^{14}$C could be a bound-state analog to the Hoyle
state resonance in $^{12}$C.

\section{Summary and outlook}

We have presented a review on the current status and understanding of microscopic
clustering in nuclei.  We began with a history of the field and then discussed recent
experimental
results on $\alpha$-conjugate nuclei, molecular structures in neutron-rich
nuclei, and constraints for {\it ab initio} theory. There has been  impressive progress in recent years clarifying clustering phenomena in $^{8,9,10}$Be, $^{10,12,13,14}$C, $^{16}$O, and several other nuclei.  However, many more precision measurements are needed, and these will provide vital benchmarks for first principles calculations. In addition to rotational bands, form factors, electromagnetic transition strengths,
 decays, and reaction cross sections, model-independent assessments of clustering
such as ANCs are also very useful in making connections to {\it ab initio}
theory.

There  are also new opportunities for discovery in exploring clustering phenomena
over a wide range of nuclear systems, from light to heavy nuclei and from the proton drip line to the neutron drip line.
One of the fundamental questions of the field is understanding how
prevalent  nuclear clustering is across the nuclear chart. This includes systems where clustering is more subtly expressed and mixed with other effects such as particle-hole excitations. Having a large empirical database of nuclear phenomena will shed light on the control parameters for nuclear cluster formation and stability.

On the theoretical side we have discussed methods used to study microscopic clustering. We reviewed the  resonating
group and generator coordinate methods, antisymmetrized molecular dynamics, Tohsaki-Horiuchi-Schuck-R{\"o}pke
wave function and container model, no-core shell model, continuum quantum Monte Carlo, and lattice effective field theory.

While there have been many significant advances in the past decade, the field of microscopic nuclear clustering theory is now just entering the era of precision calculations.  The future holds many opportunities for improvement in theory, methods or algorithms, and analysis. With the rapid growth of {\it ab initio} nuclear theory in the past few years, one great challenge for the field is to describe nuclear clustering from first principles with controlled systematic errors.  This is no easy task as recent studies have found that the interactions between nuclear clusters are very sensitive to details of the nuclear forces.

One area where all theoretical groups
may choose to invest time and effort
is on error quantification and the systematic reduction of errors.  One question relevant to all groups is how results on nuclear clustering depend on the microscopic nuclear forces utilized.  The follow-up question is how this difference can be systematically reduced by including the relevant missing physics.  For lattice calculations another important question is to estimate and reduce the size of lattice discretization errors. For methods based on finite basis truncation or variational parameter optimization, a key question is the residual dependence on the choice of truncated space or variational ansatz.  For continuum quantum Monte Carlo, the analogous question would be the dependence on wave function constraints and the trial wave function.

In addition to reproducing observed experimental data, another challenge for theoretical calculations is to compute model-independent observables that provide a quantitative measure of clustering and also serve as standard benchmarks for all different theoretical approaches. We have already mentioned ANCs for shallow bound states, but other model-independent observables could also be computed and defined for resonances as well.

We hope that our review captures some of
the excitement of the growing and vibrant field of nuclear clustering. With many open questions and challenges still remaining, we anticipate fascinating new chapters to
be written in the coming years.

%%%%%

\section*{Acknowledgments}
We acknowledge the work of our collaborators   J.~Alarc{\'o}n, D.~Du,
S.~Elhatisari, E.~Epelbaum, Y.~Funaki, N.~Klein, H.~Krebs, T.~L{\"a}hde, N.~Li, B.~Lu, T.~Luu, M.~Lyu, Z.~Ren, A.~Rokash, G.~R\"opke,
 P.~Schuck, A.~Tohsaki, C.~Xu, T.~Yamada, and B.~Zhou. We also acknowledge helpful
discussions with K. Launey, P. Navr{\'a}til, W. Nazarewicz, T. Neff, and S. Pieper.  Partial financial
support was provided by JSPS KAKENHI Grant Number 26400270, JSPS KAKENHI
Grant Number JP16K05351, Deutsche Forschungsgemeinschaft (Sino-German CRC
110), the Helmholtz Association
(Contract No. VH-VI-417), BMBF (Grant No. 05P15PCFN1), the U.S. Department
of Energy (DE-FG02-03ER41260), U.S. National Science Foundation
grant No. PHY-1307453 and by The Chinese Academy of Sciences  (CAS) President's
International Fellowship Initiative (PIFI) grant no. 2017VMA0025.  Computing resources provided by the J\"ulich
Supercomputing Centre at Forschungszentrum J{\"u}lich and RWTH Aachen.

\bibliography{full_article}

\end{document}